\documentclass[aps,twocolumn,pra,superscriptaddress,amsmath,amssymb]{revtex4-2} %% for PRL
\usepackage{dcolumn}
\usepackage{amsmath}
\usepackage{mathrsfs}
\usepackage{txfonts}
\usepackage{bm}
\usepackage[T1]{fontenc}
\usepackage{xspace}
\usepackage{ulem}
\usepackage{comment}
\usepackage{braket}
\usepackage{lipsum}
\usepackage{bbold}
\usepackage{mathtools}
\newcommand{\pdiff}[2]{\frac{\partial {#1}}{\partial {#2}}}
\newcommand{\fdiff}[2]{\frac{\delta {#1}}{\delta {#2}}}
\newcommand{\diff}[2]{\frac{d {#1}}{d {#2}}}

\ifx\pdfoutput\undefined
\usepackage[dvipdfmx]{graphicx}
\usepackage[dvipdfmx]{hyperref}
\usepackage[dvipdfmx]{color}
\usepackage[dvipdfmx]{xcolor}
\else
\usepackage{graphicx}
\usepackage{hyperref}
\usepackage{color}
\usepackage{xcolor}
\usepackage[version=3]{mhchem}
\fi
\hypersetup{
        colorlinks=true,
        citecolor=blue,
        urlcolor=blue,
        linkcolor=blue
}

\begin{document}
\let\emph\textit

\title{
    Formulation of spin Nernst effect for spin-nonconserving insulating magnets
}
\author{Shinnosuke Koyama}
\author{Joji Nasu}
\affiliation{
  Department of Physics, Tohoku University, Sendai, Miyagi 980-8578, Japan
}

\date{\today}
\begin{abstract}
The spin Nernst effect, an antisymmetric response of a spin current to a temperature gradient, has attracted attention as spin transport phenomenon arising from the topologically nontrivial band structure of carriers.  
This effect can occur not only in itinerant electron systems but also in localized electron systems that emerge due to electronic correlations.  
In such systems, elementary excitations behaving as bosons govern spin transport, and magnetic interactions originating from spin-orbit coupling can induce a topologically nontrivial band structure.  
However, such magnetic interactions can potentially break spin conservation, thereby preventing conventional spin currents from being conserved.  
In this study, starting from a general localized electron model, we formulate the spin Nernst effect in terms of a conserved spin current that remains applicable even in spin-nonconserving systems.  
To address the torque term appearing in the conserved spin current, we adopt semiclassical theory and derive an expression for the spin Nernst coefficient for bosonic systems.  
This coefficient consists of two terms originating from the Berry curvature and the quantum metric, which are distinctly different from the spin Berry curvature obtained in an approach that neglects the torque term.  
We apply our framework to two specific quantum spin models, the Kitaev-Heisenberg and Shastry-Sutherland models, and calculate the temperature dependence of the spin Nernst coefficient.  
We find that this quantity significantly differs from that obtained by neglecting the torque term in both models.  
Furthermore, we clarify that the impact of the torque term on the spin Nernst effect strongly depends on the model parameters, suggesting that our formulation based on the conserved spin current is essential for understanding this effect in insulating magnets.  
\end{abstract}
\maketitle

%%%%%%%%%%%
% Introduction

\section{Introduction}
\label{introduction}

In Mott insulators, where strong electron correlations prevent electrons from moving freely, localized degrees of freedom, such as spins and orbitals in each atom, play a crucial role in determining their low-energy properties.
At low temperatures, localized degrees of freedom exhibit various ordered states with broken symmetries, such as spin and orbital orders. Collective excitations from ordered states govern the excitation spectra and transport properties of Mott insulators.
The collective excitations are described by bosonic quasiparticles, such as magnons and orbitons, which are distinct from the original properties of electrons as fermions.
Nevertheless, peculiar states exhibited in bosonic quasiparticle systems are often found to have analogies to electronic systems. For example, the magnon band structure in insulating magnets can exhibit topologically nontrivial properties similar to that of the electronic band structure, and the magnon Berry curvature contributes to off-diagonal transport, analogous to the electronic Hall effect~\cite{katsura2010,matsumoto2011,mcclarty2022,zhang2024thermal}.
Since such quasiparticles do not possess charge degrees of freedom, heat transport can be a key observable for investigating the topological nature of bosonic quasiparticles.
In this context, the thermal Hall effect, characterized by the transverse heat flow in response to a temperature gradient, has been extensively studied in various systems with spin-orbit coupling, such as localized spin systems on pyrochlore and kagome lattices ~\cite{onose2010,ideue2012,hirschberger2015_science,hirschberger2015_prl,Akazawa2020} and Kitaev spin systems on a honeycomb lattice~\cite{nasu2017,kasahara2018_prl,kasahara2018_nature,yamashita2020,bruin2022,Hentrich2019,chern2021_prl,zhang2021_prb,czajka2023,hwang2022identification,li2022_prb,li2025}.
Furthermore, it has been suggested that triplons, which are bosonic quasiparticles associated with spin-triplet excitations from a spin-singlet state, can also exhibit the thermal Hall effect~\cite{romhanyi2015,mcclarty2017topological,Malki2017,Bhowmick2021,sun2021_triplon,suetsugu2022,esaki2024,Buzo2024}.
Thus, the study on the topological properties of bosonic quasiparticles has been actively pursued in various localized electron systems, in addition to phonon systems~\cite{sheng2006,zhang2010,zhang2011,qin2012,chen2022large,Lefrancois2022,chen2024planar,Sharma2024,Oh2025}.

Recently, the spin Nernst effect has attracted considerable attention as another topological phenomenon in insulating magnets.
This effect is introduced as the generation of a transverse spin flow by a temperature gradient and has been extensively studied for potential applications in spintronic devices.
The spin Nernst effect is considered to occur when the spin Berry curvature is nonzero, rather than the conventional Berry curvature~\cite{zyuzin2016_prl,li2020_prr,zhang2022rev}.
Unlike the thermal Hall effect, this phenomenon can arise even when time-reversal symmetry is preserved~\cite{zyuzin2016_prl,cheng2016}.
For example, since antiferromagnets generally do not break time-reversal symmetry, they do not exhibit the thermal Hall effect; however, they can potentially exhibit the spin Nernst effect.
The spin Nernst effect has been theoretically studied in several localized spin systems, including collinear and noncollinear antiferromagnets~\cite{zyuzin2016_prl,cheng2016,lee2018,li2020_prr,go2022}, in the presence of the Dzyaloshinskii-Moriya (DM) interaction.

In theoretical studies on the spin Nernst effect, it is formulated by evaluating the spin current in the bulk.
However, in experiments, the spin current cannot be directly measured, and the spin Nernst signal is usually detected indirectly using the inverse spin Hall effect~\cite{shiomi2017,zhang2022rev}.
This suggests that the spin accumulation at the edge of the sample is essential for the spin Nernst effect~\cite{Shitade2022,Shitade2022-Nernst}.
More importantly, the spin current is generally not a conserved flow, particularly in the presence of spin-orbit coupling.
The nonconserving nature of the spin current makes it difficult to define the spin current in the bulk; for example, such a nonconserved spin current can be nonzero even in the equilibrium state without a thermal gradient~\cite{rashba2003}.
Nevertheless, the formulation based on the conventional nonconserved spin current~\cite{li2020_prr} has been widely used for calculating the spin Nernst effect in insulating magnets~\cite{ma2021,fujiwara2022,go2022,to2023,klogetvedt2023,lu2024}.
To avoid such issues arising from the nonconserving nature of the spin current, another approach has been proposed, in which the spin current is introduced so as to satisfy the equation of continuity; this is called the conserved spin current~\cite{shi2006}.
In this approach, an additional contribution, referred to as a torque term, is added to the conventional current form.
This torque term originates from a nonlinear effect of the spatial gradient, making it nontrivial to use Luttinger's heat-gravity correspondence~\cite{luttinger1964} to evaluate the torque term in the presence of thermal gradients.
To address nonlinear response effects in electronic systems, the semiclassical theory and quantum kinetic theory have been developed for fermions~\cite{xiao2021conserved}.
However, theoretical frameworks for bosonic systems are still lacking, presumably due to the nonunitary nature of bosons in the Bogoliubov-de Gennes (BdG) representation, which does not conserve particle number.

In this paper, we formulate the spin Nernst effect for bosonic systems as a thermal response of the conserved spin current and apply this formulation to specific quantum spin models, whose elementary excitations are described within the linear flavor-wave theory, to demonstrate the applicability of our theory.
To this end, we first develop a semiclassical theory for bosonic systems to evaluate the expectation value of a local physical quantity up to the second order of spatial gradients.
Using the formula for the local physical quantity, we expand the spin current density up to the first order of the spatial gradient and the torque term up to the second order. Consequently, we derive the conserved spin current up to the first order of the thermal gradient by focusing on contributions independent of the relaxation time.
We find that the coefficient consists of two terms: the Berry curvature and an additional term, which is linked to the quantum metric connecting momentum space and the source field yielding the torque term.
This contrasts with the coefficient obtained by the conventional theoretical framework, which is given by the spin Berry curvature.
The additional term is given by the momentum derivative of the quantum metric. 
We show that this derivative can be avoided, which is crucial for the numerical evaluation of the spin Nernst effect.
We apply our theory to two spin-1/2 spin models: the Kitaev-Heisenberg model on a honeycomb lattice and the Shastry-Sutherland model with DM interactions.
We demonstrate that the spin Nernst coefficient significantly differs from the results obtained by the conventional theory, but the differences cannot be systematically understood as effects of introducing the torque term.
This suggests that the torque term is essential for evaluating the spin Nernst effect in bosonic systems.
Since our theory is based on a general form of localized electron models or free-boson systems, it can be applied to various bosonic systems, such as phonon systems, multipole systems, and systems with hybrid bosonic excitations originating from electron-phonon coupling, to investigate the excitation and topological properties of bosonic quasiparticles.

This paper is organized as follows.  
In the next section, we introduce a general Hamiltonian for a localized electron system with multiple degrees of freedom at each site and derive the representation of the spin Nernst conductivity.  
Section~\ref{sec:LFWT} explains the linear flavor-wave theory used to describe bosons as elementary excitations.  
The continuum limit of the bosonic system is introduced in Sec.~\ref{sec:method-continuum}.
In this limit, we define a conserved current by considering the equation of continuity in Sec.~\ref{sec:eq-cont}.
In Sec.~\ref{sec:semiclassical}, we formulate semiclassical theory for bosonic systems and derive the expectation value of a local physical quantity up to the second order of the spatial gradient.
Using this expression, we evaluate the conserved current, including a contribution from the torque term, in Sec.~\ref{sec:eval-current}.
In Sec.~\ref{sec:cond-current}, we discuss a condition for introducing a conserved current.
The expression of the spin Nernst coefficient is derived in Sec.~\ref{sec:snc}.
Section~\ref{sec:numerical-calc} presents the numerical results for the spin Nernst coefficient in two specific quantum spin models.
The results for the Kitaev-Heisenberg model and the Shastry-Sutherland model are given in Secs.~\ref{sec:KHmodel} and \ref{sec:SSmodel}, respectively.
Finally, Sec.~\ref{sec:summary} is devoted to the summary and future perspective.

\section{Model and Method}
\label{sec:Model-method}

\subsection{Linear Flavor-wave theory}
\label{sec:LFWT}

We start from the following general Hamiltonian for a localized electron system:
\begin{align}
  \label{eq:H_origin}
  \mathcal{H}
  =
  \frac{1}{2} \sum_{i,j}\sum_{\alpha\beta} J_{ij}^{\alpha\beta} O_{i}^{\alpha} O_{j}^{\beta}
  -
  \sum_{i}\sum_{\alpha} h_{i}^{\alpha} O_{i}^{\alpha},
\end{align}
where $O_{i}^{\alpha}$ is a local operator defined at site $i$ with component $\alpha$, and its dimension is denoted as $\mathscr{N}$.
The first term in Eq.~\eqref{eq:H_origin} represents the interaction between local operators $O_{i}^{\alpha}$ and $O_{j}^{\beta}$ with the interaction coefficient $J_{ij}^{\alpha\beta}$, while the second term corresponds to the one-body part with the local field $h_{i}^{\alpha}$.
In the following, we briefly review the mean-field theory and the flavor-wave theory for the Hamiltonian given in Eq.~\eqref{eq:H_origin}.
Details can be found in Refs.~\cite{nasu2021,koyama2021,nasu2022,koyama2023_prb,koyama2023_NPSM,koyama2024}.
In mean-field theory, the ground state is assumed to be a direct product of local states with a periodicity such that the unit cell includes $M$ sites.
Under this assumption, the local operator is decomposed as
\begin{align}
  \label{eq:O_decomp}
  O_{i}^{\alpha} = \delta O_{i}^{\alpha} + \braket{O^{\alpha}}_{l},
\end{align}
where $\braket{O^{\alpha}}_{l}$ represents the local expectation value of the operator $O_{i}^{\alpha}$ at site $i$, with $l=1,2,\dots, M$ denoting the sublattice index to which site $i$ belongs.
Using Eq.~\eqref{eq:O_decomp}, the Hamiltonian $\mathcal{H}$ can be decomposed into the mean-field Hamiltonian $\mathcal{H}^{\mathrm{MF}}$ and the fluctuation term $\mathcal{H}'$ as $\mathcal{H} = \mathcal{H}^{\mathrm{MF}} + \mathcal{H}'$, which are defined as 
\begin{align}
  \label{eq:HMF}
  \mathcal{H}^{\mathrm{MF}} 
  &= 
  \sum_{i} \mathcal{H}^{\mathrm{MF}}_{i} 
  - \frac{1}{2} \sum_{ll'}^{M}\sum_{i\in l}^{N_u}\sum_{j\in l'}^{N_u} J_{ij}^{\alpha\beta} \braket{O^{\alpha}}_{l} \braket{O^{\beta}}_{l'},
  \\
  \label{eq:H'}
  \mathcal{H}'
  &= 
  \frac{1}{2}\sum_{i,j} J_{ij}^{\alpha\beta} \delta O_{i}^{\alpha} \delta O_{j}^{\beta},
\end{align}
where $N_u$ is the total number of unit cells, and the local mean-field Hamiltonian $\mathcal{H}^{\mathrm{MF}}_{i}$ is given by
\begin{align}
  \label{eq:HiMF}
  \mathcal{H}_{i}^{\mathrm{MF}}
  =
  \sum_{\alpha} 
  \left(
    \sum_{l'}^{M}\sum_{j\in l'}^{N_u} \sum_{\beta} J_{ij}^{\alpha\beta} \braket{O^{\beta}}_{l'} - h_{i}^{\alpha}
  \right)
  O_{i}^{\alpha}.
\end{align}
This local Hamiltonian, which is represented by an $\mathscr{N}\times\mathscr{N}$ matrix, can be easily diagonalized, and the eigenvalues and the corresponding eigenstates are denoted as $E_{n}^{l}$ and $\ket{n; i}$, respectively, with $n=0,1,\dots,\mathscr{N}-1$.
The local operator $\delta O_{i}^{\alpha}$ is expanded using the eigenstates as
$\delta O_{i}^{\alpha}
=
\sum_{nn'=0}^{\mathscr{N}-1} X_{i}^{nn'} \left( \delta O_{l}^{\alpha} \right)_{nn'}$, where 
$
X_{i}^{nn'} = \ket{n;i}\bra{n';i}$ and 
$( \delta O_{l}^{\alpha} )_{nn'} = \braket{n;i|\delta O^{\alpha}|n';i}$.
The local projection operator $X_{i}^{nn'}$ satisfies the following commutation relation:
$[ X_{i}^{n_1n_2}, X_{j}^{n_3n_4} ]
=
\delta_{ij} \left( X_{i}^{n_1n_4}\delta_{n_2n_3} - X_{i}^{n_3n_2} \delta_{n_1n_4} \right)$.
This relation can be reproduced by introducing $\mathscr{N}-1$ bosons with the creation (annihilation) operator $a_{mi}^\dagger \ (a_{mi})$ at each site, which is known as the generalized Holstein-Primakoff transformation~\cite{joshi1999,kusunose2001,nasu2021,koyama2021,nasu2022,koyama2023_prb,koyama2023_NPSM}.
In this transformation, the local projection operators are expressed as
$X_{i}^{m0}
=
a_{mi}^\dagger \left( \mathcal{S} - \sum_{n=1}^{\mathscr{N} - 1} a_{ni}^\dagger a_{ni}\right)^{1/2}$, $X_{i}^{0m}=(X_{i}^{m0})^\dagger$, and
$X_{i}^{mm'} = a_{mi}^\dagger a_{m'i}$.
Here, $\mathcal{S}$ is defined as $\mathcal{S} = X_{i}^{00} + \sum_{m=1}^{\mathscr{N} - 1} a_{mi}^\dagger a_{mi}$.
Since a boson must occupy one of the local eigenstates at each site, $\mathcal{S} = 1$ should be satisfied.

In the case where the occupation of the local excited states is much smaller than that of the local ground state, the root in $X_{i}^{m0}$ can be expanded as $1/\mathcal{S}$, and the Hamiltonian $\mathcal{H}'$ is expanded up to terms including the products of two bosonic operators, which corresponds to the linear flavor-wave approximation~\cite{nasu2021,koyama2023_prb}.
In this approximation, the original Hamiltonian $\mathcal{H}$ is reduced to a bilinear form of the bosonic operators $\mathcal{H}_{0}$, which is symbolically expressed as
\begin{align}
  \label{eq:Hamil-A-real}
  \mathcal{H}_{0} = \frac{1}{2} \sum_{uu'}^{N_{u}} \mathcal{A}_{u}^\dagger {H}_{0,uu'}^{} \mathcal{A}_{u'}^{},
\end{align}
where ${H}_{0,uu'}$ is a $2N\times 2N$ matrix with $N=M(\mathscr{N}-1)$, and $\mathcal{A}_{u}^\dagger$ is a $2N$-dimensional bosonic operator vector defined as
\begin{align}\label{eq:definition_A}
  \mathcal{A}_{u}^{\dagger} = \left(a_{u,1}^{\dagger}, a_{u,2}^{\dagger}, \dots 
  , a_{u,N}^{\dagger} 
  , a_{u,1}^{}, a_{u,2}^{}, \dots, a_{u,N}^{} \right).
\end{align}
Note that under translational symmetry, ${H}_{0,uu'}$ depends on $\bm{r}_u-\bm{r}_{u'}$, where $\bm{r}_u$ is the representative position of the unit cell $u$.
In the original definition of the bosonic operator $a_{mi}$, it is specified by the labels of the local excited state $m=1,2,\dots,\mathscr{N}-1$ and site $i$.
Since each site is specified by the sublattice $l$ and the unit cell $u$, each bosonic operator is labeled by three indices $(m,l,u)$.
In Eq.~\eqref{eq:definition_A}, we introduce the composite label $s=(l,m)$, whose total number is $N$, and each bosonic operator is relabeled as $a_{u,s}$ using the unit cell index and this composite label.

In this study, we investigate spin currents in insulating magnets.
To describe the flow of spin within the framework of linear flavor-wave theory, it is necessary to express the spin operator in terms of bosonic operators.
Here, we consider a physical quantity expressed as the sum of local operators, $\sum_{i} O_i^\alpha$, for the component $\alpha$, or more generally, given by $O_{\rm total}=\sum_{i\alpha} c_\alpha  O_i^\alpha$, where $c_\alpha$ is a coefficient.
This quantity is expanded using the projection operator $X_{i}^{nn'}$ as $O_{\rm total}=\sum_{i\alpha}c_\alpha\sum_{nn'=0}^{\mathscr{N}-1} X_{i}^{nn'} \left( \delta O_{l}^{\alpha} \right)_{nn'}+ {\rm const.}$
In this representation, for simplicity, we neglect the off-diagonal components of $X_{i}^{nn'}$~\cite{li2020_prr}, which is justified when $[O_i^\alpha,\mathcal{H}^{\mathrm{MF}}_{i}]$ is sufficiently small at each site.
For a quantum spin model, this simplification corresponds to considering only the spin component that is parallel or antiparallel to the local spin direction in the mean-field ground state at each site.
By applying the generalized Holstein-Primakoff theory up to the bilinear form of bosonic operators, $O_{\rm total}$ is expressed as
\begin{align}\label{eq:O_total}
  O_{\rm total}&=
  \sum_{l}^{M} \sum_{i \in l}^{N_u} \sum_{m=1}^{\mathscr{N}-1}\sum_\alpha
   c_\alpha(\delta O_l^\alpha)_{mm}a^\dagger_{mi} a_{mi} + \mathrm{const.}\notag\\
 &=  \frac{1}{2}\sum_{u}^{N_u}
  \mathcal{A}^\dagger_{u} \hat{O} \mathcal{A}_{u}
  + \mathrm{const.}
\end{align}
Here, $\hat{O} = \bm{1}_{2\times 2} \otimes \bar{O}$ is a $2N\times 2N$ diagonal matrix, where $\bm{1}_{2\times 2}$ is a $2\times 2$ unit matrix, and $\bar{O}$ is an $N\times N$ diagonal matrix whose components $\bar{O}_{ss}$ are given as $(\delta O^{\alpha}_{l})_{mm}$ with $s=(l,m)$, or more generally, as a linear combination of these components with different $\alpha$.
Note that $\hat{O}$ is independent of the unit cell position $u$.
From the definition of $\hat{O}$, it satisfies the relation 
\begin{align}\label{eq:PHS-O}
  \sigma_1\hat{O}\sigma_1 = \hat{O},
\end{align}
where
$
\sigma_1 =
\begin{psmallmatrix}
0 & \bm{1}_{N\times N}\\
\bm{1}_{N\times N} &0
\end{psmallmatrix}
$
is a $2N\times 2N$ matrix with $\bm{1}_{N\times N}$ being the $N\times N$ unit matrix.

\subsection{Continuum limit}
\label{sec:method-continuum}

In this section, we introduce the continuum limit of the bosonic system.
In this limit, the bosonic Hamiltonian given by Eq.~\eqref{eq:Hamil-A-real} is expressed as
\begin{align}
  \mathcal{H}_{0} 
  &= \frac{1}{2}\sum_{\bm{\delta}} \int d\bm{r} \mathcal{A}^\dagger (\bm{r}) H_{0,\bm{\delta}} \mathcal{A} (\bm{r} + \bm{\delta})\notag\\
  \label{appeq:H0}
  &= \frac{1}{2} \int d\bm{r} \mathcal{A}^\dagger (\bm{r}) \hat{H}_{0} \mathcal{A} (\bm{r}),
\end{align}
where $\hat{H}_{0}$ is defined as
\begin{align}
  \hat{H}_{0} = \sum_{\bm{\delta}}
      H_{0,\bm{\delta}} e^{i\hat{\bm{p}}\cdot \bm{\delta}/\hbar}.
      \label{appeq:Mdelta}
\end{align}
Here, $\hat{\bm{p}}=-i\hbar \frac{\partial}{\partial \bm{r}}$ represents the momentum operator, and hence, $e^{i\hat{\bm{p}}\cdot \bm{\delta}/\hbar}$ serves as the generator of the translation of the bosonic operators by $\bm{\delta}$.
Note that the $2N\times 2N$ matrix $H_{0,\bm{\delta}}$ satisfies the relation
\begin{align}\label{eq:PHS-delta}
  \sigma_1 H_{0,\bm{\delta}} \sigma_1 = H_{0,-\bm{\delta}}^{T} =H_{0,\bm{\delta}}^*,
\end{align}
due to the particle-hole symmetry inherent in the BdG representation for bosonic systems.
Furthermore, as the continuum limit of Eq.~\eqref{eq:definition_A}, we introduce the $2N$-dimensional bosonic operator vector $\mathcal{A} (\bm{r})$, whose components are given by
\begin{align}
  \mathcal{A}_{s} (\bm{r})
  =
  \begin{cases}
          a_{s} (\bm{r}) & (s = 1,\dots , N)\\
  a_{s-N}^\dagger (\bm{r}) & (s = N+1,\dots , 2N)
  \end{cases},
\end{align}
where $a_{s}^\dagger (\bm{r})$ and $a_{s} (\bm{r})$ denote the creation and annihilation operators, respectively, satisfying the commutation relation $[ a_{s}^\dagger (\bm{r}), a_{s'}^{}(\bm{r}') ] = \delta_{ss'} \delta(\bm{r}-\bm{r}')$.
By applying the Fourier transformation to the bosonic operators as
\begin{align}\label{eq:Ar-Fourier}
  \mathcal{A}(\bm{r}) = \frac{1}{\sqrt{V}}\sum_{\bm{k}}\mathcal{A}_{\bm{k}} e^{i\bm{k}\cdot\bm{r}},
\end{align}
we obtain the Hamiltonian in momentum space as
\begin{align}
  \label{eq:conti_Hk}
  \mathcal{H}_{0} = \frac{1}{2} \sum_{\bm{k}} \mathcal{A}_{\bm{k}}^\dagger H_{0,\bm{k}} \mathcal{A}_{\bm{k}},
\end{align}
where $V$ represents the volume of the system, and $H_{0,\bm{k}} = e^{-i\bm{k}\cdot\hat{\bm{r}}} \hat{H}_{0} e^{i\bm{k}\cdot\hat{\bm{r}}}$, which is independent of $\hat{\bm{r}}$.
The summation over $\bm{k}$ is consistently performed within the first Brillouin zone in the above and subsequent expressions, unless otherwise specified.
Using the Bogoliubov transformation, the $2N\times 2N$ matrix $H_{0,\bm{k}}$ is diagonalized as
\begin{align}
  \mathcal{H}_{0}
  =
  \frac{1}{2} \sum_{\bm{k}} \mathcal{B}_{\bm{k}}^\dagger \mathcal{E}_{\bm{k}} \mathcal{B}_{\bm{k}},
\end{align}
where $\mathcal{E}_{\bm{k}}^{}=T_{\bm{k}}^\dagger H_{0,\bm{k}} T_{\bm{k}}^{}$ is the diagonal matrix, and $T_{\bm{k}}$ is a paraunitary matrix, 
satisfying $T_{\bm{k}}^{}\sigma_3 T_{\bm{k}}^\dagger = T_{\bm{k}}^{\dagger}\sigma_3 T_{\bm{k}}^{} = \sigma_3$ with 
$
\sigma_3 =
\begin{psmallmatrix}
\bm{1}_{N\times N} & 0\\
0 & -\bm{1}_{N\times N}
\end{psmallmatrix}
$~\cite{colpa}.
From Eq.~\eqref{eq:PHS-delta}, the Bloch Hamiltonian satisfies the following relation:
\begin{align}
  \sigma_{1}H_{0,\bm{k}} \sigma_{1} = H_{0,-\bm{k}}^{*},
  \label{eq:PHS}
\end{align}
and therefore, the relations $\sigma_{1}T_{\bm{k}} \sigma_{1} = T_{-\bm{k}}^{*}$ and $\sigma_{1}\mathcal{E}_{\bm{k}} \sigma_{1} = \mathcal{E}_{-\bm{k}}$ hold for the paraunitary and diagonal matrices~\cite{shindou2013,matsumoto2014,li2020_prr}.
This implies that the diagonal components of $\mathcal{E}_{\bm{k}}$ are given by $\{\varepsilon_{\bm{k},1}, \varepsilon_{\bm{k},2}, \dots, \varepsilon_{\bm{k},N}, 
\varepsilon_{-\bm{k},1}, \varepsilon_{-\bm{k},2}, \dots, \varepsilon_{-\bm{k},N}\}$.
The bosonic operator $\mathcal{B}_{\bm{k}}^{}  = T_{\bm{k}}^{-1} \mathcal{A}_{\bm{k}}^{}$ is defined as a $2N$-dimensional vector,
\begin{align}
  \mathcal{B}_{\bm{k}}^{\dagger} =\left(b_{\bm{k},1}^{\dagger}, b_{\bm{k},2}^{\dagger}, \dots, b_{\bm{k},N}^{\dagger}, 
   b_{-\bm{k},1}^{}, b_{-\bm{k},2}^{}, \dots, b_{-\bm{k},N}^{} \right),
\label{eq:Bogoliubov-Bop}
\end{align}
which satisfies the following commutation relation:
\begin{align}
    \label{eq:commu-B}
    \left[ \mathcal{B}_{\bm{k},\eta}^{}, \mathcal{B}_{\bm{k},\eta'}^\dagger \right] = \sigma_{3,\eta} \delta_{\eta,\eta'}.
\end{align}
Note that, from the paraunitary condition, we obtain $\sigma_3 H_{0,\bm{k}} T_{\bm{k}} = T_{\bm{k}}\sigma_3 \mathcal{E}_{\bm{k}}$.
This relation indicates that $T_{\bm{k}}$ consists of the right eigenvectors of $\sigma_3 H_{0,\bm{k}}$.
Similarly, based on the relation $T_{\bm{k}}^\dagger \sigma_3 \sigma_3 H_{0,\bm{k}} = \mathcal{E}_{\bm{k}} \sigma_3 T_{\bm{k}}^\dagger \sigma_3$, $T_{\bm{k}}^\dagger \sigma_3$ consists of the left eigenvectors of $\sigma_3 H_{0,\bm{k}}$.

Similar to the Hamiltonian, the continuum limit of the physical operator given in Eq.~\eqref{eq:O_total} is expressed as
\begin{align}
  O_{\rm total}&=\frac{1}{2} \int d\bm{r}\mathcal{A}^\dagger (\bm{r}) \hat{O} \mathcal{A}(\bm{r})
  =\int d\bm{r} O(\bm{r}),
\end{align}
where $O(\bm{r})$ is the local operator in the continuum limit, given by
\begin{align}
  O(\bm{r}) &= \frac{1}{2} \mathcal{A}^\dagger (\bm{r}) \hat{O} \mathcal{A}(\bm{r}).
\end{align}
Here, we omit a constant term that does not include $\mathcal{A} (\bm{r})$ and $\mathcal{A}^\dagger  (\bm{r})$ in $O(\bm{r})$, as it does not contribute to the equation of continuity, which will be introduced in the next section [see Eq.~\eqref{eq:continuity}].
Note that $\hat{O}$ does not contain the momentum operator $\hat{\bm{p}}$ because $O_{\rm total}$ is originally introduced as the sum of the local operators in Eq.~\eqref{eq:O_total}, indicating that $\hat{O}$ commutes with $\bm{r}$.

\subsection{Equation of continuity and torque term}
\label{sec:eq-cont}

Although this study focuses on spin current in insulating magnets, we generally consider the current of a local physical quantity $O(\bm{r})$ for applications to other physical quantities.  
The current is well-defined through the equation of continuity when the spatial integral $O_{\rm total}$ is conserved.  
However, if $O_{\rm total}$ does not commute with $\mathcal{H}_0$, the time derivative of $O(\bm{r})$ cannot be expressed as the divergence of the current, as demonstrated below.  
Based on the Heisenberg equation, the time derivative of $O(\bm{r})$ is given by~\cite{zyuzin2016_prl,li2020_prr}  
\begin{align}\label{eq:continuity}  
  \pdiff{O(\bm{r})}{t} = \frac{i}{\hbar} \left[ \mathcal{H}_{0}, O(\bm{r}) \right]  
  = - \bm{\nabla} \cdot \bm{j}^{O}(\bm{r}) + \tau^{O}(\bm{r}),  
\end{align}  
where $\bm{j}^{O}(\bm{r})= \mathcal{A}^\dagger (\bm{r}) \hat{\bm{j}}^{O} \mathcal{A}(\bm{r})$ and  
$\tau^{O}(\bm{r}) = \mathcal{A}^\dagger (\bm{r}) \hat{\tau}^{O} \mathcal{A}(\bm{r})$ represent the ``conventional'' current density and torque density, respectively.  
When considering the effect of collisions, an additional term appears on the right-hand side of Eq.~\eqref{eq:continuity}~\cite{culcer2004}.
In this study, we do not incorporate such a term and focus on the contributions independent of the relaxation time.  
Here, $\hat{\bm{j}}^{O}$ and $\hat{\tau}^{O}$ are the $2N\times 2N$ matrix operators, defined as  
\begin{align}  
  \label{eq:jO}  
  \hat{\bm{j}}^{O} = \frac{1}{4} \left( \hat{\bm{v}} \sigma_3 \hat{O} + \hat{O}\sigma_3\hat{\bm{v}} \right),  
  \end{align}  
where $\hat{\bm{v}} = \frac{i}{\hbar}[\hat{H}_{0}, \hat{\bm{r}}]$ is the velocity, and  
\begin{align}  
  \label{eq:tauO}  
  \hat{\tau}^{O} = \frac{i}{2\hbar} \left( \hat{H}_0\sigma_3 \hat{O} - \hat{O}\sigma_3\hat{H}_0 \right).  
\end{align}  

If the physical quantity $O_{\rm total}$ is conserved, the torque term $\tau^{O}(\bm{r})$ vanishes, and the current of $O(\bm{r})$ is given by $\bm{j}^{O}(\bm{r})$.
On the other hand, if $O_{\rm total}$ is not conserved, the torque term $\tau^{O}(\bm{r})$ is nonzero, and the current of $O(\bm{r})$ is not well-defined.
In previous theories of the spin Nernst effect in insulating magnets, the torque term has been neglected, and it has been assumed that the conventional spin current is used as the spin current.
However, the torque term is generally nonzero, particularly in spin-nonconserving systems with spin-orbit interaction, leading to topologically nontrivial magnon band structures, which generally play a crucial role in the spin Nernst effect.
It has been proposed that the torque term can be incorporated as an additional current term when it is given by the divergence of a vector as
\begin{align}\label{eq:torque-div}
  \braket{\tau^{O}(\bm{r})}=-\bm{\nabla}\cdot\bm{\pi}^{O}(\bm{r}).
\end{align}
Under this condition, Eq.~\eqref{eq:continuity} can be rewritten as
\begin{align}\label{eq:continuity-ave}
  \pdiff{\braket{O(\bm{r})}}{t} = - \bm{\nabla} \cdot \bm{\mathcal{J}}^{O}(\bm{r}),
\end{align}
where $\bm{\mathcal{J}}^{O}(\bm{r})$ is known as the conserved current, which is given by~\cite{shi2006}
\begin{align}\label{eq:conserved-current}
  \bm{\mathcal{J}}^{O}(\bm{r})=\braket{\bm{j}^{O}(\bm{r})} + \bm{\pi}^{O}(\bm{r}).
\end{align}
Since the temporal variation of $\braket{O(\bm{r})}$ is caused solely by the conserved current in Eq.~\eqref{eq:continuity-ave}, it is reasonable to consider that the boundary accumulation of the local quantity $\braket{O(\bm{r})}$ is not directly related to $\braket{\bm{j}^{O}(\bm{r})}$ but is instead attributed to the conserved current $\bm{\mathcal{J}}^{O}(\bm{r})$~\cite{shi2006}.

To formulate the spin Nernst effect, we need to evaluate the spin current in the presence of a temperature gradient, where $O$ is chosen to be the spin $\bm{S}$.
We regard the conserved spin current $\bm{\mathcal{J}}^{\bm{S}}(\bm{r})$, rather than the conventional one $\braket{\bm{j}^{\bm{S}}(\bm{r})}$, as the spin current measured in experiments.
The spin Nernst coefficient is defined as the first-order coefficient of the temperature gradient in the conserved spin current.
On the other hand, $\bm{\pi}^{\bm{S}}(\bm{r})$ is introduced such that its divergence equals the torque.
Thus, to calculate the spin Nernst coefficient, we need to evaluate the torque $\braket{\tau^{\bm{S}}(\bm{r})}$ up to the second order of the spatial gradient.
As a method for evaluating the second-order spatial gradients of physical quantities, previous studies have employed semiclassical theory~\cite{xiao2010,gao2014,xiao2021conserved} and quantum kinetic theory~\cite{sekine2017,varshney2023_prb1,varshney2023_prb2,liu2023,Mangeolle2024}.
In this study, we construct a theory for spin transport driven by a temperature gradient in bosonic systems based on semiclassical theory.

\subsection{Semiclassical theory for bosonic systems}
\label{sec:semiclassical}

To expand the torque term $\tau^{O}(\bm{r})$ up to the second order of the spatial gradient, we employ the semiclassical theory for bosonic BdG systems.
As discussed in Sec.~\ref{sec:method-continuum}, the paraunitary matrix $T_{\bm{k}}$ in the Bogoliubov transformation is composed of the right eigenvectors of $\sigma_{3}H_{0,\bm{k}}$, and $T_{\bm{k}}^\dagger \sigma_3$ consists of the left eigenvectors.
Based on this property, the eigenvalue equations for the right and left eigenvectors are given by
\begin{align}
  \sigma_{3}H_{0,\bm{k}} \ket{u_{\bm{k},\eta}^{R}} 
  =
  \bar{\varepsilon}_{\bm{k},\eta} \ket{u_{\bm{k},\eta}^{R}},
\end{align}
and
\begin{align}
  \bra{u_{\bm{k},\eta}^{L}} \sigma_{3}H_{0,\bm{k}}
  =
  \bar{\varepsilon}_{\bm{k},\eta} \bra{u_{\bm{k},\eta}^{L}},
\end{align}
respectively, in the bra-ket notation, where we define $\bar{\varepsilon}_{\bm{k},\eta} = \sigma_{3,\eta}\mathcal{E}_{\bm{k},\eta}$.
Note that in this notation, $\sigma_3$ and $H_{0,\bm{k}}$ are operators acting on quantum states represented by bras and kets.
Additionally, the relation $\bra{u^{L}_{\bm{k},\eta}} = \bra{u^{R}_{\bm{k},\eta}} \sigma_{3}$ holds.
The orthogonality and completeness relations are given by 
$\braket{u_{\bm{k},\eta}^{L}|u_{\bm{k},\eta'}^{R}} = \braket{u_{\bm{k},\eta}^{R}|\sigma_3|u_{\bm{k},\eta'}^{R}} = \sigma_{3,\eta} \delta_{\eta,\eta'}$ 
and
$\sum_{\eta}^{2N} \sigma_{3,\eta} \ket{u_{\bm{k},\eta}^{R}}\bra{u_{\bm{k},\eta}^{L}} = \sum_{\eta}^{2N} \sigma_{3,\eta} \ket{u_{\bm{k},\eta}^{R}}\bra{u_{\bm{k},\eta}^{R}} \sigma_3 = \hat{1}$,
which represents the identity operator~\cite{psaroudaki2017,li2020_prr}.
Moreover, we find the relation 
$\braket{u_{\bm{k},\eta}^{R}|H_{0,\bm{k}}|u_{\bm{k},\eta}^{R}}=\mathcal{E}_{\bm{k},\eta}$,
which suggests that the right eigenstate $\ket{u_{\bm{k},\eta}^{R}}$ can effectively be regarded as the Bloch state with branch $\eta$ for the Hamiltonian $H_{0,\bm{k}}$.

The semiclassical theory is a framework that describes the motion of wave packets formed by the superposition of Bloch states.
Using this framework, we evaluate the expectation values of $\bm{j}^{O}(\bm{r})$ and $\tau^{O}(\bm{r})$, which are generally expressed as,
\begin{align}\label{eq:AthetaA}
\bm{\theta}(\bm{r}) = \mathcal{A}^\dagger (\bm{r}) \hat{\bm{\theta}} \mathcal{A}(\bm{r}),
\end{align}
where $\hat{\bm{\theta}}$ represents the corresponding one-body operator, namely, $\hat{\theta}_\lambda=\hat{\bm{j}}^{O}_\lambda$ or $\hat{\tau}^{O}$.
To this end, we introduce virtual source fields corresponding to these physical quantities into the Hamiltonian and determine their expectation values by performing functional differentiation on the action derived from the Lagrangian~\cite{chang1996,sundaram1999,xiao2010}.  
The Hamiltonian with the virtual source field $\bm{h}(\hat{\bm{r}}, t)$ coupled to $\hat{\bm{\theta}}$ is written as  
\begin{align}  
  \label{appeq:H_semiclassical}  
  \hat{H} = \hat{H}_{0} + \frac{\hbar}{2}  
  \left[ \hat{\bm{\theta}}\cdot \bm{h}(\hat{\bm{r}}, t) + \bm{h}(\hat{\bm{r}}, t) \cdot \hat{\bm{\theta}}  \right].  
\end{align}  
After completing all calculations, we take the limit where $\bm{h}$ approaches zero.
We define the Bloch Hamiltonian $H_{\bm{q}}$ as $H_{\bm{q}} = e^{-i\bm{q}\cdot\hat{\bm{r}}} \hat{H} e^{i\bm{q}\cdot\hat{\bm{r}}}$ and introduce the right eigenstate of $\sigma_{3}H_{\bm{q}}$ as $\ket{u_{\bm{q},\eta}^{\rm tot}}$ with the eigenvalue $\bar{\varepsilon}_{\bm{q},\eta}^{\rm tot}$.
Assuming the adiabatic approximation, the $\eta$-th wave packet $\ket{W_\eta}$ is constructed solely from the $\eta$-th Bloch state as follows:
\begin{align}  
  \label{appeq:Weta_ext}  
  \ket{W_\eta}= \int d\bm{q} C_{\bm{k}_\eta,\eta}(\bm{q}) e^{i\bm{q}\cdot\hat{\bm{r}}} \ket{u_{\bm{q},\eta}^{\rm tot}},  
\end{align} 
where the integral over $\bm{q}$ is taken within the first Brillouin zone.
In the following calculations, momentum integrals will also be performed over the first Brillouin zone.
Here, $C_{\bm{k},\eta}(\bm{q})$ is the expansion coefficient, which satisfies the normalization condition $\int d\bm{q} |C_{\bm{k},\eta}(\bm{q})|^2 = 1$.
In addition to this condition, $C_{\bm{k}_\eta,\eta}(\bm{q})$ is assumed to have a sharp peak around $\bm{q} \approx \bm{k}_\eta$, which can be expressed as $\int d\bm{q} f(\bm{q})|C_{\bm{k}_\eta,\eta}(\bm{q})|^2 \approx f(\bm{k}_\eta)$ for any arbitrary function $f(\bm{q})$.
In addition, the center position of the wave packet in real space is defined as~\cite{xiao2010,zhang2019}  
\begin{align}  
  \label{eq:def_rc}  
  \bm{r}_{c,\eta} =  
  \braket{W_\eta|\sigma_3\hat{\bm{r}}|W_\eta}/\braket{W_\eta|\sigma_3|W_\eta}.  
\end{align}
The wave packet $\ket{W_\eta}$ is assumed to be well localized around $\bm{r}_{c,\eta}$ in real space and around $\bm{k}_\eta$ in momentum space.
We omit the branch label $\eta$ from $\bm{r}_{c,\eta}$ and $\bm{k}_\eta$ and denote these quantities as $\bm{r}_{c}$ and $\bm{k}$ hereafter.  

In semiclassical theory, the expectation value $\braket{\bm{\theta}(\bm{r})}$ is given by the following expression (see also Appendix~\ref{app:exp-theta-semiclassical})~\cite{culcer2004,xiao2010}:
\begin{align}
  \label{eq:ave1}
  &\braket{\bm{\theta}(\bm{r})}
  =\notag\\
  &\sum_{\eta=1}^{2N} \int d\bm{r}_c \int \frac{d\bm{k}}{(2\pi)^d} D_{\bm{k},\eta} f_{B}(\bm{r}_c,\bar{\varepsilon}^{\mathrm{tot}}_{\bm{k},\eta}) 
  \frac{\braket{W_\eta|\delta(\hat{\bm{r}} - \bm{r})\hat{\bm{\theta}}|W_\eta}}{\braket{W_{\eta}|\sigma_3|W_{\eta}}},
\end{align}
where $d$ denotes the dimension of the system, $f_{B}(\bm{r},\varepsilon)=1/(e^{\beta(\bm{r})\varepsilon}-1)$ represents the Bose distribution function, with $\beta(\bm{r})=1/k_B T(\bm{r})$ being the spatially dependent inverse temperature, and $D_{\bm{k},\eta}$ represents the density of states in the phase space of $(\bm{r}_c,\bm{k})$.
This is given by $D_{\bm{k},\eta}=1+\sum_{\lambda}\tilde{\Omega}^{k_{\lambda}r_{c,\lambda}}_{\bm{k},\eta}$,
where the mixed Berry curvature for $\mathcal{H}$ is defined as~\cite{xiao2005,dong2020}.
\begin{align}\label{eq:mixed_Berry-full}
  \tilde{\Omega}^{X_{\lambda} Y_{\lambda'}}_{\bm{k},\eta}
  &=
  -2 \mathrm{Im}
  \left[
    \sigma_{3,\eta} \braket{\partial_{X_{\lambda}}u_{\bm{k},\eta}^{\rm tot}| \sigma_3 | \partial_{Y_{\lambda'}}u_{\bm{k},\eta}^{\rm tot}}
  \right].
\end{align}
 For simplicity, we abbreviate $\frac{\partial}{\partial X_{\lambda}}$ as $\partial_{X_{\lambda}}$.

Here, we assume that the time evolution of the wave packet $\ket{W_{\eta}}$ is governed by the Schr\"odinger equation as follows:~\cite{shindou2013,lein2019}
\begin{align}\label{eq:Schrodinger}
  \left(i\hbar\pdiff{}{t}-\sigma_3\hat{H}\right)\ket{W_{\eta}} = 0.
\end{align}
In the time-dependent variational principle, the above equation is derived from the stationary condition of the action $S_\eta=\int \mathcal{L}_{\eta} dt$ as an on-shell equation, where the Lagrangian $\mathcal{L}_{\eta}$ is defined as~\cite{kramer,sundaram1999,zhang2019,dong2020,xiao2021conserved}
\begin{align}
  \label{eq:Leta1}
  \mathcal{L}_{\eta}
  =
  \frac{\braket{W_{\eta}|\sigma_3\left(i\hbar\diff{}{t}-\sigma_3\hat{H}\right)|W_{\eta}}}{\braket{W_{\eta}|\sigma_3|W_{\eta}}}.
\end{align}
The dynamics of the wave packet follows the equations of motion for $\bm{r}_{c}$ and $\bm{k}$, which are given by the on-shell Euler-Lagrange equations derived from the Lagrangian $\mathcal{L}_{\eta}$.
Using this Lagrangian, the expectation value of $\bm{\theta}(\bm{r})$ 
in Eq.~\eqref{eq:ave1} is written as~\cite{dong2020}
\begin{align}
  \label{appeq:ave2}
  &\braket{\theta_\lambda(\bm{r})}
  =\notag\\
  & -\frac{1}{\hbar}\sum_{\eta=1}^{2N}
  \int d\bm{r}_c  \int \frac{d\bm{k}}{(2\pi)^d} D_{\bm{k},\eta} f_B(\bm{r}_c,\bar{\varepsilon}^{\mathrm{tot}}_{\bm{k},\eta}) 
  \fdiff{S_\eta}{h_{\lambda}(\bm{r}, t)}\Bigg|_{\textrm{on-shell}}^{\bm{h}\to 0},
\end{align}

By expanding the Lagrangian $\mathcal{L}_{\eta}$ up to the second order in the spatial gradient around $\bm{r}_c$, we obtain the following expression for the expectation value of $\bm{\theta}(\bm{r})$:
\begin{widetext}
\begin{align}
  \braket{\theta_{\lambda}(\bm{r})}
  &\approx
  \sum_{\eta=1}^{2N}
  \int \frac{d\bm{k}}{(2\pi)^d}
  \sigma_{3,\eta}(\theta_{\bm{k},\lambda})_{\eta} f_B(\bm{r}, \bar{\varepsilon}_{\bm{k}, \eta})
  \notag\\
  &
  +\frac{1}{\hbar}
  \sum_{\lambda'}
  \sum_{\eta=1}^{2N}
  \int \frac{d\bm{k}}{(2\pi)^d}
  (\Omega^{\theta_\lambda}_{\bm{k},\eta})_{\lambda'} F_{\lambda'}(\bm{r},\bar{\varepsilon}_{\bm{k},\eta})
  -\sum_{\lambda'}
  \nabla_{\lambda'}\sum_{\eta=1}^{2N}
  \int \frac{d\bm{k}}{(2\pi)^d}
  \left\{
    (d^{\theta_\lambda}_{\bm{k},\eta})_{\lambda'} f_B(\bm{r},\bar{\varepsilon}_{\bm{k},\eta})
    + \frac{1}{\hbar}(\Omega^{\theta_\lambda}_{\bm{k},\eta})_{\lambda'} g_B (\bm{r},\bar{\varepsilon}_{\bm{k},\eta}) \right\}
  \notag\\
  &
  -\frac{1}{\hbar}\sum_{\lambda'\lambda''}
  \nabla_{\lambda'}\sum_{\eta=1}^{2N}
  \int \frac{d\bm{k}}{(2\pi)^d}
   \big(\chi_{\bm{k},\eta}^{\theta_\lambda}\big)_{\lambda'\lambda''} F_{\lambda''}(\bm{r},\bar{\varepsilon}_{\bm{k},\eta})
  + \sum_{\lambda'\lambda''} \nabla_{\lambda'}\nabla_{\lambda''}\sum_{\eta=1}^{2N}
  \int \frac{d\bm{k}}{(2\pi)^d}
  \Bigg\{
    \big(q^{\theta_\lambda}_{\bm{k},\eta}\big)_{\lambda'\lambda''} f_B(\bm{r},\bar{\varepsilon}_{\bm{k},\eta})
    + \frac{1}{\hbar} \big(\chi_{\bm{k},\eta}^{\theta_\lambda}\big)_{\lambda'\lambda''} g_B(\bm{r},\bar{\varepsilon}_{\bm{k},\eta})
    \Bigg\}
.
  \label{appeq:ave4}
\end{align}
\end{widetext}
The detailed derivation of the above expression is provided in Appendix~\ref{app:ave-theta}.
Here, $(\theta_{\bm{k},\lambda})_{\eta} = \braket{u_{\bm{k},\eta}|\theta_{\bm{k},\lambda}|u_{\bm{k},\eta}}$
 with $\theta_{\bm{k},\lambda}
= e^{-i\bm{k}\cdot\hat{\bm{r}}}\hat{\theta}_{\lambda}
e^{i\bm{k}\cdot\hat{\bm{r}}}$, and we introduce $F_{\lambda'}(\bm{r},\omega)$ as
\begin{align}
  \label{appeq:F}
  F_{\lambda'}(\bm{r},\omega)
  =
  \frac{-\nabla_{\lambda'} T}{T} \left[ \omega f_B(\bm{r},\omega) - g_B(\bm{r},\omega) \right],
\end{align}
where we assume that the spatial variation of temperature is given by $T(\bm{r})=T+\bm{r}\cdot\bm{\nabla} T$, and $g_B(\bm{r},\omega)$ is the grand potential defined as
\begin{align}
  g_B(\bm{r}, \omega) =
\frac{1}{\beta(\bm{r})} \ln \left|  1 - e^{-\beta(\bm{r}) \omega} \right|,
\end{align}
which satisfies $\partial g_B/\partial \omega = f_B(\bm{r},\omega)$.
Since $F_{\lambda'}(\bm{r},\omega)$ is proportional to the spatial gradient of temperature, the first, second, and third lines on the right-hand side of Eq.~\eqref{appeq:ave4} correspond to the zeroth, first, and second orders of the spatial gradient, respectively.
We also introduce $(\Omega^{\theta_\lambda}_{\bm{k},\eta})_{\lambda'}=\Omega^{k_{\lambda'} h_{\lambda} }_{\bm{k},\eta}\Big|_{\bm{h}\to 0}$ as the mixed Berry curvature between $\bm{k}$ and $\bm{h}$, which is defined
from the Berry curvature in the mixed space $(\bm{X},\bm{Y})$ given by 
\begin{align}\label{eq:mixed_Berry}
  \Omega^{X_{\lambda} Y_{\lambda'}}_{\bm{k},\eta}
  &=
  -2 \mathrm{Im}
  \left[
    \sigma_{3,\eta} \braket{\partial_{X_{\lambda}}u_{\bm{k},\eta}| \sigma_3 | \partial_{Y_{\lambda'}}u_{\bm{k},\eta}}
  \right],
\end{align}
where $\ket{u_{\bm{k},\eta}}$ is the Bloch state for the Hamiltonian $\hat{H}_c=\hat{H}_{0} +\hbar \hat{\bm{\theta}}\cdot \bm{h}(\bm{r}_c, t)$.
The dipole and quadrupole moments are defined as~\cite{dong2020,li2020_prr}
\begin{align}
    \label{eq:appeq:GDM_def}
    (d^{\theta_\lambda}_{\bm{k},\eta})_{\lambda'}
    =
    \frac{\mathrm{Re} \braket{W_\eta^{(0)}|(\hat{r}_{\lambda'} - r_{c,\lambda'}) \hat{\theta}_{\lambda}|W_\eta^{(0)}}}
    {\sigma_{3,\eta}},
\end{align}
and
\begin{align}
  \label{appeq:GQM_def}
  \big(q^{\theta_\lambda}_{\bm{k},\eta}\big)_{\lambda'\lambda''} =
  \frac
  {\mathrm{Re} \braket{W_\eta^{(0)}|(\hat{r}_{\lambda'} - r_{c,\lambda'}) (\hat{r}_{\lambda''} - r_{c,\lambda''})\hat{\theta}_{\lambda}|W_\eta^{(0)}}}
  {2\sigma_{3,\eta}},
\end{align}
where $\ket{W_\eta^{(0)}}$ is introduced as the first term in Eq.~\eqref{appeq:Weta_H'}.
In addition, $\big(\chi_{\bm{k},\eta}^{\theta_\lambda}\big)_{\lambda'\lambda''}$ is written as
\begin{align}
  \big(\chi_{\bm{k},\eta}^{\theta_\lambda}\big)_{\lambda'\lambda''}
  \label{appeq:chi1}
  &=
  -2\mathrm{Re} \sum_{\eta_1(\neq \eta)}^{2N} 
  \frac
  {A^{\lambda''}_{\bm{k},\eta\eta_1} \left(d^{\theta_\lambda}_{\bm{k},\eta_1\eta}\right)_{\lambda'}}
  {\bar{\varepsilon}_{\bm{k},\eta} - \bar{\varepsilon}_{\bm{k},\eta_1}}
  + \frac{\partial }{\partial k_{\lambda'}} \big(g^{\theta_\lambda}_{\bm{k},\eta}\big)_{\lambda''},
\end{align}
where $A^\lambda_{\bm{k},\eta\eta_1} = i \sigma_{3,\eta} \braket{u_{\bm{k},\eta}|\sigma_3|\partial_{k_\lambda}u_{\bm{k},\eta_1}}$ is the interband Berry connection, and 
$(d^{\theta_\lambda}_{\bm{k},\eta_1\eta})_{\lambda'}$ is defined in Eq.~\eqref{appeq:d_off_2}.
We also introduce the quantum metric in the mixed space $(\bm{k},\bm{h})$ as $\big(g^{\theta_\lambda}_{\bm{k},\eta}\big)_{\lambda'}=g^{k_{\lambda'} h_{\lambda} }_{\bm{k},\eta}\Big|_{\bm{h}\to 0}$, where $g^{X_{\lambda} Y_{\lambda'}}_{\bm{k},\eta}$ is defined as
\begin{align}
  &g^{X_{\lambda} Y_{\lambda'}}_{\bm{k},\eta}= \notag\\
  & \mathrm{Re}\left[\sigma_{3,\eta}
    \bra{\partial_{X_{\lambda}}u_{\bm{k},\eta}}\sigma_{3}\left(\hat{1}-\sigma_{3,\eta}\ket{u_{\bm{k},\eta}}\bra{u_{\bm{k},\eta}} \sigma_{3}\right)\ket{\partial_{Y_{\lambda'}} u_{\bm{k},\eta}}\right].
  \label{appeq:QMT}
\end{align}
Note that this is the real part of the quantum geometric tensor $T^{X_{\lambda} Y_{\lambda'}}_{\bm{k},\eta}$ in the mixed space $(\bm{X},\bm{Y})$ as follows:
\begin{align}
  T^{X_{\lambda} Y_{\lambda'}}_{\bm{k},\eta}=& \sigma_{3,\eta}\bra{\partial_{X_{\lambda}}u_{\bm{k},\eta}}\sigma_{3}\ket{\partial_{Y_{\lambda'}} u_{\bm{k},\eta}}\notag\\
  & 
    -\bra{\partial_{X_{\lambda}}u_{\bm{k},\eta}}\sigma_3\ket{u_{\bm{k},\eta}}\bra{u_{\bm{k},\eta}} \sigma_{3}\ket{\partial_{Y_{\lambda'}} u_{\bm{k},\eta}},
\end{align}
and its imaginary part is given by the mixed Berry curvature $\Omega^{X_{\lambda} Y_{\lambda'}}_{\bm{k},\eta}$ as $T^{X_{\lambda} Y_{\lambda'}}_{\bm{k},\eta}=g^{X_{\lambda} Y_{\lambda'}}_{\bm{k},\eta}-\frac{i}{2}\Omega^{X_{\lambda} Y_{\lambda'}}_{\bm{k},\eta}$~\cite{provost1980,Liang2023}.
The quantum metric has been discussed in relation to nonlinear phenomena~\cite{gao2014,gao2015,Das2023,Fang2024,gao2023quantum,wang2023quantum}, while it also appears in the linear response of a thermal gradient to the spin current.

\subsection{Evaluation of current using semiclassical theory}
\label{sec:eval-current}

In this section, we evaluate Eq.~\eqref{eq:continuity-ave} up to the first order of the temperature gradient using Eq.~\eqref{appeq:ave4}.
Note that the torque term $\braket{\tau^{O}(\bm{r})}$ needs to be expanded up to the second order of the spatial gradient because the second term of Eq.~\eqref{eq:conserved-current} is given such that its divergence equals $\braket{\tau^{O}(\bm{r})}$, as shown in Eq.~\eqref{eq:torque-div}.
It is also important to note that, since the local quantity $\hat{O}$ commutes with $\hat{\bm{r}}$, $O_{\bm{k}} = e^{-i\bm{k}\cdot\hat{\bm{r}}}\hat{O}e^{i\bm{k}\cdot\hat{\bm{r}}}$ is independent of $\bm{k}$ and is equivalent to $\hat{O}$.
By substituting $\bm{j}^{O}$ to $\bm{\theta}$ in Eq.~\eqref{appeq:ave4}, we obtain the expectation value of $\bm{j}^{O}(\bm{r})$ up to the first order of the spatial gradient as
\begin{widetext}
\begin{align}
  \braket{j^O_{\lambda}(\bm{r})}
  \approx&
  \sum_{\eta=1}^{2N} \int \frac{d\bm{k}}{(2\pi)^d}
  \sigma_{3,\eta} (j^{O}_{\bm{k},\lambda})_{\eta} f_B(\bm{r},\bar{\varepsilon}_{\bm{k},\eta})
  + \sum_{\lambda'}\frac{(-\nabla_{\lambda'} T)}{\hbar T}\sum_{\eta=1}^{2N}\int \frac{d\bm{k}}{(2\pi)^d}
  (\Omega^{j^{O}_\lambda}_{\bm{k},\eta})_{\lambda'} 
  [\bar{\varepsilon}_{\bm{k},\eta} f_{B}(\bm{r},\bar{\varepsilon}_{\bm{k},\eta})- g_B (\bm{r},\bar{\varepsilon}_{\bm{k},\eta})] 
  \notag\\
  &
  \label{eq:jcon_1st}
  - \sum_{\lambda'}\nabla_{\lambda'} \sum_{\eta=1}^{2N} \int \frac{d\bm{k}}{(2\pi)^d}
  \left[(d^{j^{O}_\lambda}_{\bm{k},\eta})_{\lambda'} f_B(\bm{r},\bar{\varepsilon}_{\bm{k},\eta})
  + \frac{1}{\hbar}(\Omega^{j^{O}_\lambda}_{\bm{k},\eta})_{\lambda'} g_B (\bm{r},\bar{\varepsilon}_{\bm{k},\eta}) \right].
\end{align}
Here, $(\Omega^{j^{O}_\lambda}_{\bm{k},\eta})_{\lambda'}$ and $(d^{j^{O}_\lambda}_{\bm{k},\eta})_{\lambda'}$ are calculated from Eqs.~\eqref{eq:mixed_Berry} and \eqref{eq:appeq:GDM_def}, and they can be generally written as
\begin{align}
  \label{eq:GBC-O}
    (\Omega^{\theta_\lambda}_{\bm{k},\eta})_{\lambda'}
  =
  -2\hbar^2 \sum_{\eta_1 (\neq \eta)}^{2N} \sigma_{3,\eta}\sigma_{3,\eta_1} \frac{\mathrm{Im}\left[ (v_{\bm{k},\lambda'})_{\eta\eta_1} (\theta_{\bm{k},\lambda})_{\eta_1\eta} \right]}
  {(\bar{\varepsilon}_{\bm{k},\eta}-\bar{\varepsilon}_{\bm{k},\eta_1})^2},\qquad
  (d^{\theta_\lambda}_{\bm{k},\eta})_{\lambda'}
  =
  \hbar\sum_{\eta_1(\neq \eta)}^{2N} \sigma_{3,\eta}\sigma_{3,\eta_1} \frac{\mathrm{Im}\left[ (v_{\bm{k},\lambda'})_{\eta\eta_1} (\theta_{\bm{k},\lambda})_{\eta_1\eta} \right]}{\bar{\varepsilon}_{\bm{k},\eta}-\bar{\varepsilon}_{\bm{k},\eta_1}}.
\end{align}
where we introduce $(\theta_{\bm{k},\lambda})_{\eta\eta'} = \braket{u_{\bm{k},\eta}|\theta_{\bm{k},\lambda}|u_{\bm{k},\eta'}}$.
In a similar manner, we evaluate the expectation value of $\tau^{O}(\bm{r})$ up to the second order of the spatial gradient as
\begin{align}
  \braket{\tau^{O}(\bm{r})}
  \approx&\sum_{\lambda}
  \frac{-\nabla_{\lambda} T}{\hbar T} \sum_{\eta=1}^{2N} \int \frac{d\bm{k}}{(2\pi)^d}
  (\Omega^{\tau^{O}}_{\bm{k},\eta})_{\lambda} 
  [\bar{\varepsilon}_{\bm{k},\eta} f_{B}(\bm{r},\bar{\varepsilon}_{\bm{k},\eta})-g_B (\bm{r},\bar{\varepsilon}_{\bm{k},\eta})]
  \notag\\
  & 
  - \sum_{\lambda}
  \nabla_{\lambda} \sum_{\eta=1}^{2N} \int \frac{d\bm{k}}{(2\pi)^d}
  \left[ (d^{\tau^{O}}_{\bm{k},\eta})_{\lambda} f_B(\bm{r},\bar{\varepsilon}_{\bm{k},\eta})
  + \frac{1}{\hbar} (\Omega^{\tau^{O}}_{\bm{k}, \eta})_{\lambda} g_B (\bm{r},\bar{\varepsilon}_{\bm{k},\eta}) \right]
  \notag\\
  &
  - \sum_{\lambda\lambda'}
  \nabla_{\lambda} \sum_{\eta=1}^{2N} \int \frac{d\bm{k}}{(2\pi)^d}
  \frac{-\nabla_{\lambda'} T}{\hbar T} (\chi^{\tau^{O}}_{\bm{k},\eta})_{\lambda\lambda'} 
  [\bar{\varepsilon}_{\bm{k},\eta} f_{B}(\bm{r},\bar{\varepsilon}_{\bm{k},\eta})-g_B (\bm{r},\bar{\varepsilon}_{\bm{k},\eta})]
  \notag\\
  \label{eq:tau_2nd}
  &
  +\sum_{\lambda\lambda'}
  \nabla_{\lambda} \sum_{\eta=1}^{2N} \int \frac{d\bm{k}}{(2\pi)^d}
  \nabla_{\lambda'}
  \left[ (q^{\tau^{O}}_{\bm{k},\eta})_{\lambda\lambda'} f_B(\bm{r},\bar{\varepsilon}_{\bm{k},\eta})
  + \frac{1}{\hbar} (\chi^{\tau^{O}}_{\bm{k},\eta})_{\lambda\lambda'} g_B(\bm{r},\bar{\varepsilon}_{\bm{k},\eta})  \right],
\end{align}
where $(\Omega^{\tau^{O}}_{\bm{k},\eta})_{\lambda}$ and $(d^{\tau^{O}}_{\bm{k},\eta})_{\lambda}$ are introduced by Eqs.~\eqref{eq:GBC-O}.
Moreover, $(q^{\tau^{O}}_{\bm{k},\eta})_{\lambda\lambda'}$ and $(\chi^{\tau^{O}}_{\bm{k},\eta})_{\lambda\lambda'}$ are calculated as
\begin{align}
  \label{eq:quadrupole-moment}
  \big(q^{\tau^{O}}_{\bm{k},\eta}\big)_{\lambda\lambda'}
  &=
  -(d^{j^{O}_\lambda}_{\bm{k},\eta})_{\lambda'}
  +
  \frac{1}{2}\sigma_{3,\eta} (d^{O}_{\bm{k},\eta})_{\lambda} (v_{\bm{k},\lambda'})_{\eta}
  +
  (m^{O}_{\bm{k},\eta})_{\lambda'\lambda},\\
  \label{eq:chi}
  \big(\chi^{\tau^{O}}_{\bm{k},\eta}\big)_{\lambda\lambda'}
  &=
  - (\Omega^{j^{O}_\lambda}_{\bm{k},\eta})_{\lambda'} 
  - \frac{1}{2} \sigma_{3,\eta} \Omega^{\lambda\lambda'}_{\bm{k},\eta} (O_{\bm{k}})_{\eta} 
  + \frac{1}{2} \pdiff{}{k_{\lambda'}} (d^{O}_{\bm{k},\eta})_{\lambda},
\end{align}
where $(d^{O}_{\bm{k},\eta})_{\lambda}$, $(m^{O}_{\bm{k},\eta})_{\lambda'\lambda}$ are $\Omega^{\lambda\lambda'}_{\bm{k},\eta}$ are given by
\begin{align}
(d^{O}_{\bm{k},\eta})_{\lambda}
  =
  \hbar\sum_{\eta_1(\neq \eta)}^{2N} \sigma_{3,\eta}\sigma_{3,\eta_1} \frac{\mathrm{Im}\left[ (v_{\bm{k},\lambda})_{\eta\eta_1} (O_{\bm{k}})_{\eta_1\eta} \right]}{\bar{\varepsilon}_{\bm{k},\eta}-\bar{\varepsilon}_{\bm{k},\eta_1}},
\end{align}
\begin{align}
\label{eq:orbital_tensor}
  (m^{O}_{\bm{k},\eta})_{\lambda'\lambda}
  =
  \frac{1}{2}\mathrm{Re} \sum_{\eta_1(\neq \eta)}^{2N}
  \left[ A^{\lambda'}_{\bm{k},\eta\eta_1} \sigma_{3,\eta_1} (j^{O}_{\bm{k},\lambda})_{\eta_1 \eta} - A^{\lambda}_{\bm{k},\eta\eta_1} \sigma_{3,\eta_1} (j^{O}_{\bm{k},\lambda'})_{\eta_1\eta} \right]  
  +\frac{1}{4}
  \left[ \sigma_{3,\eta} (d^{O}_{\bm{k}, \eta})_{\lambda'}  (v_{\bm{k},\lambda})_{\eta} - \sigma_{3,\eta} (d^{O}_{\bm{k},\eta})_{\lambda} (v_{\bm{k},\lambda'})_{\eta} \right],
  \end{align}
  and 
  \begin{align}
  \label{eq:BC}
  \Omega^{\lambda\lambda'}_{\bm{k},\eta}
  =
  -2 \mathrm{Im}
  \left[
    \sigma_{3,\eta} \braket{\partial_{k_{\lambda}}u_{\bm{k},\eta}| \sigma_3 | \partial_{k_{\lambda'}}u_{\bm{k},\eta}}
  \right]
  =
  -2\hbar^2\sum_{\eta_1 (\neq \eta)}^{2N} \sigma_{3,\eta}\sigma_{3,\eta_1} \frac{\mathrm{Im}\left[ (v_{\bm{k},\lambda})_{\eta\eta_1} (v_{\bm{k},\lambda'})_{\eta_1\eta} \right]}
  {(\bar{\varepsilon}_{\bm{k},\eta}-\bar{\varepsilon}_{\bm{k},\eta_1})^2},
\end{align}
respectively.
The latter is the Berry curvature in the momentum space~\cite{shindou2013}, which is antisymmetric for $\lambda$ and $\lambda'$~\cite{matsumoto2014}.
Furthermore, $(m^{O}_{\bm{k},\eta})_{\lambda'\lambda}$ is also antisymmetric for $\lambda$ and $\lambda'$.
To obtain the second term in the right-hand side of Eq.~\eqref{eq:chi}, we have used the relation
\begin{align}\label{eq:relation-g-d}
  \big(g^{\tau^O}_{\bm{k},\eta}\big)_{\lambda'}= \hbar^2
  \sum_{\eta_1(\neq \eta)}^{2N}
  \sigma_{3,\eta} \sigma_{3,\eta_1}
  \frac
  {\mathrm{Re} \left[(v_{\bm{k},\lambda'})_{\eta\eta_1} (\tau^O_{\bm{k}})_{\eta_1\eta}\right]}
  {(\bar{\varepsilon}_{\bm{k},\eta} - \bar{\varepsilon}_{\bm{k},\eta_1})^2}
  =
  \frac{\hbar}{2}\sum_{\eta_1(\neq \eta)}^{2N} \sigma_{3,\eta}\sigma_{3,\eta_1}
  \frac{\mathrm{Im}\left[(v_{\bm{k},\lambda'})_{\eta\eta_1} (O_{\bm{k}})_{\eta_1\eta}\right]}
  {\bar{\varepsilon}_{\bm{k},\eta}-\bar{\varepsilon}_{\bm{k},\eta_1}}
  =
  \frac{1}{2} (d^{O}_{\bm{k},\eta})_{\lambda'},
\end{align}
which is obtained by applying the relation
$
(\tau^{O}_{\bm{k}})_{\eta\eta_1}
= \frac{i}{2\hbar} (\bar{\varepsilon}_{\bm{k},\eta} - \bar{\varepsilon}_{\bm{k},\eta_1})
(O_{\bm{k}})_{\eta\eta_1}
$.
Note that, although $O_{\bm{k}}$ is independent of $\bm{k}$ as discussed before, $(O_{\bm{k}})_{\eta\eta'}$ depends on $\bm{k}$, originating from $u_{\bm{k},\eta}$.

Here, we decompose Eqs.~\eqref{eq:jcon_1st} and \eqref{eq:tau_2nd} into two contributions: one that exists even in the absence of a temperature gradient, denoted as $\braket{\cdot}_{\rm eq}$, and another that is proportional to the temperature gradient, denoted as $\braket{\cdot}_{\nabla T}$.
These are expressed as $\braket{\bm{j}^{O}(\bm{r})} = \braket{\bm{j}^{O}(\bm{r})}_{\mathrm{eq}} + \braket{\bm{j}^{O}(\bm{r})}_{\nabla T}$ and $\braket{\tau^{O}(\bm{r})} = \braket{\tau^{O}(\bm{r})}_{\mathrm{eq}} + \braket{\tau^{O}(\bm{r})}_{\nabla T}$, up to the first order in the temperature gradient.
In particular, $\braket{\tau^{O}(\bm{r})}_{\mathrm{eq}}$ and $\braket{\tau^{O}(\bm{r})}_{\nabla T}$ can be written as
\begin{align}
  \label{eq:tau_eq}
  \braket{\tau^{O}(\bm{r})}_{\mathrm{eq}}
  &= -\bm{\nabla} \cdot \bm{\pi}_{\mathrm{eq}}^O(\bm{r}),\\
  \label{eq:tau_T}
  \braket{\tau^{O}(\bm{r})}_{\nabla T}
  &= - \bm{\nabla} \cdot \bm{\pi}_{\nabla T}^O(\bm{r})
  - \bm{\rho}^O(\bm{r}) \cdot \bm{\nabla} T,
\end{align}
where $\bm{\pi}_{\mathrm{eq}}^O(\bm{r})$, $\bm{\pi}_{\nabla T}^O(\bm{r})$, and $\bm{\rho}^O(\bm{r})$ are given by
\begin{align}
  \pi_{\lambda,\mathrm{eq}}^O(\bm{r})
  &=
  \sum_{\eta=1}^{2N} \int \frac{d\bm{k}}{(2\pi)^d}
  \Bigg\{
  \left[ (d^{\tau^{O}}_{\bm{k},\eta})_{\lambda} f_B(\bm{r}, \bar{\varepsilon}_{\bm{k},\eta})
  + \frac{1}{\hbar}(\Omega^{\tau^{O}}_{\bm{k}, \eta})_{\lambda} g_B(\bm{r}, \bar{\varepsilon}_{\bm{k},\eta}) \right]
  -\sum_{\lambda'}\nabla_{\lambda'}
  \Big[ (q^{\tau^{O}}_{\bm{k},\eta})_{\lambda\lambda'} f_B(\bm{r}, \bar{\varepsilon}_{\bm{k},\eta})
  + \frac{1}{\hbar}(\chi^{\tau^{O}}_{\bm{k},\eta})_{\lambda\lambda'} g_B(\bm{r}, \bar{\varepsilon}_{\bm{k},\eta}) \Big]
  \Bigg\},\\
  \pi^O_{\lambda,\nabla T}(\bm{r})
  &=\sum_{\lambda'}
  \frac{(-\nabla_{\lambda'} T)}{\hbar T} \sum_{\eta=1}^{2N} \int \frac{d\bm{k}}{(2\pi)^d}
  (\chi^{\tau^{O}}_{\bm{k},\eta})_{\lambda\lambda'} 
  [\bar{\varepsilon}_{\bm{k},\eta} f_B(\bm{r},\bar{\varepsilon}_{\bm{k},\eta}) - g_B(\bm{r},\bar{\varepsilon}_{\bm{k},\eta})],
  \label{eq:P_nablaT}\\
  \rho^O_{\lambda}(\bm{r})
  &=
  \frac{1}{\hbar T} \sum_{\eta=1}^{2N} \int \frac{d\bm{k}}{(2\pi)^d}
  (\Omega^{\tau^{O}}_{\bm{k}, \eta})_{\lambda} 
  [\bar{\varepsilon}_{\bm{k},\eta} f_B(\bm{r},\bar{\varepsilon}_{\bm{k},\eta}) - g_B(\bm{r},\bar{\varepsilon}_{\bm{k},\eta})].
\end{align}
From Eq.~\eqref{eq:tau_eq}, we find that the torque term can always be expressed as a divergence in the absence of a temperature gradient.  
If $\bm{\rho}^O(\bm{r})=0$ holds in Eq.~\eqref{eq:tau_T}, the torque term can still be expressed as a divergence even in the presence of a temperature gradient.  
\end{widetext}

Let us consider the case without a temperature gradient, where the torque term can always be expressed in the form of the divergence. 
In this situation, the conserved current is written as $\bm{\mathcal{J}}_{\rm eq}^{O}(\bm{r})=\braket{\bm{j}^{O}(\bm{r})}_{\rm eq} + \bm{\pi}_{\rm eq}^{O}(\bm{r})$.
This quantity can be expressed as the rotation of a vector as
\begin{align}
  \label{eq:mathcalJ_orbital moment}
  \bm{\mathcal{J}}_{\rm eq}^{O}(\bm{r})
  =
  \bm{\nabla} \times \bm{m}^{O}(\bm{r}),
\end{align}
where $\bm{m}^{O}(\bm{r})$ is regarded as the orbital magnetization, which is given by
\begin{align}\label{eq:orbital_moment}
  &\bm{m}^{O} (\bm{r})=\notag\\
  &
  \sum_{\eta=1}^{2N} \int \frac{d\bm{k}}{(2\pi)^d}
\left[
  \bm{m}^{O}_{\bm{k},\eta} f_B(\bm{r},\bar{\varepsilon}_{\bm{k},\eta})
  + \frac{1}{2\hbar}\bm{\Omega}_{\bm{k},\eta} \sigma_{3,\eta}(O_{\bm{k}})_{\eta} g_B(\bm{r},\bar{\varepsilon}_{\bm{k},\eta})
\right].
\end{align}
Here, $\bm{m}^{O}_{\bm{k},\eta}$ and $\bm{\Omega}_{\bm{k},\eta}$ are the vectors whose components are given by $(m^{O}_{\bm{k},\eta})_{\lambda''} = \frac{1}{2}\sum_{\lambda\lambda'}\varepsilon_{\lambda\lambda'\lambda''} (m^{O}_{\bm{k},n})_{\lambda\lambda'}$, and
$
\Omega^{\lambda''}_{\bm{k},\eta} = \frac{1}{2} \sum_{\lambda\lambda'}\varepsilon_{\lambda\lambda'\lambda''} \Omega^{\lambda\lambda'}_{\bm{k},\eta}$, respectively.
The representation given in Eq.~\eqref{eq:orbital_moment} for the orbital magnetization is similar to that for fermionic systems~\cite{xiao2006,xiao2021conserved}.  
The first term in Eq.~\eqref{eq:orbital_moment} represents the contribution from the self-rotational motion of a wave packet, while the second term arises from the center-of-mass motion of a wave packet~\cite{xiao2006,xiao2010,xiao2021conserved}.  
From Eq.~\eqref{eq:mathcalJ_orbital moment}, we demonstrate that the net current vanishes in the absence of a thermal gradient.  

On the other hand, in the presence of a temperature gradient, there is an additional term in the divergence form of the torque term, as shown in Eq.~\eqref{eq:tau_T}.
Hence, from Eq.~\eqref{eq:continuity-ave}, the equation of continuity is written as
\begin{align}\label{eq:continuity-ave-nablaT}
  \pdiff{\braket{O(\bm{r})}}{t} = - \nabla \cdot \bm{\mathcal{J}}^{O}(\bm{r}) - \bm{\rho}^O(\bm{r}) \cdot \bm{\nabla} T,
\end{align}
where $\bm{\mathcal{J}}^{O}(\bm{r})=\bm{\mathcal{J}}_{\rm eq}^{O}(\bm{r})+\bm{\mathcal{J}}_{\nabla T}^{O}(\bm{r})$ with $\bm{\mathcal{J}}_{\nabla T}^{O}(\bm{r})
= \braket{\bm{j}^{O}(\bm{r})}_{\nabla T}
+ \bm{\pi}_{\nabla T}^{O}(\bm{r})$.
When the second term in Eq.~\eqref{eq:continuity-ave-nablaT} is zero, $\bm{\mathcal{J}}^{O}(\bm{r})$ can be regarded as the current of the local quantity $O(\bm{r})$.
In the following calculations, we assume that $\bm{\rho}^O(\bm{r}) = 0$.
After the calculation, we will discuss the conditions under which $\bm{\rho}^O(\bm{r}) = 0$ holds.

Since the conserved current $\bm{\mathcal{J}}^{O}(\bm{r})$ is given by the divergence of the orbital magnetization in the absence of a thermal gradient as presented in Eq.~\eqref{eq:mathcalJ_orbital moment}, the current contributing to transport phenomena driven by a temperature gradient is introduced by subtracting orbital magnetization as
\begin{align}
  \bm{\mathcal{J}}^{O}_{\mathrm{tr}}
  =
  \frac{1}{V}
  \int
  d\bm{r}
  \left[
    \bm{\mathcal{J}}^{O}(\bm{r})
    -
    \bm{\nabla} \times \bm{m}^{O}(\bm{r})
  \right],
\end{align}
where $\bm{\mathcal{J}}^{O}_{\mathrm{tr}}$ is termed transport current~\cite{cooper1997}. 
The transport current is written as 
$
\bm{\mathcal{J}}^{O}_{\mathrm{tr}}=
  \frac{1}{V}
  \int d\bm{r}
    \bm{\mathcal{J}}^{O}_{\nabla T}(\bm{r})$, and hence, we evaluate $\braket{\bm{j}^{O}(\bm{r})}_{\nabla T}$ and $\bm{\pi}_{\nabla T}^{O}(\bm{r})$.
First, we focus on $\braket{\bm{j}^{O}(\bm{r})}_{\nabla T}$, which is given by the second term in Eq.~\eqref{eq:jcon_1st} as
\begin{align}
  \braket{j^{O}_{\lambda} (\bm{r})}_{\nabla T}
  =
  \frac{1}{\hbar}\sum_{\lambda'}\sum_{\eta=1}^{2N} \int 
  \frac{d\bm{k}}{(2\pi)^d}(\Omega^{j^{O}_\lambda}_{\bm{k},\eta})_{\lambda'}
  F_{\lambda'}(\bm{r},\bar{\varepsilon}_{\bm{k},\eta}).
\end{align}
In this representation, we separate the summation over $\eta$ into two parts: $1\leq \eta \leq N$ and $N+1\leq \eta \leq 2N$ as follow:
\begin{align}
  \label{appeq:jO_nablaT1}
  \braket{j^{O}_{\lambda}(\bm{r})}_{\nabla T}=
  \frac{1}{\hbar}\sum_{\lambda'}
  &\sum_{\eta = 1}^{N} \int \frac{d\bm{k}}{(2\pi)^d}
  \Bigl[
    (\Omega^{j^{O}_\lambda}_{\bm{k},\eta})_{\lambda'}
    F_{\lambda'}(\bm{r},\bar{\varepsilon}_{\bm{k},\eta})
    \notag\\
    &  
    +
    (\Omega^{j^{O}_\lambda}_{-\bm{k},\eta+N})_{\lambda'}
    F_{\lambda'}(\bm{r},\bar{\varepsilon}_{-\bm{k},\eta+N})
  \Bigr],
\end{align}
where the integral over $\bm{k}$ is changed to $-\bm{k}$ in the second line.
From the particle-hole symmetry [Eq.~\eqref{eq:PHS}] and Eq.~\eqref{eq:PHS-O}, we have $\sigma_{1} \bm{v}_{\bm{k}} \sigma_{1} = -\bm{v}_{-\bm{k}}^{*}$ and $\sigma_{1} \bm{j}^{O}_{\bm{k}} \sigma_{1} = \bm{j}^{O*}_{-\bm{k}}$, which lead to the following relations: 
$ (\Omega^{j^{O}_\lambda}_{-\bm{k},\eta+N})_{\lambda'}
=
(\Omega^{j^{O}_\lambda}_{\bm{k},\eta})_{\lambda'}$.
Furthermore, $F_{\lambda}(\bm{r},\varepsilon)=F_{\lambda}(\bm{r},-\varepsilon)$ is satisfied, and $F_{\lambda}(\bm{r},\varepsilon)$ for $\varepsilon>0$ is rewritten as~\cite{dong2020,li2020_prr},
\begin{align}
  F_{\lambda}(\bm{r},\varepsilon)
&= \frac{-\nabla_{\lambda} T}{T}
  \int_{\varepsilon}^{\infty} \omega
  \pdiff{f_B(\bm{r},\omega)}{\omega} d\omega\notag\\
  &=
  k_B (\nabla_{\lambda} T)
  c_1[f_B(\bm{r},\varepsilon)]
  ,
\end{align}
where $c_1(f_B) = (1+f_B)\ln (1+f_B) - f_B \ln f_B$.
Using the above relations, we can rewrite Eq.~\eqref{appeq:jO_nablaT1} as
\begin{align}
  \label{appeq:jO_nablaT3}
  \braket{j^{O}_{\lambda} (\bm{r})}_{\nabla T}
  &=
  -\frac{2k_B}{\hbar V} \sum_{\lambda'} \sum_{\eta=1}^{N} \sum_{\bm{k}} 
  (\Omega^{j^{O}_\lambda}_{\bm{k},\eta})_{\lambda'}
  c_1[f_B(\varepsilon_{\bm{k},\eta})]
  (-\nabla_{\lambda'} T),
\end{align}
where we omit the spatial dependence of temperature as $T(\bm{r})\to T$ as we consider the first order of temperature gradient, and $f_B(\bm{r},\varepsilon_{\bm{k},\eta})$ is denoted as $f_B(\varepsilon_{\bm{k},\eta})$.
% The sum of $\bm{k}$ is taken over the first Brillouin zone.
% , and $V$ represents the volume of the system.
Similarly, we calculate $\bm{\pi}^{O}_{\nabla T} (\bm{r})$ given in Eq.~\eqref{eq:P_nablaT} as
\begin{align}\label{eq:Pi_nablaT}
  \pi^{O}_{\lambda,\nabla T} (\bm{r})
  =
  -\frac{2k_B}{\hbar V}\sum_{\lambda'}\sum_{\eta = 1}^{N} \sum_{\bm{k}}
  (\chi^{\tau^{O}}_{\bm{k},\eta})_{\lambda\lambda'}
  c_1[f_B(\varepsilon_{\bm{k},\eta})] (-\nabla_{\lambda'} T).
\end{align}
From Eqs.~\eqref{appeq:jO_nablaT3} and \eqref{eq:Pi_nablaT}, $\bm{\mathcal{J}}^{O}_{\mathrm{tr}}$ is represented as
\begin{align}
    &\mathcal{J}^{O}_{\lambda,\mathrm{tr}}
  =
  \notag\\
  &-\frac{2k_B}{\hbar V}\sum_{\lambda'}\sum_{\eta = 1}^{N} \sum_{\bm{k}}
  \left[
    (\Omega^{j^{S^\gamma}_\lambda}_{\bm{k},\eta} )_{\lambda'} + (\chi^{\tau^{S^\gamma}}_{\bm{k},\eta})_{\lambda\lambda'}
  \right]
  c_1[f_B(\varepsilon_{\bm{k},\eta})] (-\nabla_{\lambda'} T).
\end{align}
The first term in the square brackets originates from the conventional current and the second term is attributed to a contribution from the torque.
Finally, using Eq.~\eqref{eq:chi}, we obtain the expression for $\bm{\mathcal{J}}^{O}_{\mathrm{tr}}$ as
\begin{align}
  &\mathcal{J}^{O}_{\lambda,\mathrm{tr}}
  =
  \notag\\
  &-\frac{k_B}{\hbar V}\sum_{\lambda'}\sum_{\eta = 1}^{N} \sum_{\bm{k}}
  \left[
    \Omega^{\lambda'\lambda}_{\bm{k},\eta} (O_{\bm{k}})_{\eta}
    + \pdiff{}{k_{\lambda'}} (d^{O}_{\bm{k},\eta})_{\lambda}
  \right]
  c_1[f_B(\varepsilon_{\bm{k},\eta})] (-\nabla_{\lambda'} T).
  \label{appeq:mathcalJ_nablaT}
\end{align}

\subsection{Condition for introducing conserved current}
\label{sec:cond-current}

In this section, we discuss the condition necessary to justify the assumption $\bm{\rho}^O(\bm{r})=0$.
By performing a similar calculation to the procedure used to obtain Eq.~\eqref{eq:Pi_nablaT}, we can evaluate $\bm{\rho}^{O}(\bm{r})$ as
\begin{align}
  \rho^{O}_{\lambda}(\bm{r})
  &=
  -\frac{2k_B}{\hbar V} \sum_{\eta=1}^{N} \sum_{\bm{k}} (\Omega^{\tau^{O}}_{\bm{k}, \eta})_{\lambda} 
  c_1[f_B(\varepsilon_{\bm{k},\eta})].
\end{align}
By transforming the summation over $\bm{k}$ to cover half of the first Brillouin zone, we can rewrite the above equation as
\begin{align}
  &\rho^{O}_{\lambda}(\bm{r})
  =
  \notag\\
  &-\frac{2k_B}{\hbar V} \sum_{\eta=1}^{N} \sum_{\bm{k}}^{\mathrm{half}}
  \Biggl\{
  (\Omega^{\tau^{O}}_{\bm{k}, \eta})_{\lambda}
  c_1[f_B(\varepsilon_{\bm{k},\eta})]
  +
  (\Omega^{\tau^{O}}_{-\bm{k}, \eta})_{\lambda}
  c_1[f_B(\varepsilon_{-\bm{k},\eta})]
  \Biggr\}.
\end{align}
If the system satisfies the condition $H_{0,\bm{k}} = H_{0,-\bm{k}}$, then 
$
\varepsilon_{\bm{k},\eta} = \varepsilon_{-\bm{k},\eta}
$
and 
$
(\Omega^{\tau^{O}}_{\bm{k},\eta})_{\lambda}
=
-(\Omega^{\tau^{O}}_{-\bm{k},\eta})_{\lambda}
$
are also satisfied, leading to $\bm{\rho}^O(\bm{r})=0$~\cite{li2020_prr}.
Therefore, the relation $H_{0,\bm{k}} = H_{0,-\bm{k}}$ is a sufficient condition for satisfying the equation of continuity with $\bm{\mathcal{J}}^{O}(\bm{r})$, and we can use Eq.~\eqref{eq:snc_withtorque} for the formula of the conserved current induced by a temperature gradient.
In a previous study~\cite{shi2006}, the conserved spin current was introduced in systems with inversion symmetry, which is encompassed by the relation $H_{0,\bm{k}} = H_{0,-\bm{k}}$.

\subsection{Spin Nernst effect with torque contribution}
\label{sec:snc}

In this section, we consider the spin Nernst effect for the conserved spin current $\bm{\mathcal{J}}^{S^\gamma}(\bm{r})$, where $O$ is chosen as the $\gamma$ component of a spin operator $\bm{S}$ with $\gamma=x,y,z$, by assuming $\bm{\rho}^{S^\gamma}(\bm{r})=0$.
We introduce the response tensor $\kappa^{S^\gamma}_{\lambda\lambda'}$ for the spin current generation by a temperature gradient as
\begin{align}
  \label{eq:snc_def}
  \mathcal{J}^{S^\gamma}_{\lambda,\mathrm{tr}}
  =\kappa^{S^\gamma}_{\lambda\lambda'} (-\nabla_{\lambda'} T),
\end{align}
where the antisymmetric component of $\kappa^{S^\gamma}_{\lambda\lambda'}$ is regarded as the spin Nernst conductivity.
From Eq.~\eqref{appeq:mathcalJ_nablaT}, the response tensor $\kappa^{S^\gamma}_{\lambda\lambda'}$ is represented as
\begin{align}
  \label{eq:snc_withtorque}
  \kappa^{S^\gamma}_{\lambda\lambda'}
  &=
  -\frac{k_B}{\hbar V} \sum_{\eta=1}^{N} \sum_{\bm{k}} \big( \Upsilon^{S^\gamma}_{\bm{k},\eta} \big)_{\lambda\lambda'}
  c_1[f_B(\varepsilon_{\bm{k},\eta})],
\end{align}
where
$\big(\Upsilon^{S^{\gamma}}_{\bm{k},\eta}\big)_{\lambda\lambda'}$ is given by
\begin{align}
  \label{eq:Upsilon_chi}
  \big(\Upsilon^{S^{\gamma}}_{\bm{k},\eta}\big)_{\lambda\lambda'}
  &=
  2 (\Omega^{j^{S^\gamma}_\lambda}_{\bm{k},\eta} )_{\lambda'} + 2(\chi^{\tau^{S^\gamma}}_{\bm{k},\eta})_{\lambda\lambda'}
  \\
  &=
  -\Omega^{\lambda\lambda'}_{\bm{k},\eta} (S^{\gamma}_{\bm{k}})_{\eta} + \pdiff{}{k_{\lambda'}} (d^{S^{\gamma}}_{\bm{k},\eta})_{\lambda},
  \label{eq:Upsilon}
\end{align}
which is consistent with a formula for fermionic systems obtained the previous study~\cite{xiao2021conserved}.
From Eq.~\eqref{eq:relation-g-d}, we find 
$(d^{S^\gamma}_{\bm{k},\eta})_{\lambda}
=
2\big(g^{\tau^{S^\gamma}}_{\bm{k},\eta}\big)_{\lambda}
=
2g_{\bm{k},\eta}^{k_{\lambda}h_\gamma^\tau}$
where $h_\gamma^\tau$ is the source field for the torque term $\tau^{S^\gamma}$ in Eq.~\eqref{appeq:H_semiclassical}.
Therefore, while the first term in Eq.~\eqref{eq:Upsilon} is represented by the Berry curvature, the second term is related to the quantum metric in the mixed space $(\bm{k},\bm{h}^{\tau})$, which is the same as fermionic systems~\cite{xiao2021conserved}.

Since the second term in Eq.~\eqref{eq:Upsilon_chi} is interpreted to originate from the torque term, this contribution is not included in previous studies for the spin Nernst effect introduced on the basis of the conventional spin current~\cite{li2020_prr}.
Such earlier theoretical studies have used the formula for the response tensor based on Eq.~\eqref{appeq:jO_nablaT3}, which is explicitly given by
\begin{align}
  \label{eq:snc_withouttorque}
  \tilde{\kappa}^{S^{\gamma}}_{\lambda\lambda'}
  = -
  \frac{k_B}{\hbar V} \sum_{\eta=1}^{N}\sum_{\bm{k}}
  \big( \tilde{\Omega}^{S^\gamma}_{\bm{k},\eta} \big)_{\lambda\lambda'} c_1[f_B(\varepsilon_{\bm{k},\eta})].
\end{align}
Here, $\big( \tilde{\Omega}^{S^\gamma}_{\bm{k},\eta} \big)_{\lambda\lambda'}=2(\Omega^{j_\lambda^{S^\gamma}}_{\bm{k},\eta})_{\lambda'}$ is known as the spin Berry curvature and corresponds to the first term in Eq.~\eqref{eq:Upsilon_chi}.
From Eq.~\eqref{eq:GBC-O}, this quantity is written as,
\begin{align}
  \label{eq:Spin-Berry}
  \big( \tilde{\Omega}^{S^\gamma}_{\bm{k},\eta} \big)_{\lambda\lambda'}
  =
  -4\hbar^2 \sum_{\eta_1 (\neq \eta)}^{2N} \sigma_{3,\eta}\sigma_{3,\eta_1} \frac{\mathrm{Im}\left[ (v_{\bm{k},\lambda'})_{\eta\eta_1} (j^{S^\gamma}_{\bm{k},\lambda})_{\eta_1\eta} \right]}
  {(\bar{\varepsilon}_{\bm{k},\eta}-\bar{\varepsilon}_{\bm{k},\eta_1})^2}.
\end{align}
When the Hamiltonian commutes with the total spin operator, which is regarded as a conserved quantity, the torque term is absent, and the relations $\big( \tilde{\Omega}^{S^\gamma}_{\bm{k},\eta} \big)_{\lambda\lambda'}
=
-\Omega^{\lambda\lambda'}_{\bm{k},\eta} (S^{\gamma}_{\bm{k}})_{\eta}$ and $(d^{S^\gamma}_{\bm{k},\eta})_{\lambda}=0$ hold, which leads to the coincidence of the response tensor $\kappa^{S^\gamma}_{\lambda\lambda'}$ in Eq.~\eqref{eq:snc_withtorque} with $\tilde{\kappa}^{S^{\gamma}}_{\lambda\lambda'}$ in Eq.~\eqref{eq:snc_withouttorque}.
On the other hand, in spin-nonconserving systems, $\big( \tilde{\Omega}^{S^\gamma}_{\bm{k},\eta} \big)_{\lambda\lambda'}$ is not equal to $
-\Omega^{\lambda\lambda'}_{\bm{k},\eta} (S^{\gamma}_{\bm{k}})_{\eta}$, in addition to $(d^{S^\gamma}_{\bm{k},\eta})_{\lambda}\ne 0$.

Finally, we comment on the numerical evaluation of the response tensors $\kappa^{S^\gamma}_{\lambda\lambda'}$ and $\tilde{\kappa}^{S^{\gamma}}_{\lambda\lambda'}$.  
In our scheme, the mean fields $\braket{O^{\alpha}}_{l}$ in Eq.~\eqref{eq:HiMF} are determined through self-consistent calculations, and the one-body bosonic Hamiltonian $H_{0,\bm{k}}$ is formulated as a $2N\times 2N$ matrix using the linear flavor-wave approximation.  
This matrix is diagonalized by the Bogoliubov transformation $T_{\bm{k}}$, yielding the eigenvalues $\mathcal{E}_{\bm{k}}$.  
We also introduce the $2N\times 2N$ matrices $\hat{S}^x$, $\hat{S}^y$, and $\hat{S}^z$ from Eq.~\eqref{eq:O_total}.  
Using these quantities, the Berry curvature $\Omega^{\lambda\lambda'}_{\bm{k},\eta}$ and the spin component $(S^{\gamma}_{\bm{k}})_{\eta}$ in Eq.~\eqref{eq:Upsilon} are computed from Eq.~\eqref{eq:BC}, with $(v_{\bm{k},\lambda})_{\eta\eta'}=[T_{\bm{k}}^\dagger (\partial_{k_\lambda} H_{0,\bm{k}}) T_{\bm{k}}]_{\eta\eta'} /\hbar$ and $(S^{\gamma}_{\bm{k}})_{\eta} = [T_{\bm{k}}^\dagger  \hat{S}^\gamma T_{\bm{k}}]_{\eta\eta}$.  
On the other hand, the second term in Eq.~\eqref{eq:Upsilon} is given by the momentum derivative of the dipole moment, which is challenging to evaluate numerically.  

To avoid this difficulty, we derive the following expression for the response tensor $\kappa^{S^\gamma}_{\lambda\lambda'}$ without the momentum derivative:
\begin{align}\label{eq:derivative-d}
  \partial_{k_{\lambda'}} (d^{S^\gamma}_{\bm{k},\eta})_{\lambda}
  =
  ( D_{\bm{k},\eta}^{S^\gamma})^{\mathrm{(s)}}_{\lambda\lambda'}
  +
  ( D_{\bm{k},\eta}^{S^\gamma})^{\mathrm{(a)}}_{\lambda\lambda'}
\end{align}
where $( D_{\bm{k},\eta}^{S^\gamma})^{\mathrm{(s)}}_{\lambda\lambda'}=( D_{\bm{k},\eta}^{S^\gamma})^{\mathrm{(1)}}_{\lambda\lambda'}
+( D_{\bm{k},\eta}^{S^\gamma})^{\mathrm{(1)}}_{\lambda'\lambda}$
and
$( D_{\bm{k},\eta}^{S^\gamma})^{\mathrm{(a)}}_{\lambda\lambda'}=( D_{\bm{k},\eta}^{S^\gamma})^{\mathrm{(2)}}_{\lambda\lambda'}
-( D_{\bm{k},\eta}^{S^\gamma})^{\mathrm{(2)}}_{\lambda'\lambda}$
are the symmetric and antisymmetric parts of $\partial_{k_{\lambda'}} (d^{S^\gamma}_{\bm{k},\eta})_{\lambda}$, and
$( D_{\bm{k},\eta}^{S^\gamma})^{\mathrm{(1)}}_{\lambda\lambda'}$ and $( D_{\bm{k},\eta}^{S^\gamma})^{\mathrm{(2)}}_{\lambda\lambda'}$
are defined as
\begin{widetext}
\begin{align}\label{eq:Ds-Da}
  ( D_{\bm{k},\eta}^{S^\gamma})^{\mathrm{(1)}}_{\lambda\lambda'}
  =
  &  - \hbar^2 \mathrm{Im} \sum_{\eta_1(\neq \eta)}^{2N}
    \sigma_{3,\eta}\sigma_{3,\eta_1} [\sigma_{3,\eta} (v_{\bm{k},\lambda'})_{\eta} - \sigma_{3,\eta_1} (v_{\bm{k},\lambda'})_{\eta_1}]
    \frac{(v_{\bm{k},\lambda})_{\eta\eta_1}(S_{\bm{k}}^{\gamma})_{\eta_1\eta}}{(\bar{\varepsilon}_{\bm{k},\eta} - \bar{\varepsilon}_{\bm{k},\eta_1})^2}
  %   \notag\\
  % &
    + \frac{\hbar^2}{2} \mathrm{Im} \sum_{\eta_1(\neq \eta)}^{2N}
    \sigma_{3,\eta}\sigma_{3,\eta_1} \frac{(\mu_{\bm{k},\lambda\lambda'}^{-1})_{\eta\eta_1}(S_{\bm{k}}^{\gamma})_{\eta_1\eta}}
    {\bar{\varepsilon}_{\bm{k},\eta}-\bar{\varepsilon}_{\bm{k},\eta_1}}\notag\\
  &\qquad
  + \hbar^2\mathrm{Im} 
  \sum_{\eta_1(\neq \eta)}^{2N} \sum_{\eta_2 (\neq \eta,\eta_1)}^{2N}
  \sigma_{3,\eta}\sigma_{3,\eta_1}\sigma_{3,\eta_2}
    \frac{(v_{\bm{k},\lambda'})_{\eta\eta_2}(v_{\bm{k},\lambda})_{\eta_2\eta_1} (S_{\bm{k}}^{\gamma})_{\eta_1\eta}}
    {(\bar{\varepsilon}_{\bm{k},\eta} - \bar{\varepsilon}_{\bm{k},\eta_1})(\bar{\varepsilon}_{\bm{k},\eta} - \bar{\varepsilon}_{\bm{k},\eta_2})},\\
  %  \notag\\
  ( D_{\bm{k},\eta}^{S^\gamma})^{\mathrm{(2)}}_{\lambda\lambda'}
  =
  &-\frac{\hbar^2}{2} \mathrm{Im} \sum_{\eta_1(\neq \eta)}^{2N}
    \sigma_{3,\eta} \sigma_{3,\eta_1} [\sigma_{3,\eta} (S_{\bm{k}}^{\gamma})_{\eta} - \sigma_{3,\eta_1} (S_{\bm{k}}^{\gamma})_{\eta_1}]
    \frac{(v_{\bm{k},\lambda})_{\eta\eta_1}(v_{\bm{k},\lambda'})_{\eta_1\eta}}{(\bar{\varepsilon}_{\bm{k},\eta} - \bar{\varepsilon}_{\bm{k},\eta_1})^2}
    \notag\\
&\qquad
    + \frac{\hbar^2}{2}\mathrm{Im} \sum_{\eta_1(\neq \eta)}^{2N} \sum_{\eta_2 (\neq \eta,\eta_1)}^{2N} \sigma_{3,\eta}\sigma_{3,\eta_1}\sigma_{3,\eta_2}
    \frac{(v_{\bm{k},\lambda})_{\eta\eta_2} (S_{\bm{k}}^{\gamma})_{\eta_1\eta_2} (v_{\bm{k},\lambda'})_{\eta_2\eta}}
    {(\bar{\varepsilon}_{\bm{k},\eta}-\bar{\varepsilon}_{\bm{k},\eta_1})(\bar{\varepsilon}_{\bm{k},\eta}-\bar{\varepsilon}_{\bm{k},\eta_2})}.
\end{align}
\end{widetext}
Here, $(S^{\gamma}_{\bm{k}})_{\eta\eta'} = [T_{\bm{k}}^\dagger \hat{S}^\gamma T_{\bm{k}}]_{\eta\eta'}$, and $(\mu_{\bm{k},\lambda\lambda'}^{-1})_{\eta\eta'}=[T_{\bm{k}}^\dagger(\partial_{k_{\lambda}} \partial_{k_{\lambda'}}H_{0,\bm{k}})T_{\bm{k}}]_{\eta\eta'}/\hbar^2$ is the inverse mass tensor.
The detailed derivation of Eqs.~\eqref{eq:derivative-d} and \eqref{eq:Ds-Da} is given in the appendix~\ref{appendix:derivative}.
We can also evaluate the spin Berry curvature in Eq.~\eqref{eq:Spin-Berry} by using the the following expression of the conventional spin current: 
\begin{align}
    (j^{S^\gamma}_{\bm{k},\lambda})_{\eta\eta'}
    =\frac{1}{4}\left(v_{\bm{k},\lambda} \sigma_{3} S_{\bm{k}}^{\gamma}+
    S_{\bm{k}}^{\gamma}\sigma_{3} v_{\bm{k},\lambda}
     \right)_{\eta\eta'}. 
\end{align}

\section{Numerical calculations for spin Nernst effect}
\label{sec:numerical-calc}

In this section, we demonstrate numerical calculations for the spin Nernst effect in two specific spin models without spin conservation: the Kitaev-Heisenberg model and the Shastry-Sutherland model with DM interactions.

In the following calculations, we focus on the spin Nernst effect, and hence, we introduce the antisymmetric part of $\kappa^{S^\gamma}_{\lambda\lambda'}$ in Eq.~\eqref{eq:snc_withtorque} and $\tilde{\kappa}^{S^\gamma}_{\lambda\lambda'}$ in Eq.~\eqref{eq:snc_withouttorque} as $\kappa^{S^\gamma\mathrm{(a)}}_{\lambda\lambda'}=(\kappa^{S^\gamma}_{\lambda\lambda'}-\kappa^{S^\gamma}_{\lambda'\lambda})/2$ and $\tilde{\kappa}^{S^\gamma\mathrm{(a)}}_{\lambda\lambda'}=(\tilde{\kappa}^{S^\gamma}_{\lambda\lambda'}-\tilde{\kappa}^{S^\gamma}_{\lambda'\lambda})/2$.
Similarly, we define $\big(\Upsilon^{S^{\gamma}}_{\bm{k},\eta}\big)_{\lambda\lambda'}^{\mathrm{(a)}}$ and $\big( \tilde{\Omega}^{S^\gamma}_{\bm{k},\eta} \big)_{\lambda\lambda'}^{\mathrm{(a)}}$ as
\begin{align}\label{eq:Upsilon_a}
    \big(\Upsilon^{S^{\gamma}}_{\bm{k},\eta}\big)_{\lambda\lambda'}^{\mathrm{(a)}}&=-\Omega^{\lambda\lambda'}_{\bm{k},\eta} (S^{\gamma}_{\bm{k}})_{\eta}+
    ( D_{\bm{k},\eta}^{S^\gamma})^{\mathrm{(a)}}_{\lambda\lambda'},\\
    \big( \tilde{\Omega}^{S^\gamma}_{\bm{k},\eta} \big)_{\lambda\lambda'}^{\mathrm{(a)}}
    &=\left\{\big( \tilde{\Omega}^{S^\gamma}_{\bm{k},\eta} \big)_{\lambda\lambda'}-\big( \tilde{\Omega}^{S^\gamma}_{\bm{k},\eta} \big)_{\lambda'\lambda}\right\}/2,
\end{align}
where we use the antisymmetric property of $\Omega^{\lambda\lambda'}_{\bm{k},\eta}$ with respect to $\lambda$ and $\lambda'$.
In addition, we assume that the reduced Planck constant $\hbar$ and Boltzmann constant $k_B$ are unity for simplicity in the following calculations.
Results without antisymmetrization are provided in Appendix~\ref{app:symmetry-tensors}.

\subsection{Kitaev-Heisenberg model}
\label{sec:KHmodel}

\begin{figure}[t]
  \centering
      \includegraphics[width=\columnwidth,clip]{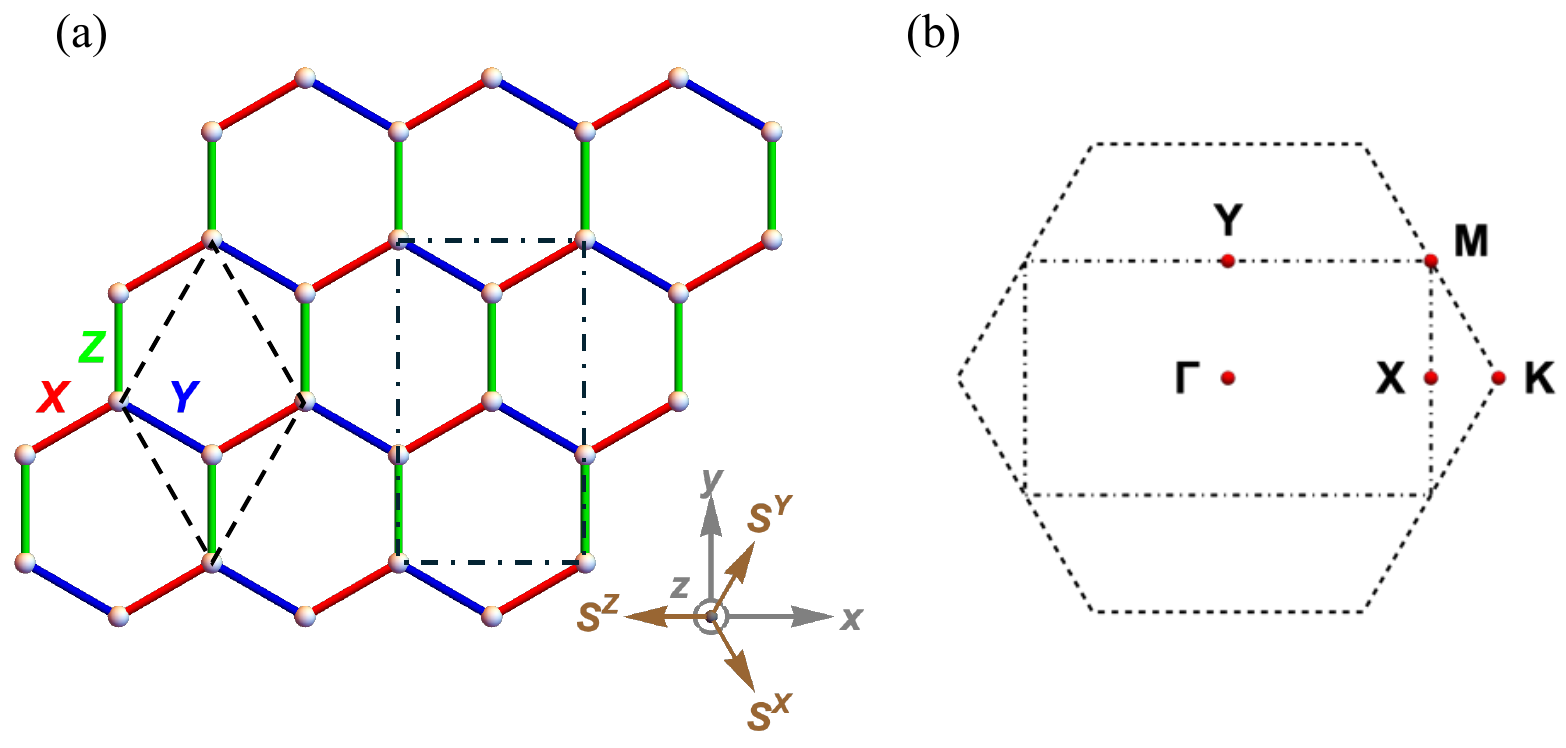}
      \caption{
(a) Schematic illustration of a honeycomb lattice on which the Kitaev model is defined.  
Red, blue, and green lines represent $X$, $Y$, and $Z$ bonds, respectively.  
The dashed line represents a unit cell of the honeycomb lattice, and the dash-dotted line denotes a magnetic unit cell including four sites.  
The latter is used in calculations for $\varphi=1.6\pi$ to realize the canted zigzag order shown in Fig.~\ref{fig:snc_kitaevheisen}(b).  
The inset represents the configurations of the real-space coordinate axes $(x, y, z)$ and the spin axes $(S^X, S^Y, S^Z)$.  
(b) First Brillouin zone shown by the dashed line.  
The dash-dotted line represents the Brillouin zone corresponding to the magnetic unit cell depicted by the dash-dotted line in panel (a). 
      }
      \label{fig:honeycomb_kitaev}
\end{figure}

We first apply our theory to the Kitaev-Heisenberg model, which is described by the following Hamiltonian~\cite{janssen2016,chaloupka2010,chaloupka2013}:
\begin{align}\label{eq:Kitaev-Heisenberg}
  \mathcal{H} = 2K \sum_{\braket{ij}_{\gamma}} S_{i}^{\gamma} S_{j}^{\gamma}
  + J \sum_{\braket{ij}} \bm{S}_{i} \cdot \bm{S}_{j}
  - \sum_{i} \bm{h} \cdot \bm{S}_{i},
\end{align}
where $\bm{S}_i=(S_i^X,S_i^Y,S_i^Z)$ represents an $S=1/2$ spin at site $i$ on a honeycomb lattice.
Here, $K$ and $J$ denote the strengths of the Kitaev and Heisenberg interactions, respectively, and $\bm{h}$ represents an external magnetic field.  
The honeycomb lattice on which this model is defined consists of three types of nearest-neighbor bonds, denoted as $\braket{ij}_{\gamma}$, where $\gamma=X,Y,Z$ [see Fig.~\ref{fig:honeycomb_kitaev}(a)].  
Note that the honeycomb lattice lies in the $xy$ plane, and we define the $z$ axis to be parallel to the [111] direction in the spin space $(S^X,S^Y,S^Z)$.  
We parametrize the Kitaev and Heisenberg interactions as $K=A\sin\varphi$ and $J=A\cos\varphi$ by introducing the parameter $\varphi$, where $A>0$ is set as the energy scale of the system.
The magnetic field is applied perpendicular to the honeycomb lattice, i.e., $\bm{h}=h(1,1,1)/\sqrt{3}$ in the spin space.

While the Kitaev model was originally proposed as an $S=1/2$ spin model exhibiting a quantum-spin-liquid ground state~\cite{kitaev,Nussinov2015,Hermanns2018rev,Knolle2019rev,takagi2019rev,Janssen_2019rev,Motome2020rev,Trebst2022rev,Matsuda2025rev_arxiv}, the Kitaev-Heisenberg model in a magnetic field is known to host a variety of magnetically ordered states due to the competition between the Kitaev and Heisenberg interactions.
Previous studies have explored the ground-state phase diagram~\cite{chaloupka2013,janssen2016,Janssen_2019rev,consoli2020,fukui2022,zou2025_arxiv} and the thermal Hall effects caused by elementary excitations~\cite{nasu2017,koyama2021,chern2021_prl,zhang2021_prb,li2022_prb,kumar2023,li2025}.
However, less is known about the spin Nernst effect in this model.
The lack of studies on spin transport phenomena is attributed to the presence of the Kitaev interaction, which does not commute with total spin operators, resulting in the emergence of torque terms.
Meanwhile, the Kitaev interaction renders the magnon bands topologically nontrivial, as it originates from spin-orbit coupling.
Thus, the Kitaev-Heisenberg model serves as a good platform for studying the spin Nernst effect in the absence of total spin conservation.

We perform a mean-field analysis to determine the magnetic order realized in the Kitaev-Heisenberg model, assuming two or four sites in the magnetic unit cell, i.e., $M=2$ or $4$, depending on the parameters $\varphi$ and $\bm{h}$.
After determining the ordered state, we construct the spin-wave Hamiltonian and diagonalize it using the Bogoliubov transformation, as presented in Sec.~\ref{sec:snc}.
We introduce the spin density operator as $S^\gamma(\bm{r}) = \frac{1}{2}\mathcal{A}^\dagger (\bm{r}) \hat{S}^\gamma \mathcal{A}(\bm{r})$ with $\hat{S}^\gamma = \bm{1}_{2\times 2} \otimes \bar{S}^{\gamma}$ being a $2N\times 2N$ diagonal matrix, where 
$\bar{S}^{\gamma}_{(l,m)} = \braket{m;i|S^{\gamma}|m;i} - \braket{0;i|S^{\gamma}|0;i}$ at site $i$ belonging to sublattice $l$.
Note that, for $S=1/2$ spin systems, the local Hilbert space is spanned by two basis states, namely, $\mathscr{N}=2$, and the index $m$ takes only one value, $m=1$, as an excited state.
In the following calculations, we examine the properties of the Kitaev-Heisenberg model for $\varphi = 3\pi/2, 1.6\pi$, where the magnetic field is fixed at $h / A = 0.1$.
For the spin Nernst effect, we apply a temperature gradient along the $y$ axis and observe the spin current flowing along the $x$ axis.

\begin{figure}[t]
  \centering
      \includegraphics[width=1\columnwidth,clip]{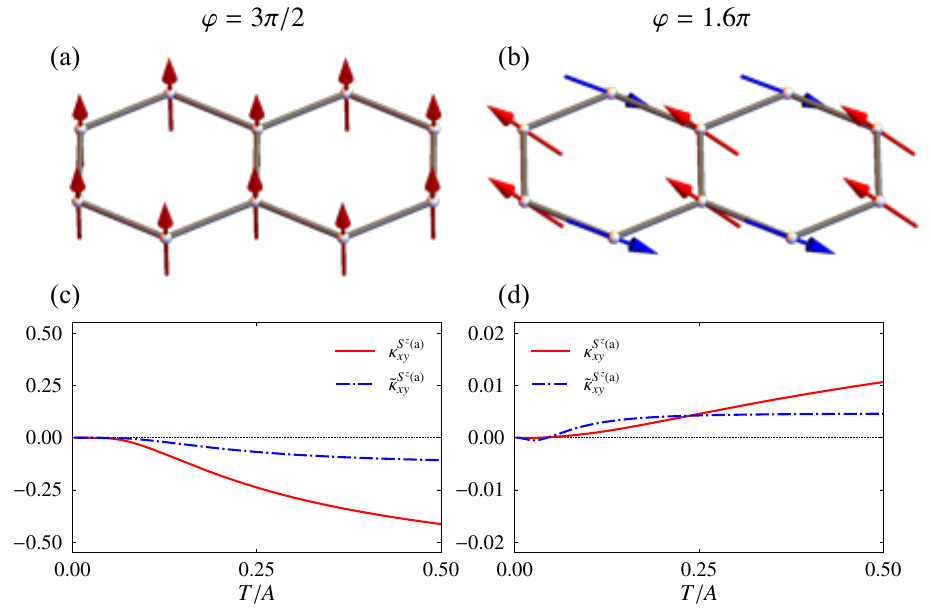}
      \caption{
        [(a),(b)] Mean-field ground states realized in the Kitaev-Heisenberg model at (a) $\varphi=3\pi/2$ and (b) $\varphi=1.6\pi$.
For the latter case, we perform four-sublattice calculations, where the magnetic unit cell is indicated by the dash-dotted line in Fig.~\ref{fig:honeycomb_kitaev}(a).
[(c),(d)] Temperature dependence of the spin Nernst coefficients $\kappa^{S^z\mathrm{(a)}}_{xy}$ and $\tilde{\kappa}^{S^z\mathrm{(a)}}_{xy}$ at (a) $\varphi=3\pi/2$ and (b) $\varphi=1.6\pi$.
The applied magnetic field is fixed to $h/A = 0.1$.
      }
      \label{fig:snc_kitaevheisen}
\end{figure}

\begin{figure}[t]
  \centering
      \includegraphics[width=1.0\columnwidth,clip]{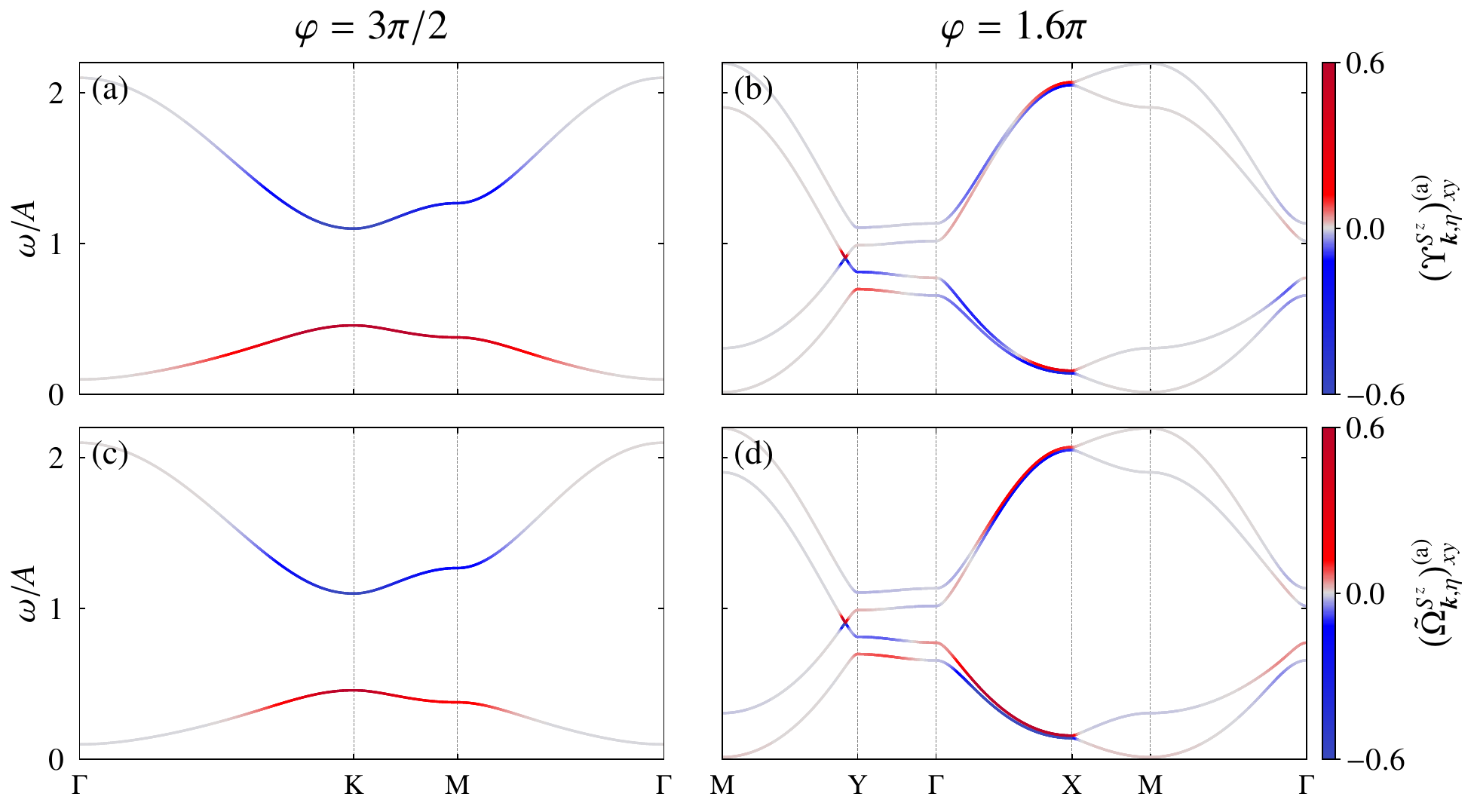}
      \caption{
[(a),(c)] Energy dispersions of magnon excitations in the Kitaev-Heisenberg model at $\varphi=3\pi/2$ under a magnetic field with $h/A = 0.1$ along the high-symmetry points in the Brillouin zone, as depicted in Fig.~\ref{fig:honeycomb_kitaev}(c).  
The color of the lines represents the values of (a) $\big(\Upsilon^{S^z}_{\bm{k},\eta}\big)_{xy}^{\mathrm{(a)}}$ and (c) $\big( \tilde{\Omega}^{S^\gamma}_{\bm{k},\eta} \big)_{xy}^{\mathrm{(a)}}$.  
[(b),(d)] Corresponding plots for $\varphi=1.6\pi$.  
      }
      \label{fig:coef_kitaevheisen}
\end{figure}

First, we consider the case of $\varphi=3\pi/2$, which corresponds to the ferromagnetic Kitaev model under a magnetic field.  
In this case, the ground state becomes a forced ferromagnetic state, where the spin moments align along the magnetic-field direction [Fig.~\ref{fig:snc_kitaevheisen}(a)], which is consistent with previous studies~\cite{janssen2016, mcclarty2018, koyama2021}.  
In this situation, there is an inversion center at the hexagonal plaquette, and consequently, the magnon Hamiltonian satisfies $H_{0,\bm{k}} = H_{0,-\bm{k}}$, allowing the use of Eq.~\eqref{eq:snc_withtorque} to evaluate the coefficient $\kappa^{S^\gamma}_{xy}$.  
We have confirmed that the spin component of $\kappa^{S^\gamma}_{xy}$ is nonzero only in the direction of the magnetization and the applied field, which is perpendicular to the honeycomb plane.  
Since all spin moments are oriented in the $z$ direction and fully polarized, $S_i^z$ commutes with the local mean-field Hamiltonian $\mathcal{H}_i^{\rm MF}$, indicating that the spin current $\bm{\mathcal{J}}^{S^z}_{\mathrm{tr}}$ is finite.  
Here, we introduce this spin component as $S^z=(S^X+S^Y+S^Z)/\sqrt{3}$, and we calculate the spin Nernst coefficient  
$\kappa^{S^z}_{xy}=
(\kappa^{S^X}_{xy}
+ \kappa^{S^Y}_{xy} +\kappa^{S^Z}_{xy}) / \sqrt{3}$.  
Note that the other spin components along the $x$ and $y$ directions are given by $S^x=(S^X+S^Y-2S^Z)/\sqrt{3}$ and $S^y=(-S^X+S^Y)/\sqrt{2}$, respectively.

The temperature dependence of the antisymmetric part $\kappa^{S^z\mathrm{(a)}}_{xy}$, namely, the spin Nernst coefficient, at $\varphi = 3\pi/2$ is shown in Fig.~\ref{fig:snc_kitaevheisen}(d) as a solid line.
We also present the spin Nernst coefficient for the conventional spin current using Eq.~\eqref{eq:snc_withouttorque}, depicted as a dash-dotted line.
Note that we have also confirmed that the symmetric components of $\kappa^{S^z}_{xy}$ and $\tilde{\kappa}^{S^z}_{xy}$ with respect to $xy$ are zero.
We find that both $\kappa^{S^z\mathrm{(a)}}_{xy}$ and $\tilde{\kappa}^{S^z\mathrm{(a)}}_{xy}$ are negative and decrease as the temperature decreases.
Moreover, they are nearly zero below $T/A \sim 0.08$.
The most significant difference between $\kappa^{S^z\mathrm{(a)}}_{xy}$ and $\tilde{\kappa}^{S^z\mathrm{(a)}}_{xy}$ is their magnitude; the absolute value of $\kappa^{S^z\mathrm{(a)}}_{xy}$ is larger than that of $\tilde{\kappa}^{S^z\mathrm{(a)}}_{xy}$, indicating that considering the conserved spin current is crucial for the spin Nernst effect in the Kitaev model under magnetic fields.
This difference originates from the coefficient $\big(\Upsilon^{S^z}_{\bm{k},\eta}\big)_{xy}^{\mathrm{(a)}}$ and the spin Berry curvature $\big(\tilde{\Omega}^{S^z}_{\bm{k},\eta}\big)_{xy}^{\mathrm{(a)}}$, as given in Eqs.~\eqref{eq:snc_withtorque} and \eqref{eq:snc_withouttorque}, respectively.
These coefficients are independent of temperature, and the temperature dependence of the spin Nernst coefficient is governed by $c_1[f_B(\varepsilon_{\bm{k},\eta})]$.
The function $c_1[f_B(\varepsilon)]$ is positive and a decreasing function of $\varepsilon$, implying that the spin Nernst coefficient is primarily influenced by low-energy magnons.
As the temperature decreases, the low-energy part of $c_1[f_B(\varepsilon)]$ is enhanced, whereas the ratio of $c_1[f_B(\varepsilon)]$ between the regions with large $\varepsilon$ and small $\varepsilon$ diminishes at high temperatures.
Therefore, to understand the temperature dependence of the spin Nernst coefficient, it is necessary to examine the magnon bands and the coefficients $\big(\Upsilon^{S^z}_{\bm{k},\eta}\big)_{\lambda\lambda'}^{\mathrm{(a)}}$ and $\big(\tilde{\Omega}^{S^z}_{\bm{k},\eta}\big)_{\lambda\lambda'}^{\mathrm{(a)}}$ in momentum space.

\begin{figure}[t]
  \centering
      \includegraphics[width=\columnwidth,clip]{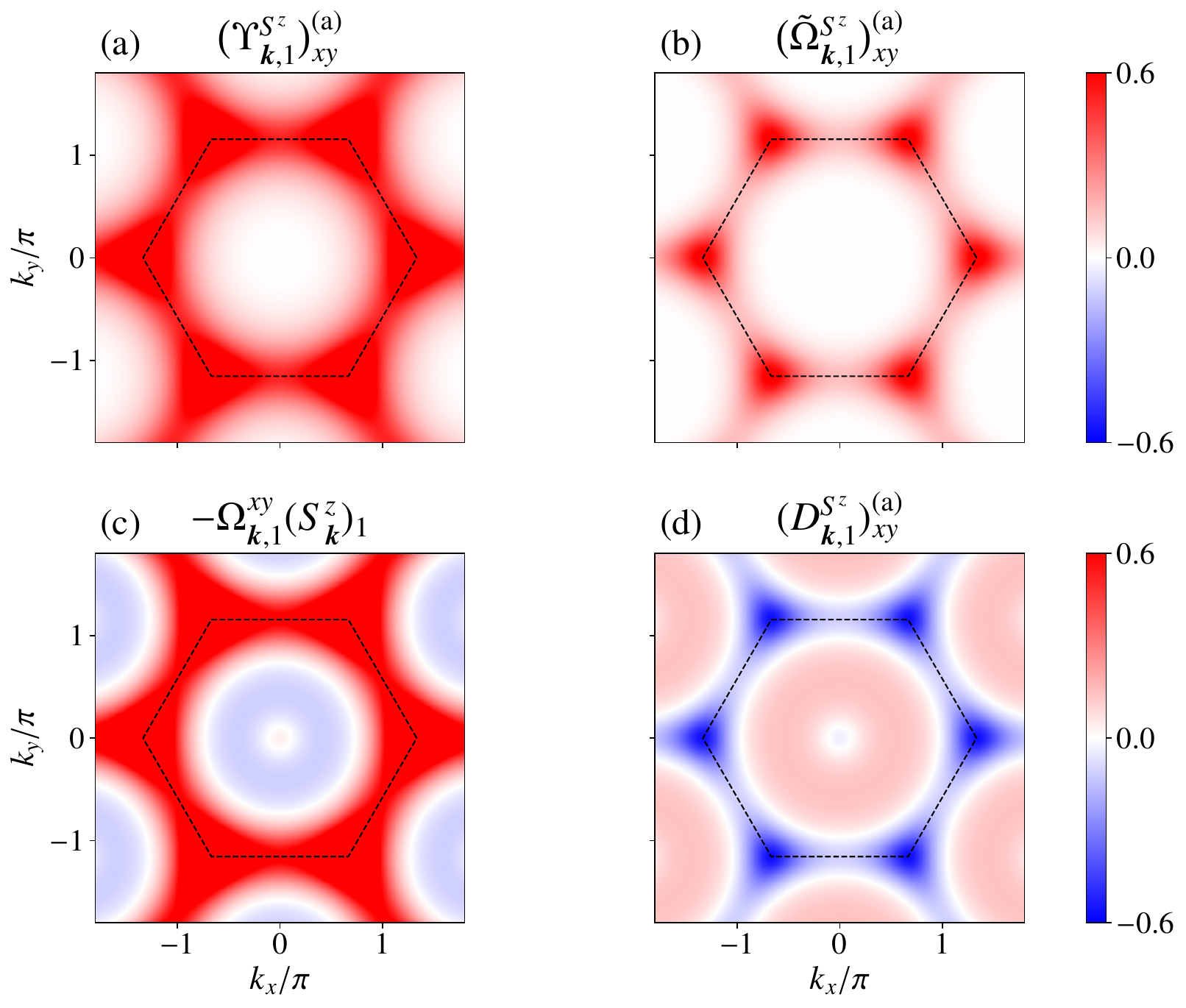}
      \caption{
Color maps of (a) $\big(\Upsilon^{S^z}_{\bm{k},\eta}\big)_{xy}^{\mathrm{(a)}}$, (b) $\big( \tilde{\Omega}^{S^\gamma}_{\bm{k},\eta} \big)_{xy}^{\mathrm{(a)}}$, (c)  $-\Omega^{xy}_{\bm{k},\eta} (S^z_{\bm{k}})_{\eta}$, and (d) $\big(D^{S^z}_{\bm{k},\eta}\big)_{xy}^{\mathrm{(a)}}$ for the lowest-energy branch with $\eta=1$ in the Kitaev-Heisenberg model at $\varphi=3\pi/2$ under a magnetic field with $h/A = 0.1$.
Dashed lines represent the first Brillouin zone of the honeycomb lattice.
      }
      \label{fig:coef_kitaevheisen_all}
\end{figure}

Figure~\ref{fig:coef_kitaevheisen}(a) shows the magnon dispersions, with the value of $\big(\Upsilon^{S^z}_{\bm{k},\eta}\big)_{xy}^{\mathrm{(a)}}$ for each branch represented by the line color.  
There are two magnon branches in the forced ferromagnetic state because the honeycomb lattice contains two sites per unit cell.  
In the low-energy branch, an excitation gap with $\omega/A \sim 0.1$ exists at the $\Gamma$ point, while the excitation energy reaches its maximum at the $\mathrm{K}$ point.  
This low-energy structure of the magnon bands is consistent with the temperature dependence of the spin Nernst coefficient shown in Fig.~\ref{fig:snc_kitaevheisen}(d), where $\kappa^{S^z\mathrm{(a)}}_{xy}$ remains nearly zero below $T/A \sim 0.08$.  
In the low-energy branch, the value of $\big(\Upsilon^{S^z}_{\bm{k},\eta}\big)_{xy}^{\mathrm{(a)}}$ is negative around the $\Gamma$ point but becomes positive in other regions of the Brillouin zone.  
To examine the momentum dependence in more detail, we present the distribution of $\big(\Upsilon^{S^z}_{\bm{k},\eta}\big)_{xy}^{\mathrm{(a)}}$ for $\eta=1$, corresponding to the lower-energy branch, on the $k_x$-$k_y$ plane in Fig.~\ref{fig:coef_kitaevheisen_all}(a).  
We find that $\big(\Upsilon^{S^z}_{\bm{k},\eta}\big)_{xy}^{\mathrm{(a)}}$ takes positive values along the Brillouin zone boundary and is suppressed in the vicinity of the $\Gamma$ point, which is consistent with the temperature dependence of $\kappa^{S^z\mathrm{(a)}}_{xy}$, where the spin Nernst coefficient is negative at high temperatures.

Let us discuss the origin of the $\bm{k}$ dependence of $\big(\Upsilon^{S^z}_{\bm{k},\eta}\big)_{xy}^{\mathrm{(a)}}$.  
As presented in Eq.~\eqref{eq:Upsilon_a}, this quantity consists of two terms: $-\Omega^{xy}_{\bm{k},\eta} (S^z_{\bm{k}})_{\eta}$, which involves the Berry curvature, and the momentum derivative of the dipole moment, $( D_{\bm{k},\eta}^{S^\gamma})^{\mathrm{(a)}}_{\lambda\lambda'}$, which can be interpreted as the quantum metric in the mixed space $(\bm{k},\bm{h}^{\tau})$, as discussed in Sec.~\ref{sec:snc}.  
Figure~\ref{fig:coef_kitaevheisen_all}(c) shows $-\Omega^{xy}_{\bm{k},\eta} (S^z_{\bm{k}})_{\eta}$ for the lower-energy branch with $\eta=1$.  
The Berry curvature $\Omega^{xy}_{\bm{k},\eta}$ has already been examined, for example, in studies of the thermal Hall effect in the Kitaev model under magnetic fields~\cite{mcclarty2018}.  
Since we assume a forced ferromagnetic state, $\hat{S}^z$, as defined in Eq.~\eqref{eq:O_total}, is equal to $-\bm{1}_{2N\times 2N}$.  
Accordingly, we have confirmed that $(S^z_{\bm{k}})_{\eta} \approx -1$ for $\eta=1$.  
Thus, the quantity in Fig.~\ref{fig:coef_kitaevheisen_all}(c) is nearly identical to the Berry curvature $\Omega^{xy}_{\bm{k},\eta}$ in the present situation.  
Near the $\Gamma$ point, the Berry curvature is negative, whereas it is positive along the boundary of the first Brillouin zone.  
The absolute value of the positive region is larger than that of the negative region, as the Chern number for the lowest-energy branch with $\eta=1$ is $+1$.  
Due to these characteristics of the Berry curvature, the thermal Hall conductivity changes its sign from positive to negative as the temperature increases~\cite{mcclarty2018}.  
On the other hand, in the spin Nernst effect, an additional term contributes besides the Berry curvature, leading to the disappearance of the negative region observed in Fig.~\ref{fig:coef_kitaevheisen_all}(c) in the $\bm{k}$ dependence of $\big(\Upsilon^{S^z}_{\bm{k},\eta}\big)^{\mathrm{(a)}}_{xy}$.  
Figure~\ref{fig:coef_kitaevheisen_all}(d) illustrates the momentum-space distribution of $( D_{\bm{k},\eta}^{S^\gamma})^{\mathrm{(a)}}_{xy}$ for the lower-energy branch with $\eta=1$.  
We find that this quantity is positive around the $\Gamma$ point and negative along the Brillouin zone boundary, which is opposite to the $\bm{k}$ dependence of $-\Omega^{xy}_{\bm{k},\eta} (S^z_{\bm{k}})_{\eta}$.  
Since the contribution from $( D_{\bm{k},\eta}^{S^\gamma})^{\mathrm{(a)}}_{\lambda\lambda'}$ is greater than that from $-\Omega^{xy}_{\bm{k},\eta} (S^z_{\bm{k}})_{\eta}$ around the $\Gamma$ point, the negative region disappears in $\big(\Upsilon^{S^z}_{\bm{k},\eta}\big)^{\mathrm{(a)}}_{xy}$, leading to the absence of a sign change in the temperature dependence of $\kappa^{S^z\mathrm{(a)}}_{xy}$.  
Thus, we conclude that $(D_{\bm{k},\eta}^{S^\gamma})^{\mathrm{(a)}}_{xy}$ plays a crucial role in yielding the difference between the thermal Hall and spin Nernst effects.

When the torque term is neglected, the spin Nernst coefficient is given by $\tilde{\kappa}^{S^z\mathrm{(a)}}_{xy}$.  
As presented in Eq.~\eqref{eq:snc_withouttorque}, this originates from the quantity $\big( \tilde{\Omega}^{S^z}_{\bm{k},\eta} \big)_{xy}^{\mathrm{(a)}}$.  
Figure~\ref{fig:coef_kitaevheisen_all}(b) shows the momentum dependence of $\big( \tilde{\Omega}^{S^z}_{\bm{k},\eta} \big)_{xy}^{\mathrm{(a)}}$ for the lower-energy branch with $\eta=1$.  
We find that it takes positive values in almost all regions of the first Brillouin zone; however, its magnitude is smaller than that of $\big(\Upsilon^{S^z}_{\bm{k},\eta}\big)_{xy}^{\mathrm{(a)}}$.  
This explains why the absolute value of $\tilde{\kappa}^{S^z\mathrm{(a)}}_{xy}$ is smaller than that of $\kappa^{S^z}_{xy}$.

\begin{figure}[t]
  \centering
      \includegraphics[width=\columnwidth,clip]{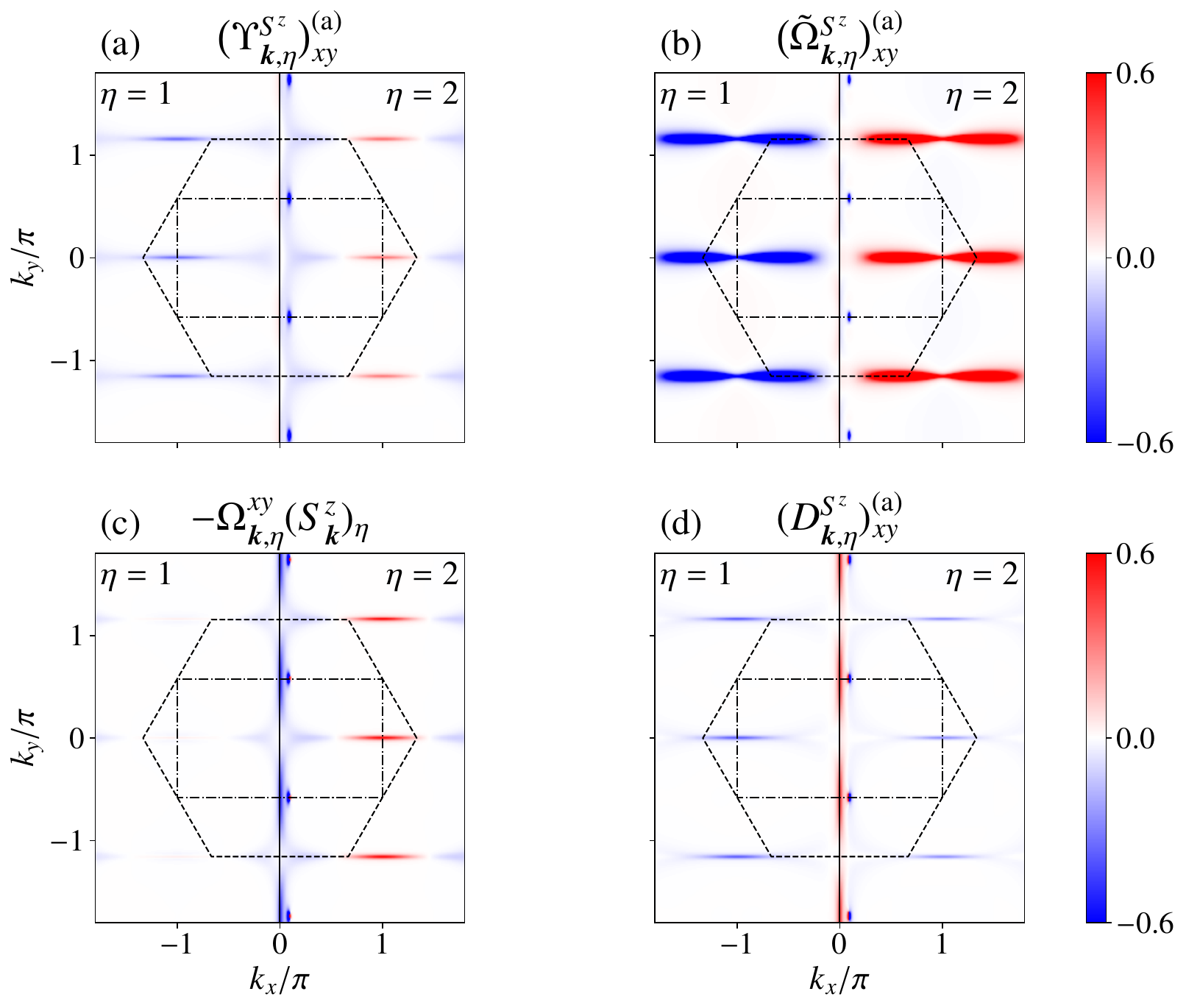}
      \caption{
Color maps of (a) $\big(\Upsilon^{S^z}_{\bm{k},\eta}\big)_{xy}^{\mathrm{(a)}}$, (b) $\big( \tilde{\Omega}^{S^\gamma}_{\bm{k},\eta} \big)_{xy}^{\mathrm{(a)}}$, (c)  $-\Omega^{xy}_{\bm{k},\eta} (S^z_{\bm{k}})_{\eta}$, and (d) $\big(D^{S^z}_{\bm{k},\eta}\big)_{xy}^{\mathrm{(a)}}$ for the lowest- and second-lowest-energy branches in the Kitaev-Heisenberg model at $\varphi=1.6\pi$ under a magnetic field with $h/A = 0.1$.
Since the $\bm{k}$ dependence of these quantities for each branch is symmetric along the $k_y$ axis, the data for the lowest-energy branch with $\eta=1$ (the second-lowest-energy branch with $\eta=2$) is plotted in the region with $k_x<0$ ($k_x>0$).
The dash-dotted lines represent the Brillouin zone corresponding to the magnetic unit cell including four sites, as also shown in Fig.~\ref{fig:honeycomb_kitaev}(b).
      }
      \label{fig:coef_kitaevheisen_all_16}
\end{figure}

Until now, we have discussed the case of $\varphi=3\pi/2$, which corresponds to the ferromagnetic Kitaev model under a magnetic field.  
Finally, we present the results in the presence of the antiferromagnetic Heisenberg interaction, where we choose $\varphi=1.6\pi$ and $h/A=0.1$.  
The magnetic ground state is a canted stripy antiferromagnetic order, as depicted in Fig.~\ref{fig:snc_kitaevheisen}(b), which is consistent with a previous study~\cite{janssen2016}.  
This magnetically ordered state with noncollinear spins forms a four-sublattice order, and the magnetic Brillouin zone is folded into half of the original Brillouin zone, as shown in Fig.~\ref{fig:honeycomb_kitaev}(b).  
In the magnetically ordered state shown in Fig.~\ref{fig:snc_kitaevheisen}(b), the red and blue spins exhibit nonequivalent canting angles, meaning that the angle between the spin moments and the magnetic field differs between the two types of spins~\cite{janssen2016,Janssen_2019rev}.  
We have numerically confirmed that $H_{0,\bm{k}} = H_{0,-\bm{k}}$ holds for the spin-wave Hamiltonian, and thus, Eq.~\eqref{eq:snc_withtorque} is applicable to the present system.

We focus on the spin current generation of the $S^z$ component as well as the case of $\varphi=3\pi/2$.
Note that the mean-field spin configuration shown in Fig.~\ref{fig:snc_kitaevheisen}(b) is not parallel to the $S^z$ direction, and hence, the local mean-field Hamiltonian at site $i$ does not commute with $S_i^z$.
In the following calculations, we omit contributions from the off-diagonal components of $S_i^z$ on the local bases given as the eigenstates of the local mean-field Hamiltonian to introduce the flow of $S^z$ [see Eq.~\eqref{eq:O_total}]~\cite{li2020_prr}.
This simplification is equivalent to considering only the component of $S^z$ projected along the direction of the local spin moment at each site in the canted stripy antiferromagnetic order.
Incorporating contributions beyond this simplification is left for future work.

Figure~\ref{fig:snc_kitaevheisen}(d) shows the temperature dependence of the spin Nernst coefficient at $\varphi=1.6\pi$.  
We find that $\kappa^{S^z\mathrm{(a)}}_{xy}$, which includes the torque term, differs significantly from $\tilde{\kappa}^{S^z\mathrm{(a)}}_{xy}$, which is obtained by neglecting the torque term.  
Although the absolute value of $\kappa^{S^z\mathrm{(a)}}_{xy}$ is much larger than that of $\tilde{\kappa}^{S^z\mathrm{(a)}}_{xy}$ for the pure Kitaev model with $\varphi=3\pi/2$ [Fig.~\ref{fig:snc_kitaevheisen}(c)], it becomes smaller than that of $\tilde{\kappa}^{S^z\mathrm{(a)}}_{xy}$ at low temperatures for the case with $\varphi=1.6\pi$ [Fig.~\ref{fig:snc_kitaevheisen}(d)].  
Moreover, while a sign change occurs in the temperature dependence of $\tilde{\kappa}^{S^z\mathrm{(a)}}_{xy}$, it is not observed in $\kappa^{S^z\mathrm{(a)}}_{xy}$.  
These results suggest that the impact of the torque term on the spin Nernst effect strongly depends on the model parameters, making it challenging to determine how this term contributes to the spin Nernst effect.  
Therefore, evaluating $\kappa^{S^z\mathrm{(a)}}_{xy}$ using Eq.~\eqref{eq:snc_withtorque} is necessary for understanding the spin Nernst effect even in other models of insulating magnets.

Let us examine the temperature dependence of $\kappa^{S^z\mathrm{(a)}}_{xy}$ and $\tilde{\kappa}^{S^z\mathrm{(a)}}_{xy}$ at $\varphi=1.6\pi$ in detail.  
As shown in Fig.~\ref{fig:snc_kitaevheisen}(d), $\kappa^{S^z\mathrm{(a)}}_{xy}$ appears to be positive as a function of temperature.  
On the other hand, $\tilde{\kappa}^{S^z\mathrm{(a)}}_{xy}$, which is the form with the torque term neglected, takes negative values at low temperatures and becomes positive as the temperature increases, which starkly contrasts with the behavior of $\kappa^{S^z\mathrm{(a)}}_{xy}$ when the torque term is included.  
Figures~\ref{fig:coef_kitaevheisen}(b) and \ref{fig:coef_kitaevheisen}(d) display the magnon bands and the values of $\big(\Upsilon^{S^z}_{\bm{k},\eta}\big)_{xy}^{\mathrm{(a)}}$ and $\big( \tilde{\Omega}^{S^\gamma}_{\bm{k},\eta} \big)_{xy}^{\mathrm{(a)}}$ for each branch, respectively.  
The excitation gap at the M point is nearly zero, and the magnon bands are nearly degenerate at the X point.  
For the lowest-energy branch, the value of $\big( \tilde{\Omega}^{S^\gamma}_{\bm{k},\eta} \big)_{xy}^{\mathrm{(a)}}$ is positive around the M point, leading to the negative value of $\tilde{\kappa}^{S^z\mathrm{(a)}}_{xy}$ at low temperatures.  
On the other hand, $\big(\Upsilon^{S^z}_{\bm{k},\eta}\big)_{xy}^{\mathrm{(a)}}$ is nearly zero at the M point.  

With increasing temperature, magnons in higher-energy branches contribute to transport properties.  
Here, we focus on the X point, where the energies of the two lower branches are close to each other [Figs.~\ref{fig:snc_kitaevheisen}(b) and \ref{fig:snc_kitaevheisen}(d)].  
The lowest-energy branch with $\eta=1$ exhibits negative values for both $\big(\Upsilon^{S^z}_{\bm{k},\eta}\big)^{\mathrm{(a)}}_{xy}$ and $\big( \tilde{\Omega}^{S^\gamma}_{\bm{k},\eta} \big)^{\mathrm{(a)}}_{xy}$ around the X point, as shown in Figs.~\ref{fig:coef_kitaevheisen}(b) and \ref{fig:coef_kitaevheisen}(d).  
This leads to a sign change in $\tilde{\kappa}^{S^z{\mathrm{(a)}}}_{xy}$ to positive values while increasing temperature.  
This momentum dependence is also observed in Figs.~\ref{fig:coef_kitaevheisen_all_16}(a) and \ref{fig:coef_kitaevheisen_all_16}(b), where the color maps of $\big(\Upsilon^{S^z}_{\bm{k},\eta}\big)^{\mathrm{(a)}}_{xy}$ and $\big( \tilde{\Omega}^{S^\gamma}_{\bm{k},\eta} \big)^{\mathrm{(a)}}_{xy}$ are plotted on the $k_x$-$k_y$ plane.  
Around the X point, the signs of $\big(\Upsilon^{S^z}_{\bm{k},\eta}\big)^{\mathrm{(a)}}_{xy}$ for these branches are opposite, and their absolute values are nearly equal, leading to the cancellation of the contribution from the X point to $\kappa^{S^z{\mathrm{(a)}}}_{xy}$ at higher temperatures.  
A similar cancellation is also expected for $\tilde{\kappa}^{S^z{\mathrm{(a)}}}_{xy}$ based on the momentum dependence of $\big( \tilde{\Omega}^{S^\gamma}_{\bm{k},\eta} \big)^{\mathrm{(a)}}_{xy}$.  
Therefore, the momentum dependence around the X point does not play a significant role in the difference between the spin Nernst coefficients with and without the torque term.  

Further increases in temperature lead to excitations of quasiparticles with higher energy.  
We focus on the momentum dependence of $\big(\Upsilon^{S^z}_{\bm{k},\eta}\big)^{\mathrm{(a)}}_{xy}$ and $\big( \tilde{\Omega}^{S^\gamma}_{\bm{k},\eta} \big)^{\mathrm{(a)}}_{xy}$ from the X point to the $\Gamma$ point, where the energies of the lower two branches increase, as shown in Figs.~\ref{fig:coef_kitaevheisen}(b) and \ref{fig:coef_kitaevheisen}(d).  
We find that $\big(\Upsilon^{S^z}_{\bm{k},\eta}\big)^{\mathrm{(a)}}_{xy}$ for $\eta=2$ changes its sign to negative while approaching the $\Gamma$ point, whereas $\big( \tilde{\Omega}^{S^\gamma}_{\bm{k},\eta} \big)^{\mathrm{(a)}}_{xy}$ for $\eta=2$ remains positive.  
These behaviors are also observed in Figs.~\ref{fig:coef_kitaevheisen_all_16}(a) and \ref{fig:coef_kitaevheisen_all_16}(b).  
We find that the Berry curvature term $-\Omega^{xy}_{\bm{k},\eta} (S^z_{\bm{k}})_{\eta}$ predominantly contributes to the negative values of $\big(\Upsilon^{S^z}_{\bm{k},\eta}\big)^{\mathrm{(a)}}_{xy}$ for $\eta=2$ around the $\Gamma$ point rather than $\big(D^{S^z}_{\bm{k},\eta}\big)_{xy}^{\mathrm{(a)}}$, as shown in Figs.~\ref{fig:coef_kitaevheisen_all_16}(c) and \ref{fig:coef_kitaevheisen_all_16}(d).  
This difference leads to larger positive values in $\kappa^{S^z{\mathrm{(a)}}}_{xy}$ compared to $\tilde{\kappa}^{S^z{\mathrm{(a)}}}_{xy}$ at higher temperatures, which is consistent with the results shown in Fig.~\ref{fig:snc_kitaevheisen}(d).

\subsection{Shastry-Sutherland model}
\label{sec:SSmodel}

\begin{figure}[t]
  \centering
      \includegraphics[width=\columnwidth,clip]{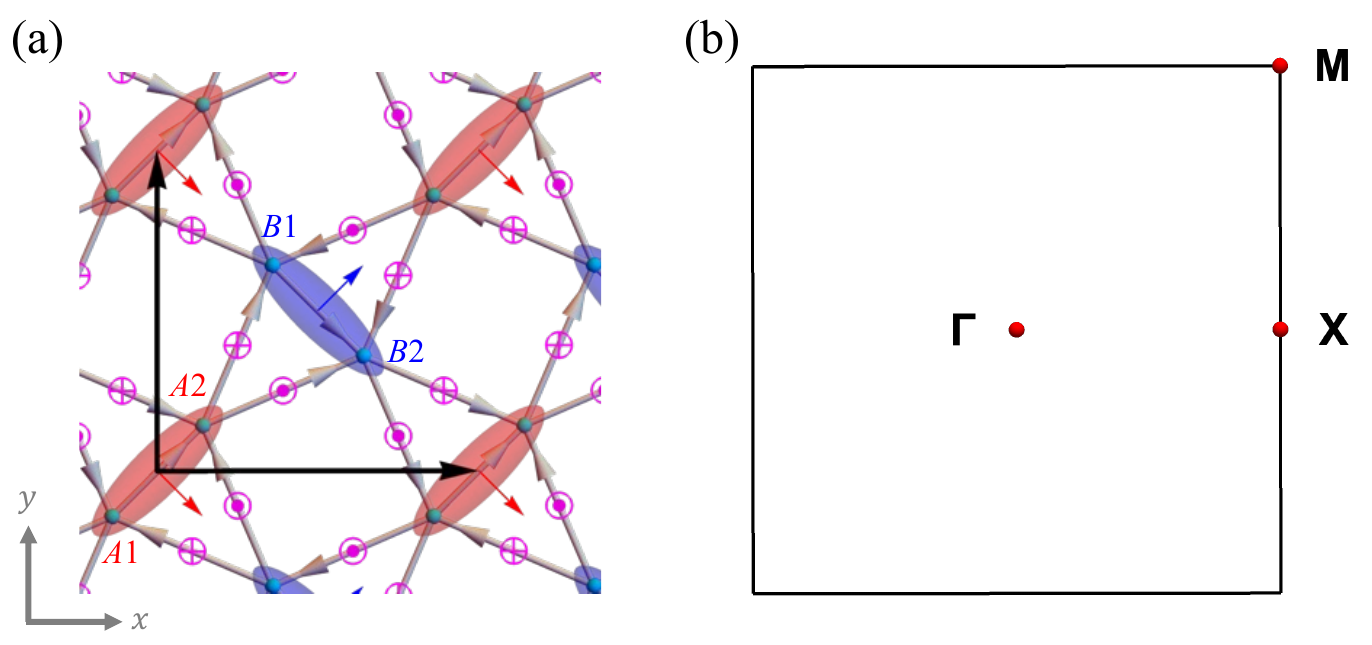}
      \caption{
(a) Schematic illustration of a lattice on which the Shastry-Sutherland model is defined.  
The red (blue) ellipse represents dimer $A$ ($B$), which contains two spin sites, $A1$ and $A2$ ($B1$ and $B2$).  
The red and blue arrows denote the DM vectors $\bm{D}_{\iota\iota'}$, where the arrows connecting two intra-dimer sites indicate the direction from $\iota$ to $\iota'$.  
Magenta symbols on next-nearest-neighbor bonds represent the DM vectors $\bm{D}'_{\iota\iota'}$, where the arrows connecting two inter-dimer sites indicate the direction from $\iota$ to $\iota'$.  
The black arrows represent the primitive translation vectors.  
(b) First Brillouin zone corresponding to the lattice presented in panel (a).  
      }
      \label{fig:ss}
\end{figure}

In this section, we consider the Shastry-Sutherland model with the DM interaction under magnetic fields~\cite{shastry1981,romhanyi2011}, which is known as a model that describes the magnetic properties of $\mathrm{Sr Cu_2 (BO_3)_2}$~\cite{kageyama1999,zayed2014}.
This model is defined on the lattice shown in Fig.~\ref{fig:ss} and is described by the following Hamiltonian:
\begin{align}
  \label{eq:H_ss}
  \mathcal{H}
  =&
  J \sum_{\braket{\iota\iota'}} \bm{S}_{\iota} \cdot \bm{S}_{\iota'}
  +
  J' \sum_{\langle\!\langle \iota\iota' \rangle\!\rangle} \bm{S}_{\iota} \cdot \bm{S}_{\iota'}
  \notag\\
  &+ \sum_{\braket{\iota\iota'}} \bm{D}_{\iota\iota'} \cdot (\bm{S}_{\iota} \times \bm{S}_{\iota'})
  + \sum_{\langle\!\langle \iota\iota' \rangle\!\rangle} \bm{D}'_{\iota\iota'} \cdot (\bm{S}_{\iota} \times \bm{S}_{\iota'})
  - h^{z}\sum_{\iota}  S_{\iota}^{z},
\end{align}
where $\bm{S}_{\iota}$ is a spin-$1/2$ operator at site $\iota$.
The parameters $J$ and $\bm{D}$ represent the Heisenberg and DM interactions between the nearest-neighbor sites $\braket{\iota\iota'}$, respectively, while $J'$ and $\bm{D}'$ correspond to those between the next-nearest-neighbor sites $\langle\!\langle \iota\iota' \rangle\!\rangle$.
The DM vectors $\bm{D}_{\iota\iota'}$ and $\bm{D}'_{\iota\iota'}$ are antisymmetric with respect to $\iota$ and $\iota'$ and are determined by the direction of the line connecting sites $\iota$ and $\iota'$, as illustrated in Fig.~\ref{fig:ss}(a).
Note that the DM vector $\bm{D}_{\iota\iota'}$ on the nearest-neighbor bonds is perpendicular to the bond direction and lies in the $xy$ plane.
As shown in Fig.~\ref{fig:ss}(a), there are two dimers, $A$ and $B$, connected by nearest-neighbor bonds in the unit cell.
In the absence of the DM interaction and magnetic fields, this model exhibits a dimer-singlet ground state, in which the spins form a singlet state on each dimer for $J'/J \lesssim 0.7$~\cite{shastry1981}.
Elementary excitations from the dimer-singlet state are described by bosonic quasiparticles known as triplons~\cite{shastry1981,romhanyi2011}.
In the presence of the DM interaction and magnetic fields, the triplon bands exhibit nontrivial topological properties, and the thermal Hall effect is theoretically predicted to occur~\cite{romhanyi2015}.
We expect that the spin Nernst effect also arises due to this property.
On the other hand, since the spin carried by the triplon is not conserved due to the DM interactions, the torque term must be taken into account.

\begin{figure*}[t]
  \centering
      \includegraphics[width=2\columnwidth,clip]{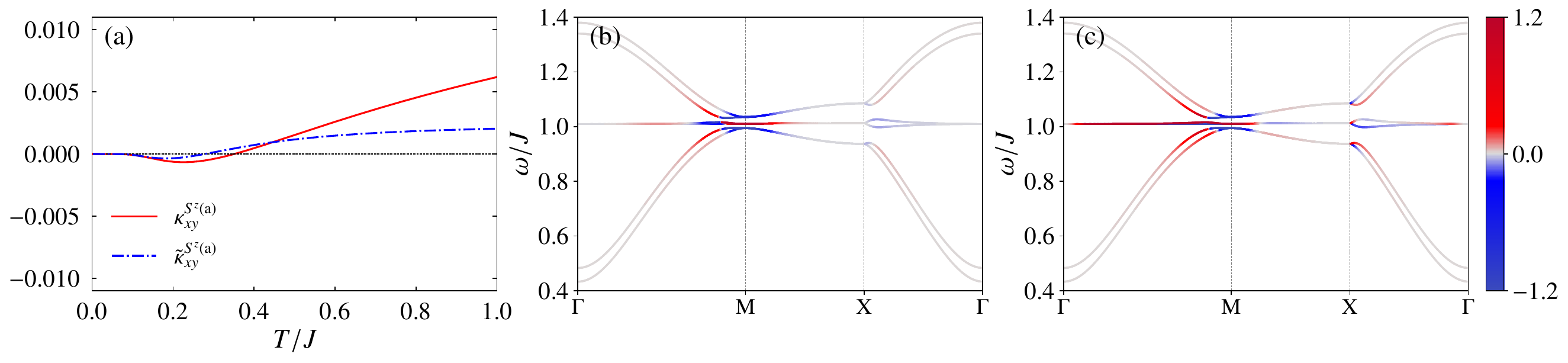}
      \caption{
       (a) Temperature dependence of the spin Nernst coefficients $\kappa^{S^z\mathrm{(a)}}_{xy}$ and $\tilde{\kappa}^{S^z\mathrm{(a)}}_{xy}$, and
     [(b),(c)] the dispersion relations of triplons in the Shastry-Sutherland model given in Eq.~\eqref{eq:H_ss}.
     In panels (b) and (c), the line colors represents $\big(\Upsilon^{S^z}_{\bm{k},\eta}\big)_{xy}^{\mathrm{(a)}}$ and $\big(\tilde{\Omega}^{S^z}_{\bm{k},\eta}\big)_{xy}^{\mathrm{(a)}}$, respectively.
     The symmetric points are shown in Fig.~\ref{fig:ss}(b).
      }
      \label{fig:snc_ss}
\end{figure*}

Since the dimer-singlet state cannot be described within mean-field theory based on the decoupling of each spin, we analyze this model using cluster mean-field theory, in which the interactions within the dimer are treated exactly, while the interactions between dimers are handled at the mean-field level.  
In this approximation, each dimer, consisting of two lattice sites, is regarded as a single site with the dimension of a local Hilbert space being $\mathscr{N}=4$.
We perform a two-sublattice ($M=2$) calculation, where the unit cell contains two dimers, $A$ and $B$, with four original spins.  
The flavor-wave theory is applicable to the dimer-singlet state based on cluster mean-field theory. 
Thus, the Hamiltonian given in Eq.~\eqref{eq:H_ss} is transformed into the bosonic BdG form in Eq.~\eqref{eq:conti_Hk} with $H_{0,\bm{k}}$.  
The bosonic quasiparticles corresponding to the three excited states of each dimer are referred to as triplons in flavor-wave theory.  
Similar to the Kitaev system discussed in the previous section, the spin Nernst effect is analyzed using the spin density operator defined as $S^\gamma(\bm{r}) = \frac{1}{2}\mathcal{A}^\dagger (\bm{r}) \hat{S}^{\gamma} \mathcal{A}(\bm{r})$, where $\hat{S}^\gamma = \bm{1}_{2\times 2} \otimes \bar{S}^{\gamma}$ is a $2N\times 2N$ diagonal matrix.  
In the present model, since each dimer contains two spins, the $N\times N$ matrix $\bar{S}^{\gamma}$ is given by the sum of the two spins in dimer $i$ as $\bar{S}^{\gamma}_{(l,m)} = \sum_{\xi=1,2} \left( \braket{m;i|S_\xi^{\gamma}|m;i} - \braket{0;i|S_\xi^{\gamma}|0;i}\right)$, where $i$ belongs to sublattice $l\in \{A,B\}$ and the spin operators are given by $S_{1}^\gamma = S^\gamma \otimes \bm{1}_{2\times 2}$ and $S_{2}^\gamma = \bm{1}_{2\times 2} \otimes S^\gamma$.

In the present calculations, we assume that the inplane components of $\bm{D}'_{ij}$ are zero, and accordingly, $\bm{D}'_{ij} = (0,0,D^{\prime z}_{ij})$.
We fix the parameters as $J'/J = 0.6$, $h^z/J = 0.02$ and $|\bm{D}_{ij}| / J = |D^{\prime z}_{ij}| / J = 0.2$ to realize a ground state close to the dimer-singlet state.
Since this state possesses inversion symmetry, our theory is applicable to this model.
Note that we have confirmed that the relation $H_{0,\bm{k}} = H_{0,-\bm{k}}$ holds.
For the spin Nernst effect, we apply a temperature gradient along the $y$-axis and measure the spin current along the $x$-axis.

Here, we examine the spin Nernst effect in the Shastry-Sutherland model.  
While the ground state is not a perfect spin-singlet state due to the DM interaction, we find that 
the commutator of the local mean-field Hamiltonian and $S_1^z+S_2^z$ is almost zero owing to the magnetic field along the $S^z$ direction.
Thus, we focus on the antisymmetric component $\kappa^{S^z\mathrm{(a)}}_{xy}$, which represents the spin Nernst coefficient.  
Figure~\ref{fig:snc_ss}(a) illustrates the temperature dependence of $\kappa^{S^z\mathrm{(a)}}_{xy}$.  
This quantity exhibits nonmonotonic behavior as a function of temperature, and its sign changes at low temperatures.  
Additionally, we plot $\tilde{\kappa}^{S^z\mathrm{(a)}}_{xy}$, which is obtained by neglecting the torque term, in Fig.~\ref{fig:snc_ss}(a).  
As shown in Fig.~\ref{fig:snc_ss}(a), these two quantities exhibit similar temperature dependence with a sign change, but the absolute value of $\kappa^{S^z\mathrm{(a)}}_{xy}$ is larger than that of $\tilde{\kappa}^{S^z\mathrm{(a)}}_{xy}$ over almost all temperature ranges.  

\begin{figure}[t]
  \centering
      \includegraphics[width=\columnwidth,clip]{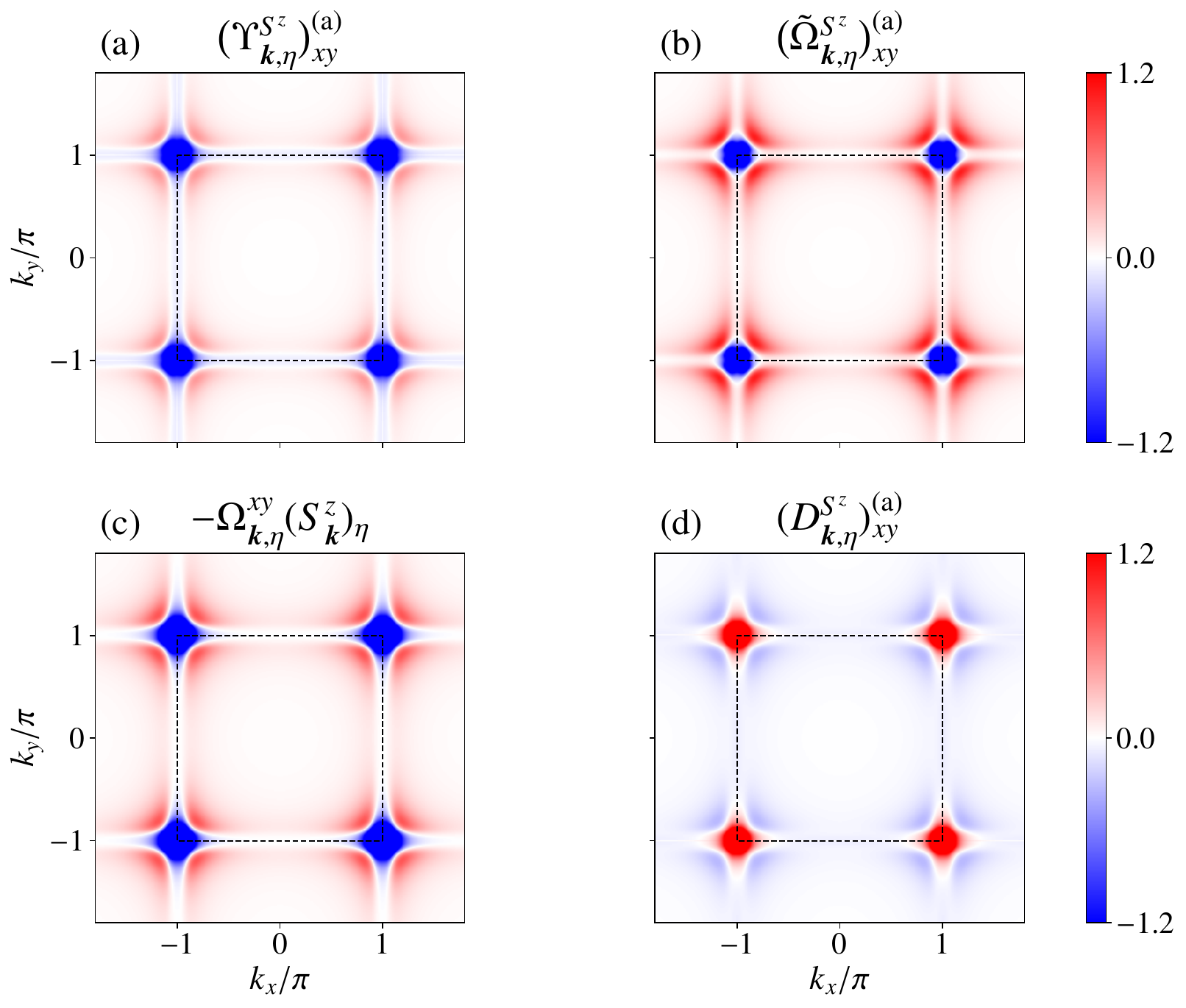}
      \caption{
      Color maps of (a) $\big(\Upsilon^{S^z}_{\bm{k},\eta}\big)_{xy}^{\mathrm{(a)}}$, (b) $\big( \tilde{\Omega}^{S^\gamma}_{\bm{k},\eta} \big)_{xy}^{\mathrm{(a)}}$, (c)  $-\Omega^{xy}_{\bm{k},\eta} (S^z_{\bm{k}})_{\eta}$, and (d) $\big(D^{S^z}_{\bm{k},\eta}\big)_{xy}^{\mathrm{(a)}}$, where
      the values in each figure are plotted as the sum of those for $\eta=1$ and $2$.
      The dashed lines represent the first Brillouin zone shown in Fig.~\ref{fig:ss}(b).
      }
      \label{fig:coef_ss}
\end{figure}

To understand the origin of the temperature dependence and the difference between $\kappa^{S^z\mathrm{(a)}}_{xy}$ and $\tilde{\kappa}^{S^z\mathrm{(a)}}_{xy}$, we examine the momentum dependence of $\big(\Upsilon^{S^z}_{\bm{k},\eta}\big)^{\mathrm{(a)}}_{xy}$ and $\big(\tilde{\Omega}^{S^z}_{\bm{k},\eta}\big)^{\mathrm{(a)}}_{xy}$.
Figures~\ref{fig:snc_ss}(b) and \ref{fig:snc_ss}(c) show the dispersion of the triplons, where the color of each branch represents the value of $\big(\Upsilon^{S^z}_{\bm{k},\eta}\big)^{\mathrm{(a)}}_{xy}$ and $\big(\tilde{\Omega}^{S^z}_{\bm{k},\eta}\big)^{\mathrm{(a)}}_{xy}$, respectively.
There are three groups of triplon branches: the lowest-energy and second-lowest-energy dispersive branches with $\eta=1$ and $2$, the nearly dispersionless bands with $\eta=3$ and $4$, and the higher-energy dispersive branches with $\eta=5$ and $6$.
In each group, the energies of the two branches are close to each other and are degenerate along the M-X line~\cite{romhanyi2011}.
Here, we discuss the momentum dependence of $\big(\Upsilon^{S^z}_{\bm{k},\eta}\big)^{\mathrm{(a)}}_{xy}$ and $\big(\tilde{\Omega}^{S^z}_{\bm{k},\eta}\big)^{\mathrm{(a)}}_{xy}$ for the lowest-energy and second-lowest-energy branches with $\eta=1$ and $2$.
Around the $\Gamma$ point, where the triplons have the lowest energy, this quantity is nearly zero.
As the momentum approaches the M point, $\big(\Upsilon^{S^z}_{\bm{k},\eta}\big)^{\mathrm{(a)}}_{xy}$ changes its sign from positive to negative, which is associated with an increase in the triplon excitation energy.
We find that it takes a significantly negative value in the vicinity of the M point and remains negative along the M-X line.
This $\bm{k}$ dependence is also observed as a color map on the $k_x$-$k_y$ plane in Fig.~\ref{fig:coef_ss}(a).
In this figure, we show $\big(\Upsilon^{S^z}_{\bm{k},1}\big)^{\mathrm{(a)}}_{xy}+\big(\Upsilon^{S^z}_{\bm{k},2}\big)^{\mathrm{(a)}}_{xy}$, as the two branches with $\eta=1$ and $2$ are degenerate along the M-X line~\footnote{
When degenerate magnon branches are present, the calculations of $\big(\Upsilon^{S^z}_{\bm{k},\eta}\big)^{\mathrm{(a)}}_{xy}$ and $\big(\tilde{\Omega}^{S^z}_{\bm{k},\eta}\big)^{\mathrm{(a)}}_{xy}$ are nontrivial because, for example, the denominator in Eq.~\eqref{eq:BC} vanishes.  
To avoid this issue, we have added a small positive constant $\delta$ to the denominators that may potentially vanish.
For example, we modify Eq.~\eqref{eq:BC} as  
$\Omega^{\lambda\lambda'}_{\bm{k},\eta} = -2\hbar^2\sum_{\eta_1 (\neq \eta)}^{2N} \sigma_{3,\eta}\sigma_{3,\eta_1} \frac{\mathrm{Im}\left[ (v_{\bm{k},\lambda})_{\eta\eta_1} (v_{\bm{k},\lambda'})_{\eta_1\eta} \right]} {(\bar{\varepsilon}_{\bm{k},\eta}-\bar{\varepsilon}_{\bm{k},\eta_1})^2+\delta^2}$.  
We have confirmed that the values of $\big(\Upsilon^{S^z}_{\bm{k},\eta}\big)^{\mathrm{(a)}}_{xy}$ and $\big(\tilde{\Omega}^{S^z}_{\bm{k},\eta}\big)^{\mathrm{(a)}}_{xy}$ remain nearly unchanged even as $\delta$ approaches $10^{-6}$
}.
From this figure, the negative value of $\kappa^{S^z\mathrm{(a)}}_{xy}$ at low temperatures is considered to originate from the positive value of $\big(\Upsilon^{S^z}_{\bm{k},\eta}\big)^{\mathrm{(a)}}_{xy}$ around the $\Gamma$ point, and the sign change of $\kappa^{S^z\mathrm{(a)}}_{xy}$ to positive with increasing temperature is attributed to thermally excited triplons with negative values of $\big(\Upsilon^{S^z}_{\bm{k},\eta}\big)^{\mathrm{(a)}}_{xy}$ around the M point.
Figures~\ref{fig:coef_ss}(c) and \ref{fig:coef_ss}(d) show the momentum dependence of the first and second terms in Eq.~\eqref{eq:Upsilon_a}, respectively, which are plotted as the sum of the two branches with $\eta=1$ and $2$.
The signs of these two quantities are opposite to each other, and the absolute value of the first term is larger than that of the second one, leading to a similar behavior of $\big(\Upsilon^{S^z}_{\bm{k},\eta}\big)^{\mathrm{(a)}}_{xy}$ and $-\Omega^{xy}_{\bm{k},\eta} (S^{z}_{\bm{k}})_{\eta}$ at low temperatures.
Thus, in this model, the spin Nernst effect is dominated by the Berry curvature term, while the $\bm{k}$ derivative of the dipole moment $( D_{\bm{k},\eta}^{S^z})^{\mathrm{(a)}}_{xy}$ plays a minor role in this effect.

To discuss the role of the torque term, we compare $\kappa^{S^z\mathrm{(a)}}_{xy}$ with $\tilde{\kappa}^{S^z\mathrm{(a)}}_{xy}$ in Fig.~\ref{fig:snc_ss}(a).  
As mentioned earlier, the absolute value of $\tilde{\kappa}^{S^z\mathrm{(a)}}_{xy}$, which excludes the torque term, is smaller than that of $\kappa^{S^z\mathrm{(a)}}_{xy}$, which includes the torque term, at almost all temperatures.  
To elucidate this difference, we plot the momentum dependence of $\big(\tilde{\Omega}^{S^z}_{\bm{k},\eta}\big)^{\mathrm{(a)}}_{xy}$ in Figs.~\ref{fig:snc_ss}(c) and \ref{fig:coef_ss}(b).  
While $\big(\Upsilon^{S^z}_{\bm{k},\eta}\big)^{\mathrm{(a)}}_{xy}$ is negative along the M-X line [Fig.~\ref{fig:snc_ss}(b)], $\big(\tilde{\Omega}^{S^z}_{\bm{k},\eta}\big)^{\mathrm{(a)}}_{xy}$ exhibits both positive and negative values along this line [Fig.~\ref{fig:snc_ss}(c)].  
Moreover, the positive values of $\big(\tilde{\Omega}^{S^z}_{\bm{k},\eta}\big)^{\mathrm{(a)}}_{xy}$ are larger in magnitude than those of $\big(\Upsilon^{S^z}_{\bm{k},\eta}\big)^{\mathrm{(a)}}_{xy}$ over a broader region.  
This momentum dependence suggests that the pronounced cancellation between the positive and negative components results in the suppression of the temperature dependence of $\tilde{\kappa}^{S^z\mathrm{(a)}}_{xy}$.  
On the other hand, the region where $\big(\Upsilon^{S^z}_{\bm{k},\eta}\big)^{\mathrm{(a)}}_{xy}$ takes negative values is larger than that of $\big(\tilde{\Omega}^{S^z}_{\bm{k},\eta}\big)^{\mathrm{(a)}}_{xy}$, leading to a greater absolute value of $\kappa^{S^z\mathrm{(a)}}_{xy}$ compared to $\tilde{\kappa}^{S^z\mathrm{(a)}}_{xy}$ at higher temperatures.

\section{Summary and outlook}
\label{sec:summary}

In summary, we have investigated the spin Nernst effect in insulating magnets without spin conservation.
We begin with a general localized electron model and apply the linear flavor-wave approximation to derive a free-boson model in the BdG form for elementary excitations.
We then formulate a spin transport theory for bosonic systems, which is applicable even to systems with no spin conservation due to, for example, DM interactions.
We introduce the conserved spin current, which includes a contribution from the torque term, and derive the spin Nernst coefficient using semiclassical theory by expanding up to the second order of the spatial gradient in bosonic systems.
The spin Nernst coefficient is expressed as a sum of the Berry curvature term and the momentum derivative of the dipole moment, which corresponds to the quantum metric in the mixed space of the momentum and virtual field for the torque.
We also show the representation avoiding the momentum derivative, which enables numerical calculations without finite difference methods.
We apply this theory to the Kitaev-Heisenberg model and the Shastry-Sutherland model, which are known to exhibit a nontrivial topological band structure in elementary excitations.
In both cases, the temperature dependence of the spin Nernst coefficient is significantly different from that obtained by neglecting the torque term, the latter of which corresponds to the conventional theory based on the spin Berry curvature.
The present results indicate that contributions from the quantum metric are also important in the spin Nernst effect in addition to the Berry curvature in momentum space.
Nevertheless, the impact of the torque term on the spin Nernst effect strongly depends on the model parameters, suggesting that our formulation based on the conserved spin current is essential for understanding this effect in insulating magnets.

Our theory can be applied not only to various insulating magnets with conventional magnons including those in altermagnets~\cite{hoyer2025} but also to systems with multipole degrees of freedom~\cite{kusunose2001,nasu2021,nasu2022}.
For example, while the orbital Nernst effect has been theoretically proposed as a phenomenon applicable to orbitronics~\cite{Go2024magnon}, it can also be described by the present theory by replacing the spin current with the orbital current.
Our approach can also be applied to quantum spin liquids without any long-range orders if one could use the Schwinger boson method~\cite{Arovas1988}.
Furthermore, the present theory could potentially be extended to magnon-phonon coupled systems~\cite{Nasu2013} since magnon-phonon hybrid excitations can be described by bosonic quasiparticles within the flavor-wave theory.
In our calculations, we do not consider impurity scattering effects, such as side jump and skew scattering, which are also expected to play a significant role in the spin Nernst effect in real materials~\cite{xiao2021conserved}.
These effects could be incorporated into our theory by introducing a relaxation time into the Boltzmann equation or equation of motion, as in previous studies employing semiclassical theory~\cite{Sinitsyn2006,Xiao2017,Atencia2022}.
In addition to extrinsic effects such as impurities, magnon-magnon interactions~\cite{harris1971,zhitomirsky1999,costa2000,chernyshev2006,chernyshev2009,mourigal2010,mourigal2013,zhitomirsky2013,maksimov2016_prb,maksimov2016_prl,chernyshev2016,winter2017_nc,mcclarty2018,mcclarty2019,kim2019,smit2020,mook2020,mook2021,maksimov2020,maksimov2022,bai2023,Sourounis2024}, inevitably present as higher-order terms in the Holstein-Primakoff transformation, are also expected to contribute to the spin Nernst effect in insulating magnets.
Since this contribution can be described using the Green function technique~\cite{koyama2023_prb,koyama2023_NPSM,koyama2024,habel2024}, a full quantum-mechanical formulation beyond the semiclassical approach using the Green function technique is required, which remains a challenge for future studies.
In this study, we have focused on the case where the torque term is expressed as the divergence of a vector field, but such situations are not always realized in real materials.
Constructing a general theory that includes torque terms not expressible in divergence form is another important future direction.
Recent studies addressing this issue have proposed theories evaluating spin accumulation instead of spin current~\cite{Shitade2022,Shitade2022-Nernst,Shitade2025_arxiv}.
Relations between torque terms and spin accumulation should be clarified in the future.
Finally, we comment on possible applications of the present theory to other transport phenomena in insulating magnets.
Since our theory addresses nonlinear terms of spatial gradients, it could also be applied to nonlinear thermal Hall and spin Nernst effects~\cite{Kondo2022,varshney2023_prb1,varshney2023_prb2} as well as strain-induced Hall effects~\cite{Jiang2024_arxiv,Takahashi2025_arxiv} in insulating magnets, which are expected to be described by incorporating higher-order spatial gradient terms.

\begin{acknowledgments}
  The authors thank A.~Shitade and A.~Ono for fruitful discussions.
  Parts of the numerical calculations were performed in the supercomputing
  systems in ISSP, the University of Tokyo.
  This work was supported by Grant-in-Aid for Scientific Research from
  JSPS, KAKENHI Grant No.~JP20H00122, JP22H01175, JP23H01129, JP23H04865, JP24K00563,
  and by JST, the establishment of university fellowships towards the creation of science technology innovation, Grant Number JPMJFS2102.
  \end{acknowledgments}

\appendix

\begin{widetext}

\section{Expectation value in semiclassical theory}
\label{app:exp-theta-semiclassical}

In this appendix, we explain why the expectation value of a physical quantity can be evaluated using Eq.~\eqref{eq:ave1} under the spatially dependent Hamiltonian $\mathcal{H}$ given in Eq.~\eqref{appeq:H_semiclassical}.
To elucidate this, we begin with the Hamiltonian $\mathcal{H}_0$ for a homogeneous system.
By employing the paraunitary matrix $T_{\bm{k}}$, $\mathcal{A}_s(\bm{r})$ in Eq.~\eqref{eq:Ar-Fourier} can be expressed as
\begin{align}\label{eq:app-Ar}
   \mathcal{A}_s(\bm{r})
    =\frac{1}{\sqrt{V}}\sum_{\bm{k}}\sum_{\eta=1}^{2N} (T_{\bm{k}})_{s\eta} \mathcal{B}_{\bm{k},\eta}e^{i\bm{k}\cdot\bm{r}}.
\end{align}
The expectation value $\langle\theta(\bm{r}) \rangle$ is obtained by substituting the above equation into Eq.~\eqref{eq:AthetaA}.
Note that $\hat{\theta}$ in Eq.~\eqref{eq:AthetaA} acts as a differential operator with respect to $\bm{r}$ on $\mathcal{A}(\bm{r})$.
To explicitly indicate the variable on which $\hat{\theta}$ operates, we denote it as $\hat{\theta}_{[\bm{r}]}$, and rewrite Eq.~\eqref{eq:AthetaA} as
\begin{align}
\theta(\bm{r}) = \mathcal{A}^\dagger (\bm{r}) \hat{\theta}_{[\bm{r}]} \mathcal{A}(\bm{r})
=\int d\bm{r}' \mathcal{A}^\dagger (\bm{r}') \delta(\bm{r}-\bm{r}') \hat{\theta}_{[\bm{r}']} \mathcal{A}(\bm{r}').
\end{align}
Then, $\langle\theta(\bm{r}) \rangle$ is given by
\begin{align}
   \langle\theta(\bm{r}) \rangle
&=\frac{1}{V}\int d\bm{r}'
\sum_{\bm{k},\bm{k}'}\sum_{\eta,\eta'=1}^{2N}\sum_{s,s'}
\langle
\mathcal{B}_{\bm{k},\eta}^\dagger\mathcal{B}_{\bm{k}',\eta'}
\rangle 
(T_{\bm{k}}^\dagger)_{\eta s}  e^{-i\bm{k}\cdot\bm{r}'}
\Bigl(\delta(\bm{r}'-\bm{r}) \hat{\theta}_{[\bm{r}']}\Bigr)_{ss'} 
e^{i\bm{k}'\cdot\bm{r}'}(T_{\bm{k}'})_{s'\eta'}.
\end{align}
Here, using the relation $\braket{\mathcal{B}_{\bm{k},\eta}^\dagger \mathcal{B}_{\bm{k}',\eta'}} = \sigma_{3,\eta} f_B(\bar{\varepsilon}_{\bm{k},\eta})\delta_{\bm{k},\bm{k}'}\delta_{\eta,\eta'}$, the above equation can be rewritten as
\begin{align}
   \langle\theta(\bm{r}) \rangle
=\frac{1}{V}\int d\bm{r}'
\sum_{\bm{k}}\sum_{\eta=1}^{2N}\sum_{s,s'}
\sigma_{3,\eta} f_B(\bar{\varepsilon}_{\bm{k},\eta})
(T_{\bm{k}}^\dagger)_{\eta s}  e^{-i\bm{k}\cdot\bm{r}'}
\Bigl(\delta(\bm{r}'-\bm{r}) \hat{\theta}_{[\bm{r}']}\Bigr)_{ss'} 
e^{i\bm{k}\cdot\bm{r}'}(T_{\bm{k}})_{s'\eta}.
\end{align}
Here, we introduce the Bloch wave function as $e^{i\bm{k}\cdot\bm{r}}(T_{\bm{k}})_{s\eta}=\braket{\bm{r}, s|\psi_{\bm{k},\eta}}$.
Then, the expectation value can be calculated as
\begin{align}
\langle\theta(\bm{r}) \rangle
&=\frac{1}{V}\int d\bm{r}'
\sum_{\bm{k}}\sum_{\eta=1}^{2N}\sum_{s}
\sigma_{3,\eta} f_B(\bar{\varepsilon}_{\bm{k},\eta})
(T_{\bm{k}}^\dagger)_{\eta s}  e^{-i\bm{k}\cdot\bm{r}'}
\bra{\bm{r}',s}
\delta(\hat{\bm{r}}-\bm{r}) \hat{\theta}
\ket{\psi_{\bm{k},\eta}}
\notag\\
&
=\frac{1}{V}
\sum_{\bm{k}}\sum_{\eta=1}^{2N}
\sigma_{3,\eta} f_B(\bar{\varepsilon}_{\bm{k},\eta})
\bra{\psi_{\bm{k},\eta}}
\delta(\hat{\bm{r}}-\bm{r}) \hat{\theta}
\ket{\psi_{\bm{k},\eta}}.
\end{align}
This form is consistent with that given in Ref.~\cite{li2020_prr}.
On the other hand, the normalization of $\ket{\psi_{\bm{k},\eta}}$ is given by $\bra{\psi_{\bm{k},\eta}} \sigma_3 \ket{\psi_{\bm{k},\eta}}= V\sigma_{3,\eta}$.
Finally, $\langle\theta(\bm{r}) \rangle$ is expressed as
\begin{align}
\langle\theta(\bm{r}) \rangle
=
\sum_{\bm{k}}\sum_{\eta=1}^{2N}
f_B(\bar{\varepsilon}_{\bm{k},\eta})
\frac{
\bra{\psi_{\bm{k},\eta}}
\delta(\hat{\bm{r}}-\bm{r}) \hat{\theta}
\ket{\psi_{\bm{k},\eta}}}
{\bra{\psi_{\bm{k},\eta}}
\sigma_3
\ket{\psi_{\bm{k},\eta}}}.
\end{align}
In the limit of continuous wavenumber, the above expression becomes
\begin{align}\label{appeq:theta-r}
\langle\theta(\bm{r}) \rangle
=
\sum_{\eta=1}^{2N}V\int \frac{d\bm{k}}{(2\pi)^d} 
f_B(\bar{\varepsilon}_{\bm{k},\eta})
\frac{
\bra{\psi_{\bm{k},\eta}}
\delta(\hat{\bm{r}}-\bm{r}) \hat{\theta}
\ket{\psi_{\bm{k},\eta}}}
{\bra{\psi_{\bm{k},\eta}}
\sigma_3
\ket{\psi_{\bm{k},\eta}}}.
\end{align}
This expression is not directly applicable to the spatially dependent Hamiltonian $\mathcal{H}$ given in Eq.~\eqref{appeq:H_semiclassical}.
Here, the expectation value is evaluated using a semiclassical approximation.
This approximation is implemented by replacing the Bloch state $\ket{\psi_{\bm{k},\eta}}$ with the wave packet $\ket{W_{\eta}}$ given in Eq.~\eqref{appeq:Weta_ext}, whose position and momentum centers are located at $\bm{r}_c$ and $\bm{k}$, and by replacing the integration over momentum with an integration over phase space $(\bm{r}_c,\bm{k})$~\cite{culcer2004,xiao2010}.
By applying the semiclassical approximation to Eq.~\eqref{appeq:theta-r}, we finally obtain Eq.~\eqref{eq:ave1}.

\section{Derivation of Eq.~\eqref{appeq:ave4}}
\label{app:ave-theta}

In this appendix, we present the derivation of Eq.~\eqref{appeq:ave4} in detail.
We start from Eq.~\eqref{appeq:H_semiclassical} and expand the virtual field $\bm{h}(\hat{\bm{r}}, t)$ around $\bm{r}_c(t)$ as
\begin{align}
  h_\lambda(\bm{r}_c + (\hat{\bm{r}} - \bm{r}_c), t)
  =
  h_{\lambda}(\bm{r}_c) + \sum_{\lambda'}(\hat{r}_{\lambda'} - r_{c,\lambda'}) \partial_{r_{c,\lambda'}} h_{\lambda}(\bm{r}_c, t)
  + \frac{1}{2} \sum_{\lambda'\lambda''}(\hat{r}_{\lambda'} - r_{c,\lambda'}) (\hat{r}_{\lambda''} - r_{c,\lambda''})
  \partial_{r_{c,\lambda'}} \partial_{r_{c,\lambda''}} h_{\lambda}(\bm{r}_c, t)
  + \cdots
\end{align}
From this expansion, Eq.~\eqref{appeq:H_semiclassical} can be written as
\begin{align}\label{eq:H_semiclassical_expanded}
  \hat{H} = \hat{H}_c + \hat{H}' + \hat{H}'' + \cdots,
\end{align}
where $\hat{H}_c = \hat{H}_{0} + \hat{\bm{\theta}} \cdot \bm{h}(\bm{r}_c, t)$.
The terms $\hat{H}'$ and $\hat{H}''$ are the first and second orders of the spatial gradient, which are represented as
\begin{align}
  \label{appeq:H'}
  \hat{H}' &=
  \frac{\hbar}{2}\sum_{\lambda\lambda'} \left\{ \hat{\theta}_{\lambda}, (\hat{r}_{\lambda'} - r_{c,\lambda'}) \right\}  \partial_{r_{c,\lambda'}} h_{\lambda}\\
  \label{appeq:H''}
  \hat{H}'' &=
  \frac{\hbar}{2}\sum_{\lambda\lambda'\lambda''} \left\{ \hat{\theta}_{\lambda}, \frac{1}{2}(\hat{r}_{\lambda'} - r_{c,\lambda'})  (\hat{r}_{\lambda''} - r_{c,\lambda''})\right\} 
  \partial_{r_{c,\lambda'}} \partial_{r_{c,\lambda''}} h_{\lambda},
\end{align}
respectively.
Here, $\{A, B \}$ denotes the anti-commutator of $A$ and $B$.
We take the limit of $\bm{h}\to 0$, $\partial_{r_{c,\lambda}} \bm{h} \to 0$, $\partial_{t} \bm{h} \to 0$, and $\partial_{r_{c,\lambda}} \partial_{r_{c,\lambda'}} \bm{h} \to 0$ after calculating derivatives with respect to the virtual field.

Here, we regard $H_{c,\bm{q}}$ as a unperturbed Hamiltonian and the other terms, such as $\hat{H}'$ and $\hat{H}''$, as perturbations in Eq.~\eqref{eq:H_semiclassical_expanded}.
The eigenstate of the unperturbed Hamiltonian $H_{c,\bm{q}}$ is denoted as $\ket{u_{\bm{q},\eta}(\bm{r}_c)}$, which satisfies $\sigma_{3} H_{c,\bm{q}} \ket{u_{\bm{q},\eta}(\bm{r}_c)} = \bar{\varepsilon}_{\bm{k},\eta}(\bm{r}_c) \ket{u_{\bm{q},\eta}(\bm{r}_c)}$.
Since $\hat{H}_c$ possesses $\bm{r}_c$ as a parameter, its eigenstate $\ket{u_{\bm{q},\eta}(\bm{r}_c)}$ and eigenenergy $\bar{\varepsilon}_{\bm{k},\eta}(\bm{r}_c)$ depend on $\bm{r}_c$.
In the following, we omit the argument $\bm{r}_c$ for simplicity.
Since we are interested in the second order of the spatial gradient of physical quantities, the wave function is demanded to be expanded up to the first order of the spatial gradient.
By considering $\hat{H}'$ as a perturbation, we expand the wave function $\ket{W_\eta}$ in Eq.~\eqref{appeq:Weta_ext} as~\cite{gao2014,xiao2021conserved,li2024}
\begin{align}
  \label{appeq:Weta_H'}
  \ket{W_\eta}
  \approx
  \int d\bm{q} e^{i\bm{q}\cdot\hat{\bm{r}}}
  \left[ C_{\bm{k},\eta}^{(0)}(\bm{q}) \ket{u_{\bm{q},\eta}}
  + \sum_{\eta_1(\neq \eta)}^{2N} C_{\bm{k},\eta_1}(\bm{q}) \ket{u_{\bm{q},\eta_1}} \right],
\end{align}
where the coefficient $C_{\bm{k},\eta}(\bm{q})$ with $\eta$ in the first term is denoted as $C_{\bm{k},\eta}^{(0)}(\bm{q})$ to emphasize that it corresponds to the zeroth order in the spatial gradient of the wave packet.
Here, the second term is a correction due to the perturbation $\hat{H}'$, and hence $C_{\bm{k},\eta_1}(\bm{q})$ is the first order of the spatial gradient.
The coefficient $C_{\bm{k},\eta_1}(\bm{q})$ can be evaluated from the Schr\"odinger equation given in Eq.~\eqref{eq:Schrodinger}.
Up to the first order of the spatial gradient, the left-hand side of Eq.~\eqref{eq:Schrodinger} is calculated as
\begin{align}
  \braket{u_{\bm{q},\eta_1}|\sigma_{3} e^{-\bm{q}\cdot\hat{\bm{r}}}i\hbar \partial_t|W_\eta}
  \approx
  \sum_{\lambda\lambda'}
  \sigma_{3,\eta_1} A^{h_{\lambda}}_{\bm{q},\eta_1\eta} \sigma_{3,\eta}\hbar(v_{\bm{k},\lambda'})_{\eta} C_{\bm{k},\eta}^{(0)}(\bm{q})  \partial_{r_{c,\lambda'}} h_{\lambda}
  + \sigma_{3,\eta_1} \bar{\varepsilon}_{\bm{q},\eta} C_{\bm{k},\eta_1}(\bm{q}),
  \label{appeq:schro_l2}
\end{align}
where $A^{h_\lambda}_{\bm{k},\eta\eta'} = i \sigma_{3,\eta} \braket{u_{\bm{k},\eta}|\sigma_3|\partial_{h_\lambda}u_{\bm{k},\eta'}}$, and we use the relation $\partial_t h_\lambda = \sum_{\lambda'}\dot{r}_{c,\lambda'} (\partial_{r_{c,\lambda'}} h_{\lambda})$, where $\dot{r}_{c,\lambda'}$ is evaluated up to the zeroth order of the spatial gradient of $\bm{r}_c$.
Under this condition, $ \dot{r}_{c,\lambda'}$ is written as $\partial_t r_{c,\lambda'} = \sigma_{3,\eta}(v_{\bm{k},\lambda'})_{\eta}$, which is obtained from the equation of motion of $\bm{r}_c$ given in Eq.~\eqref{appeq:EoM_r} as discussed later.
Similarly, the right-hand side of the Schr\"odinger equation is calculated within the same order as
\begin{align}
  \bra{u_{\bm{q},\eta_1}} \sigma_{3} e^{-i\bm{q}\cdot\hat{\bm{r}}} \sigma_{3} \hat{H} \ket{W_\eta}\approx
  \sigma_{3,\eta_1}\bar{\varepsilon}_{\bm{q},\eta_1} C_{\bm{k},\eta_1}(\bm{q})
  +
  \int d\bm{q}' C_{\bm{k},\eta}^{(0)}(\bm{q}')
  \braket{u_{\bm{q},\eta_1}|e^{i(\bm{q}'-\bm{q})\cdot\hat{\bm{r}}} H'_{\bm{q}'}|u_{\bm{q}',\eta}}.
  \label{appeq:schro_r2}
\end{align}
From Eqs.~\eqref{appeq:schro_l2} and \eqref{appeq:schro_r2}, we obtain the expression of $C_{\bm{k},\eta_1}(\bm{q})$  as
\begin{align}
  C_{\bm{k},\eta_1}(\bm{q})
  =
  -i\sum_{\lambda\lambda'}
  \left[ \left( i \partial_{q_{\lambda'}} + A^{\lambda'}_{\bm{q},\eta} - r_{c,\lambda'} \right) C_{\bm{k},\eta}^{(0)}(\bm{q}) \right] 
  A^{h_{\lambda}}_{\bm{q},\eta_1 \eta} \partial_{r_{c,\lambda'}} h_{\lambda}
  +\sum_{\lambda\lambda'}
  C_{\bm{k},\eta}^{(0)}(\bm{q}) \frac{
    (d^{\theta_\lambda}_{\bm{k},\eta_1\eta})_{\lambda'}  
  }{\bar{\varepsilon}_{\bm{q}, \eta}-\bar{\varepsilon}_{\bm{q}, \eta_1}} 
  \partial_{r_{c,\lambda'}} h_{\lambda},
  \label{appeq:Cketa_1}
\end{align}
where we used the relation $
\sigma_{3,\eta_1}(\theta_{\bm{q},\lambda})_{\eta_1\eta}
=
-\frac{i}{\hbar} (\bar{\varepsilon}_{\bm{q},\eta} - \bar{\varepsilon}_{\bm{q},\eta_1})A^{h_\lambda}_{\bm{q},\eta_1\eta}
$.
Here, the Berry connection is given by $A^\lambda_{\bm{k},\eta} = i \sigma_{3,\eta} \braket{u_{\bm{k},\eta}|\sigma_3|\partial_{k_\lambda}u_{\bm{k},\eta}}$, and $(d^{\theta_\lambda}_{\bm{k},\eta_1\eta})_{\lambda'}$ is defined as
\begin{align}
  (d^{\theta_\lambda}_{\bm{k},\eta_1\eta})_{\lambda'} 
  =
  \frac{\hbar}{2}
  \left[ i \partial_{q_{\lambda'}} + A^{\lambda'}_{\bm{q},\eta_1} - A^{\lambda'}_{\bm{q},\eta} \right]
  \sigma_{3,\eta_1}(\theta_{\bm{q},\lambda})_{\eta_1 \eta}
  + \frac{\hbar}{2}\sum_{\eta_2(\neq \eta_1)}^{2N} A^{\lambda'}_{\bm{q},\eta_1\eta_2} \sigma_{3,\eta_2}(\theta_{\bm{q},\lambda})_{\eta_2\eta}
  + \frac{\hbar}{2}\sum_{\eta_2(\neq \eta)}^{2N} \sigma_{3,\eta_1}(\theta_{\bm{q},\lambda})_{\eta_1\eta_2}  A^{\lambda'}_{\bm{q},\eta_2\eta}
  - \hbar A^{h_{\lambda}}_{\bm{q},\eta_1 \eta}\sigma_{3,\eta}(v_{\bm{k},\lambda'})_{\eta}.
  \label{appeq:d_off_2}
\end{align}

In the next step, we evaluate the center position of the wave packet, $\bm{r}_c$, which is defined in Eq.~\eqref{eq:def_rc}.  
By expanding $\bm{r}_c$ up to the first order of the spatial gradient using the wave function $\ket{W_\eta}$ given in Eq.~\eqref{appeq:Weta_H'}, we obtain  
\begin{align}  
  \label{appeq:rc_withfirstpert}  
  \bm{r}_c \approx \bm{r}_c^{(0)} + \bm{a}^{}_{\bm{k},\eta},  
\end{align}  
where $\bm{r}_c^{(0)}$ represents the zeroth-order term in the spatial gradient expansion of the center position and is given by  
\begin{align}  
  \label{appeq:rc_nopert}  
  \bm{r}_c^{(0)}  
  =  
  \pdiff{\phi_\eta}{\bm{k}} + \bm{A}_{\bm{k},\eta}.  
\end{align}  
Here, we introduce $\phi_\eta(\bm{k})$ as $C_{\bm{k},\eta}^{(0)}(\bm{q}) = |C_{\bm{k},\eta}^{(0)}(\bm{q})|e^{-i\phi_\eta(\bm{q})}$, and $\bm{a}_{\bm{k},\eta}$ is defined as
\begin{align}
  \label{appeq:a_shift}
  \bm{a}_{\bm{k},\eta} = 2 \mathrm{Re} \int d\bm{q} \sum_{\eta_1(\neq \eta)}^{2N} 
  C_{\bm{k},\eta}^{(0)*}(\bm{q}) C_{\bm{k},\eta_1}(\bm{q}) 
  \bm{A}_{\bm{q},\eta \eta_1},
\end{align}
which corresponds to the shift of the center position due to the first order of the spatial gradient~\cite{gao2014}.
Substituting the expression of $\bm{a}_{\bm{k},\eta}$ given in Eq.~\eqref{appeq:Cketa_1}, we obtain the following equation:
\begin{align}
  \label{appeq:alambda''_shift}
  a^{\lambda''}_{\bm{k},\eta}
  =\sum_{\lambda\lambda'}
  \left[
    2\mathrm{Re} \sum_{\eta_1(\neq \eta)}^{2N} 
    \frac
    {A^{\lambda''}_{\bm{k},\eta\eta_1} 
    (d^{\theta_\lambda}_{\bm{k},\eta_1\eta})_{\lambda'}}
    {\bar{\varepsilon}_{\bm{k},\eta} - \bar{\varepsilon}_{\bm{k},\eta_1}}
    - \partial_{k_{\lambda'}} 
    \big(g^{\theta_\lambda}_{\bm{k},\eta}\big)_{\lambda''}
  \right] \partial_{r_{c,\lambda'}} h_{\lambda},
\end{align}
where $\big(g^{\theta_\lambda}_{\bm{k},\eta}\big)_{\lambda'}$ is written by the quantum metric as $\big(g^{\theta_\lambda}_{\bm{k},\eta}\big)_{\lambda'}=g^{k_{\lambda'} h_{\lambda} }_{\bm{k},\eta}$, which is introduced in Eq.~\eqref{appeq:QMT}.

Here, we evaluate the Lagrangian $\mathcal{L}_{\eta}$, which is given in Eq.~\eqref{eq:Leta1}, up to the second order of the spatial gradient.
The Lagrangian is written as
\begin{align}
    \label{appeq:Leta}
    \mathcal{L}_\eta
    = \frac{\braket{W_\eta|i\hbar\sigma_3\diff{}{t}|W_\eta}}{\braket{W_\eta|\sigma_3|W_\eta}}
    -\frac{\braket{W_\eta|\sigma_3(\sigma_3 \hat{H})|W_\eta}}{\braket{W_\eta|\sigma_3|W_\eta}},
\end{align}
where the second term is equal to $\bar{\varepsilon}_{\bm{k},\eta}^{\mathrm{tot}}$, which is introduced in Sec.~\ref{sec:semiclassical}.
The denominator of the Lagrangian is calculated as
\begin{align}
  \braket{W_\eta|\sigma_3|W_\eta}
  \approx
  \sigma_{3,\eta} \int d\bm{q} \left[ |C_{\bm{k},\eta}^{(0)}(\bm{q})|^2 + \sum_{\eta_1(\neq \eta)}^{2N}\sigma_{3,\eta}\sigma_{3,\eta_1} |C_{\bm{k},\eta_1}(\bm{q})|^2 \right]
  = \sigma_{3,\eta} (1 + \delta)^{-2},
\end{align}
where $-2\delta = \int d\bm{q} \sum_{\eta_1(\neq \eta)}^{2N} \sigma_{3,\eta}\sigma_{3,\eta_1} |C_{\bm{k},\eta_1}(\bm{q})|^2$, which is a contribution from the second order of the spatial gradient.
Using this result, we evaluate the first term of the Lagrangian up to the second order of the spatial gradient as
\begin{align}
  \frac{\braket{W_\eta|i\hbar\sigma_3\diff{}{t}|W_\eta}}{\braket{W_\eta|\sigma_3|W_\eta}}
  \approx&
  \hbar\diff{\phi_\eta}{t}
  -\hbar \sum_{\lambda}\dot{k}_{\lambda} {r}_{c,\lambda} + \hbar \sum_{\lambda}\dot{k}_{\lambda} \tilde{A}^{\lambda}_{\bm{k},\eta}
  + \hbar \sum_{\lambda}\dot{r}_{c,\lambda} \tilde{A}^{r_{c,\lambda}}_{\bm{k},\eta}
  + \hbar \tilde{A}^t_{\bm{k},\eta}\notag\\
  &
  + 2\delta i\hbar \int d\bm{q} C^{(0)*}_{\bm{k},\eta}(\bm{q}) [\partial_{t}C_{\bm{k},\eta}^{(0)}(\bm{q})]
  + i\hbar \int d\bm{q} \sum_{\eta_1(\neq \eta)}^{2N} \sigma_{3, \eta}\sigma_{3,\eta_1}
  C_{\bm{k},\eta_1}^{*}(\bm{q}) [\partial_{t}C_{\bm{k},\eta_1}(\bm{q})].
  \label{appeq:dt in Leta}
\end{align}
Here, we introduce 
$\tilde{A}^{\lambda}_{\bm{k},\eta}
=
A^{\lambda}_{\bm{k},\eta} + a^{\lambda}_{\bm{k},\eta}$.
Similarly, we define 
$\tilde{A}^{X}_{\bm{k},\eta}
=A^{X}_{\bm{k},\eta} + a^{X}_{\bm{k},\eta}$
 with 
$A^{X}_{\bm{k},\eta}
= i\sigma_{3,\eta} \braket{u_{\bm{k},\eta}|\sigma_3|\partial_{X} u_{\bm{k},\eta}}
$ and 
$a^{X}_{\bm{k},\eta} 
= 
2 \mathrm{Re} \int d\bm{q} \sum_{\eta_1(\neq \eta)}^{2N} C_{\bm{k},\eta}^{(0)*}(\bm{q}) C_{\bm{k},\eta_1}(\bm{q}) A^{X}_{\bm{k},\eta\eta_1}$
for $X=r_{c,\lambda}$, $h_{\lambda}$, and $t$.
From the Berry connection $\tilde{A}^{X}_{\bm{k},\eta}$, we define the Berry curvature incorporating corrections by the perturbation as
\begin{align}
  \label{appeq:GBC_withpert}
  \tilde{\Omega}^{X_{\lambda}Y_{\lambda'}}_{\bm{k},\eta}
  =
  \pdiff{}{X_{\lambda}} \tilde{A}^{Y_{\lambda'}}_{\bm{k},\eta} 
  -
  \pdiff{}{Y_{\lambda'}} \tilde{A}^{X_{\lambda}}_{\bm{k},\eta}.
\end{align}
This is written as $\tilde{\Omega}^{X_{\lambda}Y_{\lambda'}}_{\bm{k},\eta}
=
\Omega^{X_{\lambda}Y_{\lambda'}}_{\bm{k},\eta}
+ \delta\Omega^{X_{\lambda}Y_{\lambda'}}_{\bm{k},\eta}$, where $
\Omega^{X_{\lambda}Y_{\lambda'}}_{\bm{k},\eta}
=
\pdiff{}{X_{\lambda}} A^{Y_{\lambda'}}_{\bm{k},\eta} 
-
\pdiff{}{Y_{\lambda'}} A^{X_{\lambda}}_{\bm{k},\eta}
$ is the Berry curvature without perturbation and $\delta\Omega^{X_{\lambda}Y_{\lambda'}}_{\bm{k},\eta}
=
\pdiff{}{X_{\lambda}} a^{Y_{\lambda'}}_{\bm{k},\eta}
-
\pdiff{}{Y_{\lambda'}} a^{X_{\lambda}}_{\bm{k},\eta}$ is a correction term due to the perturbation.
Similarly, the second term in Eq.~\eqref{appeq:Leta}, namely $\bar{\varepsilon}_{\bm{k},\eta}^{\mathrm{tot}}$, is evaluated up to the second order of the spatial gradient as
\begin{align}
  \bar{\varepsilon}_{\bm{k},\eta}^{\mathrm{tot}}
  \approx&
  \sigma_{3,\eta}\braket{W_\eta^{(0)}|\hat{H}_{c}|W_\eta^{(0)}} + \sigma_{3,\eta}\braket{W_\eta^{(0)}|\hat{H}'|W_\eta^{(0)}}\notag\\
  &+ \sigma_{3,\eta}\braket{W_\eta^{(1)}|\hat{H}_{c}|W_\eta^{(1)}} + 2\sigma_{3,\eta}\mathrm{Re} \braket{W_\eta^{(0)}|\hat{H}'|W_\eta^{(1)}} + \sigma_{3,\eta}\braket{W_\eta^{(0)}|\hat{H}''|W_\eta^{(0)}}
  + 2\delta \sigma_{3,\eta}\braket{W_\eta^{(0)}|\hat{H}_{c}|W_\eta^{(0)}},
  \label{appeq:enetot_2nd}
\end{align}
where $\ket{W_\eta^{(0)}}$ and $\ket{W_\eta^{(1)}}$ are the zeroth and first order in the spatial gradient of the wave packet.
In present study, we are interested in the second derivative of the spatial gradient, $\partial_{r_{c,\lambda}} \partial_{r_{c,\lambda'}}$ and not in the square of first derivatives, such as $(\partial_{r_{c,\lambda'}} h_{\lambda}) (\partial_{r_{c,\lambda''}} h_{\lambda'''})$, among the second order of the spatial gradient.
From this consideration, 
the terms containing $|C_{\bm{k},\eta_1}(\bm{q})|^2 \propto (\partial_{r_{c,\lambda'}} h_{\lambda}) (\partial_{r_{c,\lambda''}} h_{\lambda'''})$ in Eqs.~\eqref{appeq:dt in Leta} and \eqref{appeq:enetot_2nd} are omitted.
Consequently, we can simplify $\mathcal{L}_\eta$ up to the second order of the spatial gradient to~\cite{sundaram1999}
\begin{align}
  \label{appeq:Leta3}
  \mathcal{L}_{\eta}
  &\approx
  \hbar \sum_{\lambda}k_{\lambda} \dot{r}_{c,\lambda} + \hbar \sum_{\lambda}\dot{k}_{\lambda} \tilde{A}^{\lambda}_{\bm{k},\eta}
  + \hbar \sum_{\lambda}\dot{r}_{c,\lambda} \tilde{A}^{r_{c,\lambda}}_{\bm{k},\eta} 
  + \hbar \tilde{A}^t_{\bm{k},\eta}
  - \bar{\varepsilon}_{\bm{k},\eta}^{\mathrm{tot}},
\end{align}
where $\bar{\varepsilon}_{\bm{k},\eta}^{\mathrm{tot}}$ is rewritten as
\begin{align}
  \bar{\varepsilon}_{\bm{k},\eta}^{\mathrm{tot}}
  \approx&
  \bar{\varepsilon}_{\bm{k},\eta} + \sigma_{3,\eta}\braket{W_\eta^{(0)}|\hat{H}'|W_\eta^{(0)}} + \sigma_{3,\eta}\braket{W_\eta^{(0)}|\hat{H}''|W_\eta^{(0)}}.
  \label{appeq:enetot_2nd_simeq}
\end{align}
Note that we omitted the total time derivative term in Eq.~\eqref{appeq:Leta3}.

As we have obtained the Lagrangian $\mathcal{L}_{\eta}$ in Eq.~\eqref{appeq:Leta3}, we can determine the equation of motion from the following Lagrange equation with respect to $\bm{r}_c$:
\begin{align}
  \diff{}{t} \left( \pdiff{\mathcal{L}_{\eta}}{\dot{r}_{c,\lambda}} \right) - \pdiff{\mathcal{L}_{\eta}}{r_{c,\lambda}}
  =
  0
\end{align}
This gives the equation of motion for $\bm{r}_c$ as
\begin{align}
  \label{appeq:EoM_r}
  \dot{r}_{c,\lambda}
  &=
  \frac{1}{\hbar} \pdiff{\bar{\varepsilon}^{\mathrm{tot}}_{\bm{k},\eta}}{k_{\lambda}}
  - \tilde{\Omega}^{k_{\lambda} \mathcal{T}}_{\bm{k},\eta},
\end{align}
where $\tilde{\Omega}^{X_{\lambda} \mathcal{T}}_{\bm{k},\eta}$ is defined as
\begin{align}
  \label{appeq:GBC_totaltimediff}
  \tilde{\Omega}^{X_{\lambda} \mathcal{T}}_{\bm{k},\eta}
  =
  \sum_{\lambda'}\tilde{\Omega}^{X_{\lambda} r_{c,\lambda'}}_{\bm{k},\eta} \dot{r}_{c,\lambda'}
  + \sum_{\lambda'}\tilde{\Omega}^{X_{\lambda} k_{\lambda'}}_{\bm{k},\eta} \dot{k}_{\lambda'}
  + \tilde{\Omega}^{X_{\lambda} t}_{\bm{k},\eta}.
\end{align}
In a similar manner, we can derive the equation of motion for $\bm{k}$ as
\begin{align}
  \label{appeq:EoM_k}
  \dot{k}_{\lambda}
  &=
  -\frac{1}{\hbar} \pdiff{\bar{\varepsilon}^{\mathrm{tot}}_{\bm{k},\eta}}{r_{c,\lambda}}
  + \tilde{\Omega}^{r_{c,\lambda} \mathcal{T}}_{\bm{k},\eta}.
\end{align}

To further proceed a calculation for the Lagrangian $\mathcal{L}_{\eta}$ in Eq.~\eqref{appeq:Leta3}, we evaluate $\bar{\varepsilon}^{\mathrm{tot}}_{\bm{k},\eta}$ given in Eq.~\eqref{appeq:enetot_2nd_simeq}.
Here, we calculate the second and third terms in Eq.~\eqref{appeq:enetot_2nd_simeq}.
By using the expression of the dipole moment given in Eq.~\eqref{eq:appeq:GDM_def}, the second term is written as
\begin{align}
  \sigma_{3,\eta}\braket{W_\eta^{(0)}|\hat{H}'|W_\eta^{(0)}}
  &=
  \hbar \sum_{\lambda\lambda'}
  (d^{\theta_\lambda}_{\bm{k},\eta})_{\lambda'}
   \partial_{r_{c,\lambda'}} h_{\lambda}.
\end{align}
Since the correction of the Berry connection in Eq.~\eqref{appeq:rc_withfirstpert} satisfies $\bm{a}_{\bm{k},\eta} \propto \partial_{r_{c,\lambda'}} h_{\lambda}$, the dipole moment is approximated as
$(d^{\theta_\lambda}_{\bm{k},\eta})_{\lambda'}
   \partial_{r_{c,\lambda'}} h_{\lambda} \approx \mathrm{Re} \ \sigma_{3,\eta} \braket{W_\eta^{(0)}|(\hat{r}_{\lambda'} - r_{c,\lambda'}^{(0)}) \hat{\theta}_{\lambda}|W_\eta^{(0)}} 
\partial_{r_{c,\lambda'}} h_{\lambda}$ by neglecting the square of the first order of spatial derivatives.
Using this approximation, we can evaluate the dipole moment as
   \begin{align}
    (d^{\theta_\lambda}_{\bm{k},\eta})_{\lambda'}
    \approx \frac{1}{\hbar}
    \mathrm{Im}  \  \sigma_{3,\eta}
    \bra{\partial_{k_{\lambda'}}u_{\bm{k},\eta}}\sigma_{3}
    \left( \bar{\varepsilon}_{\bm{k},\eta} -  \sigma_3 H_{c,\bm{k}} \right) \ket{\partial_{h_{\lambda}} u_{\bm{k},\eta}}
  \end{align}
By taking the zero limit of the virtual field, we obtain
\begin{align}
  (d^{\theta_\lambda}_{\bm{k},\eta})_{\lambda'}
  \approx \hbar
  \sum_{\eta_1(\neq \eta)}^{2N} \sigma_{3,\eta}\sigma_{3,\eta_1} 
  \frac{\mathrm{Im}\left[ (v_{\bm{k},\lambda'})_{\eta\eta_1} (\theta_{\bm{k},\lambda})_{\eta_1\eta} \right]}{\bar{\varepsilon}_{\bm{k},\eta}-\bar{\varepsilon}_{\bm{k},\eta_1}},
\end{align}
which corresponds to Eq.~\eqref{eq:GBC-O} in the main text.

Next, we calculate the third term in Eq.~\eqref{appeq:enetot_2nd_simeq}.
Using the expression of the quadrupole moment given in Eq.~\eqref{appeq:GQM_def}, we rewrite the third term in Eq.~\eqref{appeq:enetot_2nd_simeq} as
\begin{align}
  \label{appeq:enetot_3}
  \sigma_{3,\eta}\braket{W_{\eta}^{(0)}|\hat{H}''|W_{\eta}^{(0)}}
  =
  \hbar \sum_{\lambda\lambda'\lambda''}
  \big(q^{\theta_\lambda}_{\bm{k},\eta}\big)_{\lambda'\lambda''}
   \partial_{r_{c,\lambda'}} \partial_{r_{c,\lambda''}} h_{\lambda}.
\end{align}
As we focus on contributions up to the second order of the spatial gradient, the quadrupole moment is approximated up to the zeroth order of the spatial gradient, namely, $\bm{r}_c$ in Eq.~\eqref{appeq:GQM_def} can be replaced to $\bm{r}_c^{(0)}$.
By applying this approximation, the quadrupole moment is written as
\begin{align}
  2\big(q^{\theta_\lambda}_{\bm{k},\eta}\big)_{\lambda'\lambda''}
  \approx
  \mathrm{Re} \ \sigma_{3,\eta} \braket{W_\eta^{(0)}|\hat{r}_{\lambda'} \hat{r}_{\lambda''} \hat{\theta}^{\lambda}|W_\eta^{(0)}} 
  - r_{c,\lambda'}^{(0)} (d^{\theta_\lambda}_{\bm{k},\eta})_{\lambda''}
  - r_{c,\lambda''}^{(0)} (d^{\theta_\lambda}_{\bm{k},\eta})_{\lambda'}
  - r_{c,\lambda'}^{(0)} r_{c,\lambda''}^{(0)} \sigma_{3,\eta}\braket{W_\eta^{(0)}|\hat{\theta}^{\lambda}|W_\eta^{(0)}}.
  \label{appeq:GQM2}
\end{align}
The first term in Eq.~\eqref{appeq:GQM2} is evaluated as
\begin{align}
  \mathrm{Re} \ \sigma_{3,\eta} \braket{W_\eta^{(0)}|\hat{r}_{\lambda'} \hat{r}_{\lambda''} \hat{\theta}^{\lambda}|W_\eta^{(0)}}
=&
  -\mathrm{Re} \int d\bm{q} [\partial_{q_{\lambda'}} \partial_{q_{\lambda''}} C_{\bm{k},\eta}^{(0)*}(\bm{q})] C_{\bm{k},\eta}^{(0)}(\bm{q}) \sigma_{3,\eta}(\theta_{\bm{q},\lambda})_{\eta}
  + \pdiff{\phi_\eta}{k_{\lambda''}} \mathrm{Re} \sum_{\eta_1=1}^{2N} A^{\lambda'}_{\bm{k},\eta\eta_1} \sigma_{3,\eta_1}(\theta_{\bm{k},\lambda})_{\eta_1\eta}\notag\\
  &
  + \pdiff{\phi_\eta}{k_{\lambda'}} \mathrm{Re} \sum_{\eta_1=1}^{2N} A^{\lambda''}_{\bm{k},\eta\eta_1} \sigma_{3,\eta_1}(\theta_{\bm{k},\lambda})_{\eta_1\eta}
  +\frac{1}{2}\mathrm{Re}\left\{ \sigma_{3,\eta} \braket{\partial_{k_{\lambda''}}u_{\bm{k},\eta}|\Big(\partial_{k_{\lambda'}}\theta_{\bm{k},\lambda}|u_{\bm{k},\eta}}\Big)  
  + (\lambda' \leftrightarrow \lambda'')\right\}
  \label{appeq:rrtheta}
\end{align}
The first term of Eq.~\eqref{appeq:rrtheta} is expanded as
\begin{align}
  &-\mathrm{Re} \int d\bm{q} [\partial_{q_{\lambda'}} \partial_{q_{\lambda''}} C_{\bm{k},\eta}^{(0)*}(\bm{q})] C_{\bm{k},\eta}^{(0)}(\bm{q}) \sigma_{3,\eta}(\theta_{\bm{q},\lambda})_{\eta}
  \notag\\
  &\quad=
  \pdiff{\phi_\eta}{k_{\lambda'}}\pdiff{\phi_\eta}{k_{\lambda''}} \sigma_{3,\eta}(\theta_{\bm{k},\lambda})_{\eta} 
  - \frac{1}{2} \sigma_{3,\eta} \partial_{k_{\lambda'}} \partial_{k_{\lambda''}}(\theta_{\bm{k},\lambda})_{\eta}
  - \int d\bm{q} \left[\partial_{q_{\lambda'}} |C_{\bm{k},\eta}^{(0)}(\bm{q})|\right] \left[\partial_{q_{\lambda''}} |C_{\bm{k},\eta}^{(0)}(\bm{q})|\right] \sigma_{3,\eta}(\theta_{\bm{q},\lambda})_{\eta}.
  \label{appeq:neg_shape}
\end{align}
The last term depends on the shape of the wave packet, but the envelope $|C_{\bm{k},\eta}(\bm{q})|$ has a peak at $\bm{q} = \bm{k}$, which suggests that the contribution of $\partial_{q_{\lambda}} |C_{\bm{k},\eta}^{(0)}(\bm{q})|$ at $\bm{q} = \bm{k}$ is nearly zero.
Consequently, we neglect the last term in Eq.~\eqref{appeq:neg_shape}.
Proceeding further calculations in Eq.~\eqref{appeq:GQM2} in a similar manner to those in Ref.~\cite{xiao2021conserved}, we obtain
\begin{align}
  2\big(q^{\theta_\lambda}_{\bm{k},\eta}\big)_{\lambda'\lambda''}
  &=
  - \frac{1}{2}\partial_{k_{\lambda'}} \partial_{k_{\lambda''}} \sigma_{3,\eta}(\theta_{\bm{k},\lambda})_{\eta}
  + \mathrm{Re} \sum_{\eta_1(\neq \eta)}^{2N} \sum_{\eta_2(\neq \eta)}^{2N} A^{\lambda'}_{\bm{k},\eta\eta_1} 
  \sigma_{3,\eta_1}(\theta_{\bm{k},\lambda})_{\eta_1\eta_2} A^{\lambda''}_{\bm{k},\eta_2\eta}
  + \frac{1}{2}\left[ \mathrm{Re} \braket{\partial_{k_{\lambda'}} u_{\bm{k},\eta}|(\partial_{k_{\lambda''}} \theta_{\bm{k},\lambda})|u_{\bm{k},n}} + (\lambda'\leftrightarrow\lambda'') \right].
  \label{appeq:small_qm}
\end{align}
This expression is used to derive Eq.~\eqref{eq:quadrupole-moment} in the main text.
From the above results, $\bar{\varepsilon}^{\mathrm{tot}}_{\bm{k},\eta}$ is written as
$\bar{\varepsilon}^{\mathrm{tot}}_{\bm{k},\eta}
\approx\bar{\varepsilon}_{\bm{k},\eta}
+ \hbar \sum_{\lambda\lambda'}
(d^{\theta_\lambda}_{\bm{k},\eta})_{\lambda'}
 \partial_{r_{c,\lambda'}} h_{\lambda}
  + \hbar \sum_{\lambda\lambda'\lambda''}\big(q^{\theta_\lambda}_{\bm{k},\eta}\big)_{\lambda'\lambda''} \partial_{r_{c,\lambda'}} \partial_{r_{c,\lambda''}} h_{\lambda}$, and thereby,
we obtain expression of the Lagrangian $\mathcal{L}_{\eta}$ up to the second order of the spatial gradient as 
\begin{align}\label{appeq:Leta-full}
  \mathcal{L}_{\eta}
  \approx
  \hbar \sum_{\lambda}k_{\lambda} \dot{r}_{c,\lambda} + \hbar \sum_{\lambda}\dot{k}_{\lambda} \tilde{A}^{\lambda}_{\bm{k},\eta}
  + \hbar \sum_{\lambda}\dot{r}_{c,\lambda} \tilde{A}^{r_{c,\lambda}}_{\bm{k},\eta} 
  + \hbar \tilde{A}^t_{\bm{k},\eta}
  -\bar{\varepsilon}_{\bm{k},\eta}
  - \hbar \sum_{\lambda\lambda'}
  (d^{\theta_\lambda}_{\bm{k},\eta})_{\lambda'}
   \partial_{r_{c,\lambda'}} h_{\lambda}
    - \hbar \sum_{\lambda\lambda'\lambda''}\big(q^{\theta_\lambda}_{\bm{k},\eta}\big)_{\lambda'\lambda''} \partial_{r_{c,\lambda'}} \partial_{r_{c,\lambda''}} h_{\lambda}.
\end{align}

Finally, let us evaluate Eq.~\eqref{appeq:ave2} in the main text using the Lagrangian in Eq.~\eqref{appeq:Leta-full}.
The functional derivative of the action $\mathcal{S}_{\eta}$ with respect to $h_{\lambda}$ is given by
\begin{align}
  \fdiff{\mathcal{S}_{\eta}}{h_{\lambda}(\bm{r}, t)}\Bigg|_{\textrm{on-shell}}
  &\approx
  \pdiff{\mathcal{L}_\eta}{h_{\lambda}}\Bigg|_{(\bm{r}_c, t)} \delta(\bm{r} - \bm{r}_c) 
  + \sum_{\lambda'}\pdiff{\mathcal{L}_\eta}{(\partial_{r_{c,\lambda'}} h_{\lambda})}\Bigg|_{(\bm{r}_c, t)} \partial_{r_{c,\lambda'}}\delta(\bm{r}-\bm{r}_c)\notag\\
  &
  - \diff{}{t}\left[ \pdiff{\mathcal{L}_\eta}{(\partial_t h_{\lambda})}\Bigg|_{(\bm{r}_c, t)} \delta(\bm{r} - \bm{r}_c) \right]
  + \sum_{\lambda'\lambda''}\pdiff{\mathcal{L}_\eta}{(\partial_{r_{c,\lambda'}} \partial_{r_{c,\lambda''}}h_{\lambda})}\Bigg|_{(\bm{r}_c, t)} \partial_{r_{c,\lambda'}} \partial_{r_{c,\lambda''}} \delta(\bm{r} - \bm{r}_c)
\end{align}
Using this expression, we can evaluate the part of Eq.~\eqref{appeq:ave2} in the main text as
\begin{align}
  &\int d\bm{r}_c  \int \frac{d\bm{k}}{(2\pi)^d} D_{\bm{k},\eta} f_B(\bm{r}_c,\bar{\varepsilon}^{\mathrm{tot}}_{\bm{k},\eta}) 
  \fdiff{S_\eta}{h_{\lambda}(\bm{r}, t)}\Bigg|_{\textrm{on-shell}}\notag\\
  &\quad\approx
  \int \frac{d\bm{k}}{(2\pi)^d} D_{\bm{k},\eta} f_B(\bm{r},\bar{\varepsilon}^{\mathrm{tot}}_{\bm{k},\eta}) \pdiff{\mathcal{L}_\eta}{h_{\lambda}}
   - \sum_{\lambda'}\frac{\partial}{\partial r_{\lambda'}}  \int \frac{d\bm{k}}{(2\pi)^d} D_{\bm{k},\eta} f_B(\bm{r},\bar{\varepsilon}^{\mathrm{tot}}_{\bm{k},\eta}) \pdiff{\mathcal{L}_\eta}{(\partial_{ r_{\lambda'}} h_{\lambda})}
  - \int \frac{d\bm{k}}{(2\pi)^d} D_{\bm{k},\eta} f_B(\bm{r},\bar{\varepsilon}^{\mathrm{tot}}_{\bm{k},\eta}) \diff{}{t} \pdiff{\mathcal{L}_\eta}{(\partial_t h_{\lambda})}\notag\\
  &\quad+
  \sum_{\lambda'}\frac{\partial}{\partial r_{\lambda'}} \int \frac{d\bm{k}}{(2\pi)^d} D_{\bm{k},\eta} f_B(\bm{r},\bar{\varepsilon}^{\mathrm{tot}}_{\bm{k},\eta}) \pdiff{\mathcal{L}_\eta}{(\partial_t h_{\lambda})} \dot{r}_{\lambda'}
  + \sum_{\lambda'\lambda''}\frac{\partial^2}{\partial r_{\lambda'}\partial r_{\lambda''}} \int \frac{d\bm{k}}{(2\pi)^d} D_{\bm{k},\eta} f_B(\bm{r},\bar{\varepsilon}^{\mathrm{tot}}_{\bm{k},\eta}) \pdiff{\mathcal{L}_\eta}{(\partial_{ r_{\lambda'}}\partial_{ r_{\lambda''}}h_{\lambda})}.
\end{align}
By carrying out the partial derivatives in the above expression, Eq.~\eqref{appeq:ave2} is rewritten as~\cite{xiao2021thermoelectric,xiao2021conserved}
\begin{align}
  \braket{\theta_{\lambda}(\bm{r})}
  \approx \frac{1}{\hbar}\sum_{\eta=1}^{2N}
  &\Bigg\{
    \int \frac{d\bm{k}}{(2\pi)^d} D_{\bm{k},\eta} f_B(\bm{r},\bar{\varepsilon}^{\mathrm{tot}}_{\bm{k},\eta}) \left( \partial_{h_{\lambda}} \bar{\varepsilon}^{\mathrm{tot}}_{\bm{k},\eta}
  + \hbar\tilde{\Omega}^{h_{\lambda}\mathcal{T}}_{\bm{k},\eta}  \right) \notag\\
  &
  -
  \sum_{\lambda'}\frac{\partial}{\partial r_{\lambda'}} \int \frac{d\bm{k}}{(2\pi)^d} D_{\bm{k},\eta} f_B(\bm{r},\bar{\varepsilon}^{\mathrm{tot}}_{\bm{k},\eta}) \hbar
  \left[ 
    (d^{\theta_\lambda}_{\bm{k},\eta})_{\lambda'} - \sum_{\lambda''}\dot{k}_{\lambda''} \pdiff{a^{\lambda''}_{\bm{k},\eta}}{(\partial_{r_{\lambda'}} h_{\lambda})}
    - \sum_{\lambda''}\left(\sum_{\lambda'''}\dot{r}_{\lambda'''}  \frac{\partial h_{\lambda''}}{\partial r_{\lambda'''}} + \frac{\partial h_{\lambda''}}{\partial t}\right) \pdiff{a^{h_{\lambda''}}_{\bm{k},\eta}}{(\partial_{r_{\lambda'}} h_{\lambda})}
  \right]\notag\\
  &
  + \sum_{\lambda'\lambda''}\frac{\partial^2}{\partial r_{\lambda'}\partial r_{\lambda''}} \int \frac{d\bm{k}}{(2\pi)^d} D_{\bm{k},\eta} f_B(\bm{r},\bar{\varepsilon}^{\mathrm{tot}}_{\bm{k},\eta}) \hbar \big(q^{\theta_\lambda}_{\bm{k},\eta}\big)_{\lambda'\lambda''}
  \Bigg\}\Bigg|_{\bm{h}\to 0}.
  \label{appeq:ave3}
\end{align}
The expression of Eq.~\eqref{appeq:ave3} resembles that for the fermionic version obtained in Refs.~\cite{sundaram1999,dong2020,xiao2021thermoelectric,xiao2021conserved}.
By performing calculations in a similar manner as in the fermionic case, we obtain Eq.~\eqref{appeq:ave4} in the main text.

\section{Evaluation of momentum derivative of dipole moment}
\label{appendix:derivative}

In this appendix, we show Eq.~\eqref{eq:derivative-d} in the main text.
First, by using the relation
$(d^{S^\gamma}_{\bm{k},\eta})_{\lambda}
= \mathrm{Re} \sum_{\eta_1(\neq \eta)}^{2N} A^{\lambda}_{\bm{k},\eta\eta_1}\sigma_{3, \eta_1} (S^{\gamma}_{\bm{k}})_{\eta_1 \eta}
$ with interband Berry connection $A^{\lambda}_{\bm{k},\eta\eta_1}$, the left-hand side of Eq.~\eqref{eq:derivative-d} can be written as
\begin{align}
  \label{appeq:bibun}
  \partial_{k_{\lambda'}} (d^{S^\gamma}_{\bm{k},\eta})_{\lambda}
  =
  \mathrm{Re} \sum_{\eta_1(\neq \eta)}^{2N} \left(\partial_{k_{\lambda'}} A^{\lambda}_{\bm{k},\eta\eta_1}\right) \sigma_{3, \eta_1}(S^{\gamma}_{\bm{k}})_{\eta_1 \eta}
  +
  \mathrm{Re} \sum_{\eta_1(\neq \eta)}^{2N} A^{\lambda}_{\bm{k},\eta\eta_1} \sigma_{3, \eta_1}\left[ \partial_{k_{\lambda'}} (S^{\gamma}_{\bm{k}})_{\eta_1 \eta}\right].
\end{align}
Here, we carry out the $\bm{k}$-derivative in Eq.~\eqref{appeq:bibun}.
For the first term on the right-hand side of Eq.~\eqref{appeq:bibun}, we use the relation 
$A^{\lambda}_{\bm{k},\eta\eta_1} = i \hbar\sigma_{3, \eta}(v_{\bm{k},\lambda})_{\eta\eta_1}/ (\bar{\varepsilon}_{\bm{k},\eta_1} - \bar{\varepsilon}_{\bm{k},\eta})$ for $\eta_1 \neq \eta$.
Then, we obtain
\begin{align}
  \partial_{k_{\lambda'}} A^{\lambda}_{\bm{k},\eta\eta_1}
 = 
 \hbar^2 [\sigma_{3, \eta}(v_{\bm{k},\lambda'})_{\eta\eta} - \sigma_{3, \eta_1}(v_{\bm{k},\lambda'})_{\eta_1\eta_1}] \frac{i \sigma_{3, \eta}(v_{\bm{k},\lambda})_{\eta\eta_1}}{(\bar{\varepsilon}_{\bm{k},\eta} - \bar{\varepsilon}_{\bm{k},\eta_1})^2}
  +
   \frac{i\hbar\sigma_{3, \eta}[\partial_{k_{\lambda'}} (v_{\bm{k},\lambda})_{\eta\eta_1}]}{\bar{\varepsilon}_{\bm{k},\eta_1} - \bar{\varepsilon}_{\bm{k},\eta}}.
  \label{appeq:bibun1}
\end{align}
Here, the second term of Eq.~\eqref{appeq:bibun1} is written as
\begin{align}
  \sigma_{3,\eta}\partial_{k_{\lambda'}}(v_{\bm{k},\lambda})_{\eta\eta_1}
  &=
  \sigma_{3,\eta}\braket{\partial_{k_{\lambda'}}u_{\bm{k},\eta}|v_{\bm{k},\lambda}|u_{\bm{k},\eta_1}} + \sigma_{3,\eta}\braket{u_{\bm{k},\eta}|v_{\bm{k},\lambda}|\partial_{k_{\lambda'}}u_{\bm{k},\eta_1}}
  + \frac{1}{\hbar}\sigma_{3,\eta}\braket{u_{\bm{k},\eta}|(\partial_{k_{\lambda}} \partial_{k_{\lambda'}}H_{0,\bm{k}})|u_{\bm{k},\eta_1}} \notag\\
  &=
  \frac{1}{\hbar}\sum_{\eta_2 (\neq \eta,\eta_1)}^{2N}
  (\bar{\varepsilon}_{\bm{k},\eta_1}-\bar{\varepsilon}_{\bm{k},\eta_2}) A^{\lambda'}_{\bm{k},\eta\eta_2} A^{\lambda}_{\bm{k},\eta_2 \eta_1}
  + \frac{1}{\hbar}(\bar{\varepsilon}_{\bm{k},\eta_1}-\bar{\varepsilon}_{\bm{k},\eta}) A^{\lambda'}_{\bm{k},\eta} A^{\lambda}_{\bm{k},\eta \eta_1}
  - \frac{\hbar^2\sigma_{3,\eta}\sigma_{3,\eta_1} (v_{\bm{k},\lambda'})_{\eta\eta_1} (v_{\bm{k},\lambda})_{\eta_1\eta_1}}{\bar{\varepsilon}_{\bm{k},\eta_1}-\bar{\varepsilon}_{\bm{k},\eta}}\notag\\
  &\quad 
  - \frac{1}{\hbar}\sum_{\eta_2 (\neq \eta,\eta_1)}^{2N}
  (\bar{\varepsilon}_{\bm{k},\eta_2}-\bar{\varepsilon}_{\bm{k},\eta}) A^{\lambda}_{\bm{k},\eta\eta_2} A^{\lambda'}_{\bm{k},\eta_2 \eta_1}
  + \frac{ \hbar^2(v_{\bm{k},\lambda})_{\eta\eta} (v_{\bm{k},\lambda'})_{\eta \eta_1}}{\bar{\varepsilon}_{\bm{k},\eta_1}-\bar{\varepsilon}_{\bm{k},\eta}}
  \nonumber\\
  &\quad - \frac{1}{\hbar}(\bar{\varepsilon}_{\bm{k},\eta_1} - \bar{\varepsilon}_{\bm{k},\eta}) A^{\lambda}_{\bm{k},\eta\eta_1} A^{\lambda'}_{\bm{k},\eta_1}
  + \frac{1}{\hbar}\sigma_{3,\eta}\braket{u_\eta|(\partial_{k_{\lambda}} \partial_{k_{\lambda'}}H_{0,\bm{k}})|u_{\eta_1}}.
\end{align}
Substituting this expression into the second term of Eq.~\eqref{appeq:bibun1}, we obtain
\begin{align}\label{appeq:bibun2}
  \partial_{k_{\lambda'}} A^{\lambda}_{\bm{k},\eta\eta_1}
  =&
  \left\{
    i\hbar^2[\sigma_{3,\eta}(v_{\bm{k},\lambda'})_{\eta\eta} - \sigma_{3,\eta_1}(v_{\bm{k},\lambda'})_{\eta_1\eta_1}] \frac{\sigma_{3,\eta}(v_{\bm{k},\lambda})_{\eta\eta_1}}{(\bar{\varepsilon}_{\bm{k},\eta} - \bar{\varepsilon}_{\bm{k},\eta_1})^2}
    + \frac{i}{2} \frac{\sigma_{3,\eta}(\partial_{k_{\lambda}} \partial_{k_{\lambda'}}H_{0,\bm{k}})_{\eta\eta_1}}{\bar{\varepsilon}_{\bm{k},\eta_1}-\bar{\varepsilon}_{\bm{k},\eta}} + (\lambda \leftrightarrow \lambda')
  \right\}
  \notag\\
  &
  + i\sum_{\eta_2 (\neq \eta,\eta_1)}^{2N}
  \frac{\bar{\varepsilon}_{\bm{k},\eta_1}-\bar{\varepsilon}_{\bm{k},\eta_2}}{\bar{\varepsilon}_{\bm{k},\eta_1} - \bar{\varepsilon}_{\bm{k},\eta}}
  A^{\lambda'}_{\bm{k},\eta\eta_2} A^{\lambda}_{\bm{k},\eta_2 \eta_1} 
  - i\sum_{\eta_2 (\neq \eta,\eta_1)}^{2N}
  \frac{\bar{\varepsilon}_{\bm{k},\eta_2}-\bar{\varepsilon}_{\bm{k},\eta}}{\bar{\varepsilon}_{\bm{k},\eta_1}-\bar{\varepsilon}_{\bm{k},\eta}}
  A^{\lambda}_{\bm{k},\eta\eta_2} A^{\lambda'}_{\bm{k},\eta_2 \eta_1} 
  + i A^{\lambda'}_{\bm{k},\eta} A^{\lambda}_{\bm{k},\eta \eta_1}
  - i A^{\lambda}_{\bm{k},\eta\eta_1} A^{\lambda'}_{\bm{k},\eta_1}.
\end{align}

Next, we consider the second term on the right-hand side of Eq.~\eqref{appeq:bibun}.
For $\eta_1 \neq \eta$, we have
\begin{align}
  \sigma_{3,\eta_1}\partial_{k_{\lambda'}} (S_{\bm{k}}^{\gamma})_{\eta_1\eta}
  &=
  i \sum_{\eta_2(\neq \eta,\eta_1)}^{2N}
  A^{\lambda'}_{\bm{k},\eta_1\eta_2} \sigma_{3,\eta_2}(S_{\bm{k}}^{\gamma})_{\eta_2\eta}
  + i A^{\lambda'}_{\bm{k},\eta_1\eta} \sigma_{3,\eta}(S_{\bm{k}}^{\gamma})_{\eta}
  + i A^{\lambda'}_{\bm{k},\eta_1}\sigma_{3,\eta_1} (S_{\bm{k}}^{\gamma})_{\eta_1\eta}
  \notag\\
  &\quad
  - i \sum_{\eta_2(\neq \eta,\eta_1)}^{2N}
  \sigma_{3,\eta_1}(S_{\bm{k}}^{\gamma})_{\eta_1 \eta_2} A^{\lambda'}_{\bm{k},\eta_2\eta} - i \sigma_{3,\eta_1}(S_{\bm{k}}^{\gamma})_{\eta_1\eta} A^{\lambda'}_{\bm{k},\eta}
  - i \sigma_{3,\eta_1}(S_{\bm{k}}^{\gamma})_{\eta_1} A^{\lambda'}_{\bm{k},\eta_1\eta},
\end{align}
Substituting this expression and Eq.~\eqref{appeq:bibun2} into Eq.~\eqref{appeq:bibun}, we obtain Eq.~\eqref{eq:derivative-d} with Eq.~\eqref{eq:Ds-Da}.

\end{widetext}

\section{Response tensors without antisymmetrization}
\label{app:symmetry-tensors}

\begin{figure}[t]
  \centering
      \includegraphics[width=\columnwidth,clip]{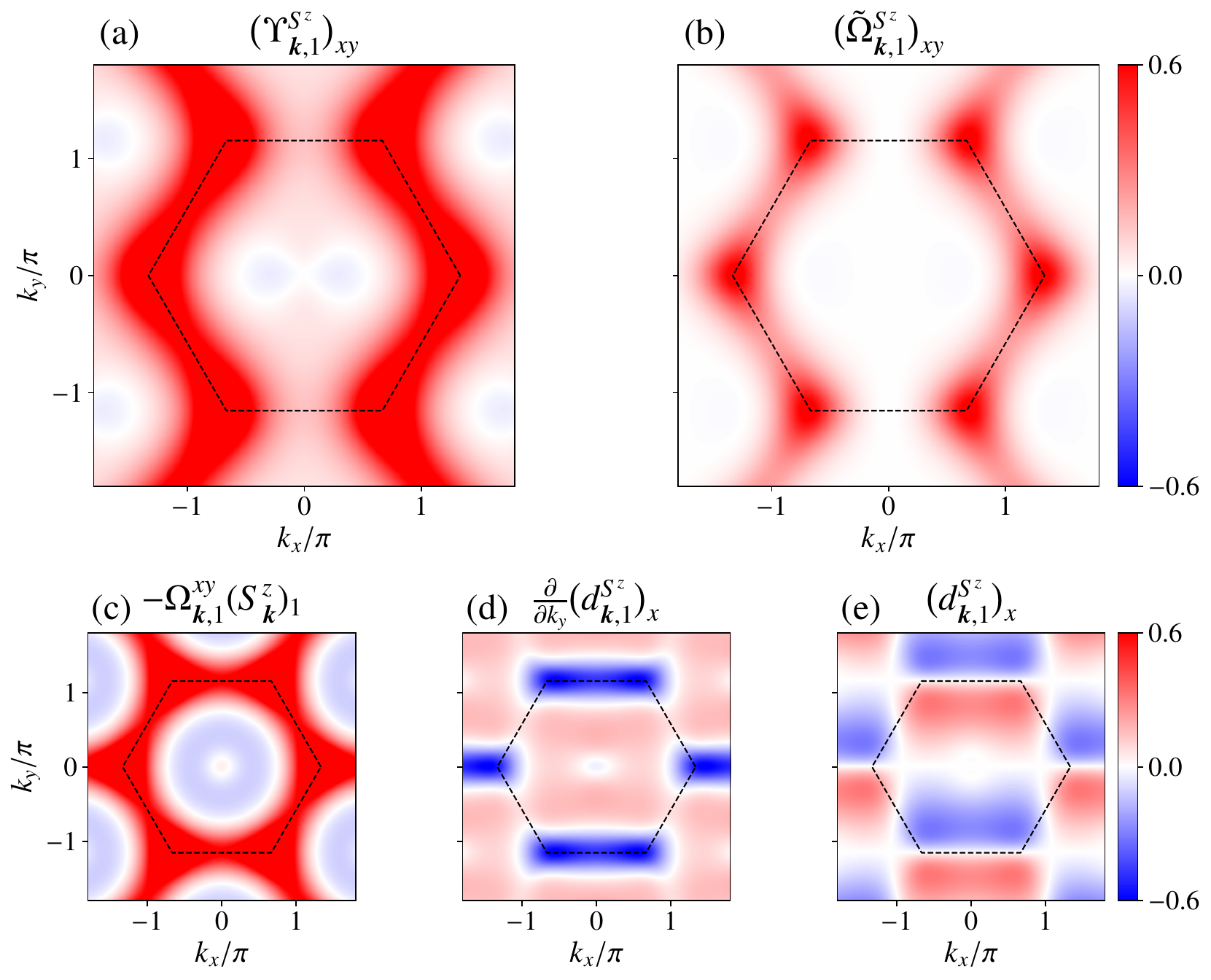}
      % 幅を\columnwidthの係数で指定する
      \caption{
Color maps of (a) $\big(\Upsilon^{S^z}_{\bm{k},\eta}\big)_{xy}$, (b) $\big( \tilde{\Omega}^{S^\gamma}_{\bm{k},\eta} \big)_{xy}$, (c)  $-\Omega^{xy}_{\bm{k},\eta} (S^z_{\bm{k}})_{\eta}$, (d) 
$\pdiff{}{k_{y}} (d^{S^{z}}_{\bm{k},\eta})_{x}$, and (e) $(d^{S^{z}}_{\bm{k},\eta})_{x}$
for the lowest-energy branch with $\eta=1$ in the Kitaev-Heisenberg model at $\varphi=3\pi/2$ under a magnetic field with $h/A = 0.1$.
Panel (c) is the same as that in Fig.~\ref{fig:coef_kitaevheisen_all}.
Dashed lines represent the first Brillouin zone of the honeycomb lattice.
      }
      \label{fig:coef_kitaevheisen_app}
\end{figure}

In the main text, we present the data after antisymmetrization because we focus on the spin Nernst effect.
In this appendix, we show the momentum dependence of $\big(\Upsilon^{S^z}_{\bm{k},\eta}\big)_{xy}$ and $\big( \tilde{\Omega}^{S^\gamma}_{\bm{k},\eta} \big)_{xy}$, which are defined in Eqs.~\eqref{eq:Upsilon} and \eqref{eq:Spin-Berry}, respectively, for reference.
Here, we focus on the results of the Kitaev model corresponding to the case of $\varphi=3\pi/2$ under a magnetic field with $h/A=0.1$, based on the Hamiltonian given in Eq.~\eqref{eq:Kitaev-Heisenberg}.
Figure~\ref{fig:coef_kitaevheisen_app}(a) shows the momentum-space distribution of $\big(\Upsilon^{S^z}_{\bm{k},\eta}\big)_{xy}$ for $\eta=1$, which corresponds to the lower-energy branch in Fig.~\ref{fig:coef_kitaevheisen_all}(a).
As shown in this figure, the negative region located around the $\Gamma$ point appears only along the $k_x$ axis, while it takes positive values along the $k_y$ axis.
We also find that $\big(\Upsilon^{S^z}_{\bm{k},\eta}\big)_{xy}$ takes significantly larger positive values along the Brillouin zone boundary compared to the region along the line at $k_x=0$.

To discuss the origin of the momentum-space distribution of $\big(\Upsilon^{S^z}_{\bm{k},\eta}\big)_{xy}$, we present $-\Omega^{xy}_{\bm{k},\eta} (S^z_{\bm{k}})_{\eta}$ including the Berry curvature and the momentum derivative of the dipole moment $(d^{S^{z}}_{\bm{k},\eta})_{x}$ in Figs.~\ref{fig:coef_kitaevheisen_app}(c) and \ref{fig:coef_kitaevheisen_app}(d), respectively, the sum of which equals to $\big(\Upsilon^{S^z}_{\bm{k},\eta}\big)_{xy}$ as given in Eq.~\eqref{eq:Upsilon}.  
Note that the dipole moment $(d^{S^{z}}_{\bm{k},\eta})_{x}$ corresponds to the quantum metric in the mixed space $(\bm{k},\bm{h}^{\tau})$, as discussed in Sec.~\ref{sec:snc}.  
Although $-\Omega^{xy}_{\bm{k},\eta} (S^z_{\bm{k}})_{\eta}$ exhibits six-fold rotational symmetry, Fig.~\ref{fig:coef_kitaevheisen_app}(d) does not exhibit this symmetry.  
Thus, the absence of six-fold rotational symmetry in $\big(\Upsilon^{S^z}_{\bm{k},\eta}\big)_{xy}$ originates from the momentum derivative of the dipole moment.  
A similar momentum dependence to that of $\big(\Upsilon^{S^z}_{\bm{k},\eta}\big)_{xy}$ is also observed in $\big( \tilde{\Omega}^{S^\gamma}_{\bm{k},\eta} \big)_{xy}$, as shown in Fig.~\ref{fig:coef_kitaevheisen_app}(b).  

Here, we discuss the symmetries of $\big(\Upsilon^{S^z}_{\bm{k},\eta}\big)_{xy}$ and $\big( \tilde{\Omega}^{S^\gamma}_{\bm{k},\eta} \big)_{xy}$ in momentum space.
In the present calculations, we assume that the Bloch Hamiltonian of the system possesses the symmetry $H_{0,\bm{k}} = H_{0,-\bm{k}}$.
This directly leads to the following relations: $\big( \tilde{\Omega}^{S^\gamma}_{\bm{k},\eta} \big)_{xy}=\big( \tilde{\Omega}^{S^\gamma}_{-\bm{k},\eta} \big)_{xy}$, $\Omega^{xy}_{\bm{k},\eta} (S^z_{\bm{k}})_{\eta}=\Omega^{xy}_{-\bm{k},\eta} (S^z_{-\bm{k}})_{\eta}$, and $(d^{S^z}_{\bm{k},\eta})_x = -(d^{S^z}_{-\bm{k},\eta})_x$.
Furthermore, we focus on the successive operations of a mirror reflection with respect to the $zx$ plane and the time-reversal operation.
The former induces the transformation $(x,y,z)\to (x,-y,z)$ in real space, and the two successive operations result in the transformation $(S^x,S^y,S^z)\to (-S^x,S^y,-S^z)\to (S^x,-S^y,S^z)$ in spin space.
Here, the spin Hamiltonian given in Eq.~\eqref{eq:Kitaev-Heisenberg} remains invariant under these two successive operations, and we assume that $\braket{S^y}=0$ for the local mean fields.
In this case, the spin-wave Hamiltonian also preserves this symmetry.
When the components of an $S=1/2$ spin are expressed as Pauli matrices in the conventional manner, the transformation $(S^x,S^y,S^z)\to (S^x,-S^y,S^z)$ in spin space corresponds to the complex conjugation of the real-space Hamiltonian $H_{0,uu'}$ given in Eq.~\eqref{eq:Hamil-A-real} or $H_{0,\bm{\delta}}$ in Eq.~\eqref{appeq:Mdelta} in the continuum limit.
The two successive operations in both real and spin spaces result in the transformation $H_{0,\bm{k}}\to H_{0,(-k_x,k_y)}^*$ for the spin-wave Hamiltonian.
Thus, the Hamiltonian satisfies the relation $H_{0,\bm{k}}= H_{0,(-k_x,k_y)}^*$.
Using this invariant property of the Hamiltonian in momentum space, we obtain $(v_{\bm{k},x})_{\eta\eta'}=-(v_{(-k_x,k_y),x})_{\eta\eta'}$ and $(v_{\bm{k},y})_{\eta\eta'}=(v_{(-k_x,k_y),y})_{\eta\eta'}$, which leads to the relations $\big( \tilde{\Omega}^{S^\gamma}_{\bm{k},\eta} \big)_{xy} = \big( \tilde{\Omega}^{S^\gamma}_{(-k_x,k_y),\eta} \big)_{xy}$, $\Omega^{xy}_{\bm{k},\eta} (S^z_{\bm{k}})_{\eta}=\Omega^{xy}_{(-k_x,k_y),\eta} (S^z_{(-k_x,k_y)})_{\eta}$, and $(d^{S^z}_{\bm{k},\eta})_x = (d^{S^z}_{(-k_x,k_y),\eta})_x$.
From these relations, we conclude that $\big( \tilde{\Omega}^{S^\gamma}_{\bm{k},\eta} \big)_{xy}$ and $\Omega^{xy}_{\bm{k},\eta} (S^z_{\bm{k}})_{\eta}$ are symmetric with respect to the $k_x$ and $k_y$ axes, while $(d^{S^z}_{\bm{k},\eta})_x$ is antisymmetric (symmetric) with respect to the $k_x$ ($k_y$) axis.
We note that, although the dipole moment $\sum_{\bm{k}} \big(d^{S^z}_{\bm{k}, \eta}\big)_{x} = 0$ holds for each branch, its $k_y$ derivative is nonzero, and hence, the second term of Eq.~\eqref{eq:Upsilon} can contribute to $\kappa^{S^z}_{xy}$.

\bibliography{./refs}
\end{document}